\documentclass[12pt]{article}

\usepackage[margin=1in]{geometry}
\usepackage{amsmath,amssymb}
\usepackage{graphicx}
\usepackage{subcaption}
\usepackage{booktabs}
\usepackage{float}
\usepackage{hyperref}
\usepackage{multirow}
\usepackage{natbib}

\title{Deep Learning Models for Fractional Option Pricing}
\author{
Hana Sagor\textsuperscript{1,2} \and
Edward L. Boone\textsuperscript{1} \and
Ryad A. Ghanam\textsuperscript{3}
}
\date{}

\begin{document}

\maketitle
\begin{center}
\small
\textsuperscript{1}Department of Statistical Sciences and Operations Research, Virginia Commonwealth University, Richmond, Virginia, USA\\
\textsuperscript{2}University of Jeddah, Jeddah, Saudi Arabia\\
\textsuperscript{3}VCUarts Qatar, Virginia Commonwealth University, Doha, Qatar\\[0.5em]
\textit{Corresponding author: Hana Sagor (sagorh@vcu.edu)}
\end{center}

\begin{abstract}
This paper investigates whether neural networks can infer fractional option prices and the Hurst parameter directly from sequences of fractional Gaussian returns, without separately estimating the latent parameters for each new sequence. Feedforward neural networks (FNNs) and long short-term memory (LSTM) networks are trained to jointly predict European call prices, put prices, and the Hurst parameter from 252-return sequences. A controlled simulation varies the total sample size from 5,000 to 500,000 sequences and evaluates point prediction, empirical residual-based prediction intervals, and computational performance. The neural-network models are also compared on a common simulated test set with Bayesian joint estimation and exact Gaussian maximum likelihood estimation. In that four-method comparison, the LSTM achieves the lowest Hurst-parameter root mean squared error (0.033) and option-price root mean squared error (1.305). After training, neural-network predictions are substantially faster than repeated parameter estimation, although the methods quantify uncertainty in different ways. Applications to five financial return series illustrate differences between data-driven and model-based fractional price estimates. The results support sequence-based learning as a computationally efficient complement to statistical inference under the assumed fractional model; they do not establish out-of-distribution accuracy or accuracy against observed traded-option prices.
\end{abstract}



\section{Introduction}
\label{sec:introduction}

European option pricing is commonly formulated through a stochastic model for the underlying asset. The classical Black--Scholes framework provides a tractable benchmark, but its Brownian-motion specification imposes independent increments and constant volatility \citep{BlackScholes1973,Merton1973}. Fractional Brownian motion (fBM) offers a distinct dependence structure: its increments, fractional Gaussian noise (fGn), are governed by a Hurst parameter $H$, with $H=0.5$ corresponding to the Brownian case and $H>0.5$ representing persistent increments \citep{Mandelbrot1968,SamorodnitskyTaqqu1994}. Whether such dependence is empirically appropriate depends on the asset and sampling scale; long memory should not be assumed to characterize all financial returns \citep{lo1991long}. Moreover, the persistent-return setting studied here differs from rough-volatility models, which commonly consider a Hurst parameter below one half for volatility rather than returns \citep{gatheral2018volatility}.

Fractional extensions of option-pricing formulas introduce a role for $H$ in the time scaling of the model-implied price \citep{Elliott2003,HuOksendal2003}. These formulas require careful financial interpretation. For $H\neq 0.5$, fBM is not a semimartingale, and frictionless fractional markets can admit arbitrage under some trading assumptions \citep{Rogers1997,Cheridito2003,bjork2005note}. Accordingly, the prices considered in this paper are conditional on the chosen fractional pricing specification, not claims of arbitrage-free market valuations under unrestricted trading.

A central computational difficulty is that $H$ and volatility are generally unknown. Prior work by Sagor and coauthors investigated simulation-based bias correction for method-of-moments estimation of $H$ and Bayesian joint estimation of $H$ and volatility, with uncertainty propagated to fractional option prices \citep{Sagor2025,Sagor2026}. In these approaches, each new return sequence ordinarily requires statistical parameter estimation before pricing. When many sequences must be processed, the repeated estimation stage can be costly.

Neural networks offer an alternative computational route. Early studies established their potential for approximating option-pricing relationships \citep{hutchinson1994nonparametric,garcia2000pricing}; later work developed deep-learning pricing and surrogate approaches \citep{ruf2020neural,liu2019pricing,chen2026deep}. Feedforward neural networks (FNNs) can learn nonlinear maps from fixed-dimensional inputs, while long short-term memory (LSTM) networks explicitly process ordered observations using recurrent memory mechanisms \citep{cybenko1989approximation,hochreiter1997long}. Recent option-pricing applications have examined sequence-based models, including LSTMs \citep{pimentel2026option,fritzsch2026predicting}. Fractional-derivative neural pricing equations have also been studied, although that formulation differs from the fBM/fGn return-dependence model used here \citep{song2024calculations}.

The present paper examines a different input--output problem from the common practice of feeding known volatility or other pricing parameters directly into a network. Each network receives only a sequence of 252 simulated fGn returns as its stochastic input and jointly predicts a European call price, a European put price, and $H$. The volatility scale and temporal dependence must therefore be inferred implicitly from the return sequence. The FNN treats the sequence as a vector; the LSTM processes its temporal order. Both architectures are trained and evaluated using common simulated datasets and a consistent prediction protocol.

The paper makes four contributions. First, it formulates fractional option pricing as a direct sequence-to-output prediction task, avoiding a separate explicit estimation step at prediction time. Second, it compares FNN and LSTM architectures as the total simulated dataset increases from 5,000 to 500,000 sequences, thereby documenting the accuracy--sample-size relationship. Third, it compares neural-network predictions with Bayesian joint estimation and exact Gaussian maximum likelihood on a common simulated test set, considering prediction accuracy, interval coverage and width, and computational cost. Fourth, it examines how the methods behave on five empirical financial return series, while distinguishing comparisons of model-implied prices from validation against observed traded-option prices.

The purpose is not to establish that neural networks universally outperform statistical estimators. Bayesian and likelihood-based approaches retain explicit parameter interpretation and their own inferential foundations; neural-network residual-based intervals have a different interpretation from posterior credible intervals and likelihood-based confidence intervals. In addition, the controlled training and testing design does not establish performance beyond the simulated parameter ranges, fixed option contract, or assumed fGn family. The contribution is instead to characterize the conditions under which direct sequence-based prediction offers an accurate and computationally efficient complement to model-based estimation.

The remainder of the paper presents the fractional simulation and pricing design, describes the FNN and LSTM architectures and statistical benchmarks, reports simulation and uncertainty results, and examines the empirical applications before discussing limitations and directions for further research.

\section{Simulation Study}
\label{sec:simulation_study}

In order to better understand the properties and predictive
performance of the proposed neural-network approaches, a
simulation study is conducted. By generating data under known
fractional model parameters, the study provides a controlled
framework for evaluating the accuracy of the predicted option
prices and Hurst parameter. Specifically, the simulation study
aims to address the following research questions:

\begin{enumerate}
    \item How accurately can the feedforward neural network
    (FNN) and long short-term memory (LSTM) models predict
    European call prices, put prices, and the Hurst parameter?

    \item How does the predictive performance of the FNN
    and LSTM models compare?

    \item How does the training sample size affect the
    prediction accuracy of the neural-network models?

    \item How do the neural-network approaches compare
    with traditional statistical estimation methods,
    including Bayesian estimation and maximum likelihood
    estimation (MLE)?
\end{enumerate}

\subsection{Experimental Design}
\label{sec:experimental_design}

This section describes the experimental framework used to evaluate the
feedforward neural network (FNN) and long short-term memory (LSTM)
models. Both models were trained and evaluated using the same simulated
datasets, data partitions, target variables, and performance measures.
This common experimental framework provides a consistent basis for
comparison between the two architectures. 

\subsubsection{Prediction Task}
\label{subsec:prediction_task}

The objective of the deep learning models is to learn the nonlinear
relationship between a sequence of simulated returns and three target
quantities associated with the fractional Black--Scholes model. For each
observation, the input consists of a sequence of \(252\) fractional
Gaussian noise returns,

\[
\mathbf{x}_i =
\left(
r_{i,1}, r_{i,2}, \ldots, r_{i,252}
\right),
\]

where \(i\) denotes the observation index. The corresponding output
vector is

\[
\mathbf{y}_i =
\left(
C_i, P_i, H_i
\right),
\]

where \(C_i\) is the fractional call option price, \(P_i\) is the
fractional put option price, and \(H_i\) is the Hurst parameter.

Accordingly, both the FNN and LSTM models perform a multivariate
regression task in which the \(252\)-dimensional return sequence is
mapped simultaneously to the call price, put price, and Hurst
parameter. Although volatility is an underlying determinant of the
simulated option prices, it is not included as a separate
neural-network output. Instead, the FNN and LSTM learn the pricing
mapping directly from the complete return sequence and therefore
extract information about both volatility and temporal dependence
implicitly from the observed returns. Consequently, the neural-network
models do not require separate estimation of \(H\) and \(\sigma\)
before generating the corresponding option-price predictions.

This direct sequence-to-output formulation differs from the Bayesian
and exact Gaussian maximum likelihood approaches considered later in
the four-method comparison. Those procedures explicitly estimate the
parameters of the fractional Gaussian model and subsequently propagate
the estimated \(H\) and volatility through the fractional
Black--Scholes pricing model. The neural-network models instead learn
this mapping during training, allowing option prices and the Hurst
parameter to be obtained directly from a new return sequence through a
single forward pass.

\subsubsection{Simulation Framework}
\label{subsec:simulation_framework}

For each simulated observation, the Hurst parameter, volatility, and drift
were independently generated according to

\[
H \sim U(0.50,0.95), \qquad
\sigma \sim U(0.10,1.00), \qquad
\mu \sim U(-0.05,0.05).
\]

For a given parameter combination, a fractional Gaussian noise sequence
of length \(252\) was generated using the Davies--Harte algorithm
\cite{davies1987tests}. The simulated daily return process was constructed as

\[
r_t =
\frac{\mu}{252}
+
\sigma
\left(\frac{1}{252}\right)^H X_t,
\]

where \(X_t\) denotes unit-variance fractional Gaussian noise with Hurst
parameter \(H\), and \(\sigma\) represents the volatility scale used in
the fractional return model.

Under this specification, the covariance matrix of the daily return
sequence is

\[
\operatorname{Cov}(\mathbf{r})
=
\sigma^2
\left(\frac{1}{252}\right)^{2H}
R(H),
\]

where \(R(H)\) denotes the fractional Gaussian noise correlation matrix.
Equivalently, if the return process is parameterized in terms of a
daily fGn scale,

\[
\sigma_{\mathrm{daily}}
=
\sigma\,252^{-H},
\]

then the volatility scale used in the simulation and fractional
Black--Scholes pricing model is recovered as

\[
\sigma
=
\sigma_{\mathrm{daily}}\,252^H.
\]

This fractional scaling follows from the self-similarity properties of
fractional Brownian motion and the corresponding scaling of its
increments \cite{Mandelbrot1968,SamorodnitskyTaqqu1994}.

This fractional scaling is important because the usual
\(\sqrt{252}\) annualization is obtained only in the special case
\(H=0.5\). The Bayesian and exact Gaussian MLE procedures used later in
the comparison estimate volatility on the daily fGn scale and then
apply the transformation
\(\sigma=\sigma_{\mathrm{daily}}252^H\) before computing fractional
option prices, ensuring consistency with the data-generating process.

The Davies--Harte algorithm was adopted for the large-scale simulation
study because it permits efficient generation of fractional Gaussian
noise sequences while preserving the covariance structure implied by
the specified Hurst parameter \cite{davies1987tests}. This computational efficiency was particularly
important for constructing the master dataset of \(500{,}000\)
observations.

For each simulated return sequence, the corresponding European call and
put option prices were calculated using the fractional Black--Scholes
pricing framework with fixed option parameters \cite{Elliott2003,HuOksendal2003,Sagor2025,Sagor2026}.

\[
S_0=100,\qquad
K=98,\qquad
r=0.05,\qquad
T=\frac{90}{365}.
\]

Thus, each simulated observation consists of a 252-day return sequence
together with its associated call price, put price, and true Hurst
parameter. Because the option parameters are held fixed across the
simulation, variation in the target option prices arises from variation
in the simulated fractional return dynamics, particularly the volatility
scale and Hurst parameter.

\subsubsection{Simulation Sample Sizes}
\label{subsec:sample_sizes}

To examine the effect of the amount of simulated data on predictive
performance, six nested experiments were conducted using total sample
sizes of

\[
N \in
\left\{
5{,}000,\,
25{,}000,\,
50{,}000,\,
100{,}000,\,
200{,}000,\,
500{,}000
\right\}.
\]

Here, \(N\) denotes the total number of simulated observations in each
experiment rather than the number of observations used exclusively for
model training. A single master dataset containing \(500{,}000\)
observations was generated, and the smaller experimental datasets were
constructed as nested subsets of this master dataset. This design allows
the effect of increasing sample size to be examined under a common
simulation framework while maintaining consistent data generation across
experiments.

The use of six sample sizes makes it possible to assess whether predictive
accuracy continues to improve beyond \(N=100{,}000\), and whether the
additional gains diminish as the total simulated sample size increases.

\subsubsection{Training, Validation, and Test Sets}
\label{subsec:data_partition}

For each sample size, the data were partitioned into training,
validation, and test sets. The proportions used were

\[
64\% \text{ training}, \qquad
16\% \text{ validation}, \qquad
20\% \text{ testing}.
\]

Therefore, for a dataset containing \(N\) observations, the approximate
numbers of observations in each partition were

\[
N_{\mathrm{train}} = 0.64N,
\qquad
N_{\mathrm{validation}} = 0.16N,
\qquad
N_{\mathrm{test}} = 0.20N.
\]

Table~\ref{tab:data_partitions} presents the resulting partition sizes
for the six experiments.

\begin{table}[H]
\centering
\caption{Training, validation, and test sample sizes used in the
experiments.}
\label{tab:data_partitions}
\begin{tabular}{rrrr}
\toprule
Total Sample Size & Training Set & Validation Set & Test Set \\
\midrule
5,000   & 3,200   & 800    & 1,000   \\
25,000  & 16,000  & 4,000  & 5,000   \\
50,000  & 32,000  & 8,000  & 10,000  \\
100,000 & 64,000  & 16,000 & 20,000  \\
200,000 & 128,000 & 32,000 & 40,000  \\
500,000 & 320,000 & 80,000 & 100,000 \\
\bottomrule
\end{tabular}
\end{table}

The training set was used to estimate the neural network parameters. The
validation set was used to monitor generalization performance during
training and to determine when training should be stopped. The test set
was kept separate from the model-fitting process and was used only for
the final performance evaluation.

The same data partitions were used for the FNN and LSTM models within
each experiment. Consequently, differences in test performance can be
attributed primarily to the model architectures rather than to
differences in the observations used for training or evaluation.

\subsubsection{Data Preprocessing}
\label{subsec:data_preprocessing}

The input returns and output variables have different numerical scales.
Consequently, data standardization was applied before model training.
For a variable \(z\), the standardized value was calculated as

\[
z_i^{*}
=
\frac{z_i-\bar{z}_{\mathrm{train}}}
{s_{\mathrm{train}}},
\]

where \(\bar{z}_{\mathrm{train}}\) and \(s_{\mathrm{train}}\) denote the
mean and standard deviation computed using only the training set.

The training-set means and standard deviations were subsequently used
to transform the validation and test sets. This procedure prevents
information from the validation or test observations from entering the
model-training process.

For the FNN model, each return sequence was represented as a
\(252\)-dimensional input vector. For the LSTM model, the same input was
reshaped as a sequence with \(252\) time steps and one feature per time
step. Thus, the LSTM input had dimensions

\[
N \times 252 \times 1.
\]

After prediction, the standardized model outputs were transformed back
to their original scales before the performance measures were
calculated.


\subsection{Model Training and Architectures}

This subsection describes the training procedures and neural-network
architectures used to predict European call prices, put prices, and
the Hurst parameter from simulated fractional Gaussian return
sequences. It first outlines the model training procedure and then
presents the feedforward neural network (FNN) and long short-term
memory (LSTM) architectures used in the simulation study.

\subsubsection{Model Training Procedure}
\label{subsec:training_procedure}

The FNN and LSTM models were trained using the Adam optimization
algorithm \cite{KingmaBa2015} with an initial learning rate of

\[
\eta = 0.001.
\]

The mean squared error was used as the training loss function because
the problem involves the simultaneous prediction of three continuous
output variables. For a batch containing \(m\) observations, the loss
function can be expressed as

\[
\mathcal{L}
=
\frac{1}{3m}
\sum_{i=1}^{m}
\sum_{j=1}^{3}
\left(
y_{ij}^{*}-\widehat{y}_{ij}^{*}
\right)^2,
\]

where \(y_{ij}^{*}\) and \(\widehat{y}_{ij}^{*}\) are the standardized
true and predicted values, respectively.

Training was permitted to continue for a maximum of \(250\) epochs,
although the actual number of completed epochs was determined by the
early-stopping criterion.
The mean absolute error was also monitored during training. To reduce overfitting, dropout regularization \cite{Srivastava2014} and validation-based callbacks were employed. Early stopping terminated
training when the validation loss failed to improve for a specified
number of epochs. A learning-rate reduction callback decreased the
learning rate when validation performance reached a plateau.

In addition, a model-checkpoint procedure saved the parameter values
associated with the lowest validation loss. The best-performing model,
rather than the model obtained during the final training epoch, was
therefore used to generate the test-set predictions.

The same optimizer, initial learning rate, loss function, data
partitions, and evaluation procedure were applied to both architectures.
However, the number of completed epochs could differ between models and
sample sizes because early stopping was determined separately for each
experiment.

\subsubsection{Neural Network Architectures}
\label{subsec:model_architectures}

Two neural network architectures were considered: a feedforward neural
network (FNN) and a long short-term memory (LSTM) network. Both models
were designed to map a sequence of \(252\) simulated daily returns to
three continuous outputs: the call option price, the put option price,
and the Hurst parameter.

The FNN treated the \(252\) returns as a fixed-dimensional input vector,
consistent with the standard feedforward function-approximation
framework \cite{hornik1989multilayer}.
Its architecture consisted of three fully connected hidden layers with
\(128\), \(64\), and \(32\) units, respectively. Dropout with a rate of
\(0.20\) was applied after the first and second hidden layers. Rectified
linear unit (ReLU) activation functions \cite{NairHinton2010} were used in the hidden layers,
and the final output layer contained three units with linear activation.

In contrast, the LSTM model treated each return sequence as an ordered
time series of \(252\) time steps with one feature per time step,
following the recurrent memory architecture introduced by
Hochreiter and Schmidhuber \cite{hochreiter1997long}. The
network consisted of an LSTM layer with \(64\) units, followed by
dropout with a rate of \(0.20\), a fully connected layer with \(32\)
ReLU units, a second dropout layer with a rate of \(0.20\), and a
three-unit output layer with linear activation.

The use of these two architectures permits comparison between a
feedforward model that processes the complete return sequence as a
fixed input vector and a recurrent model that explicitly processes the
returns according to their temporal ordering. The same three prediction
targets and common experimental framework were used for both
architectures to facilitate direct comparison of their predictive
performance.

The FNN contained \(42{,}819\) trainable parameters, whereas the LSTM
contained \(19{,}075\) trainable parameters. Thus, the LSTM used
approximately \(55\%\) fewer trainable parameters than the FNN despite
explicitly modeling the sequential structure of the return series.
This difference provides an additional basis for comparing the two
architectures in terms of predictive performance and parameter
efficiency.

\subsection{Performance Measures}
\label{subsec:performance_measures}

The predictive performance of each model was evaluated separately for
the call option price, put option price, and Hurst parameter. Four
performance measures were considered: root mean squared error, mean
absolute error, Pearson correlation, and the coefficient of
determination.


The root mean squared error is defined as

\[
\mathrm{RMSE}
=
\sqrt{
\frac{1}{n_{\mathrm{test}}}
\sum_{i=1}^{n_{\mathrm{test}}}
\left(
y_i-\widehat{y}_i
\right)^2
}.
\]

The RMSE assigns greater weight to large prediction errors. Lower RMSE
values indicate better predictive performance.


The mean absolute error is given by

\[
\mathrm{MAE}
=
\frac{1}{n_{\mathrm{test}}}
\sum_{i=1}^{n_{\mathrm{test}}}
\left|
y_i-\widehat{y}_i
\right|.
\]

The MAE represents the average absolute difference between the true and
predicted values. As with the RMSE, lower values indicate more accurate
predictions.


The Pearson correlation coefficient between the true and predicted
values is defined as

\[
\mathrm{Cor}(y,\widehat{y})
=
\frac{
\sum_{i=1}^{n_{\mathrm{test}}}
(y_i-\bar{y})
(\widehat{y}_i-\overline{\widehat{y}})
}{
\sqrt{
\sum_{i=1}^{n_{\mathrm{test}}}
(y_i-\bar{y})^2
}
\sqrt{
\sum_{i=1}^{n_{\mathrm{test}}}
(\widehat{y}_i-\overline{\widehat{y}})^2
}
}.
\]

A correlation close to one indicates a strong positive linear
association between the true and predicted quantities.


The coefficient of determination is calculated as

\[
R^2
=
1-
\frac{
\sum_{i=1}^{n_{\mathrm{test}}}
(y_i-\widehat{y}_i)^2
}{
\sum_{i=1}^{n_{\mathrm{test}}}
(y_i-\bar{y})^2
}.
\]

The \(R^2\) value measures the proportion of test-set variation
explained by the model. Values closer to one indicate stronger
predictive performance.

\subsection{Diagnostic Evaluation}
\label{subsec:diagnostic_evaluation}

In addition to the numerical performance measures, several graphical
diagnostics were examined. Training and validation loss curves were
used to evaluate convergence and detect potential overfitting.
Scatterplots of true versus predicted values were used to assess the
agreement between the model predictions and the simulated targets.

Residuals were calculated as

\[
e_i = y_i-\widehat{y}_i.
\]

Residual plots were examined to identify systematic prediction patterns,
changes in residual variability, or possible model bias. Histograms of
the residuals were also considered to evaluate their concentration,
symmetry, and centering around zero.

To avoid excessive repetition, numerical performance measures are reported
for all six sample sizes, whereas detailed graphical diagnostics are
presented primarily for the final experiment with \(N=500{,}000\).
This experiment contains \(320{,}000\) training observations,
\(80{,}000\) validation observations, and an independent test set of
\(100{,}000\) observations, and represents the final neural-network models
used in the subsequent empirical applications.

\subsection{Model Comparison Strategy}
\label{subsec:comparison_strategy}

The simulation analysis was conducted in two stages. First, the FNN and
LSTM models were evaluated across all six simulated sample sizes using
test-set RMSE, MAE, correlation, and \(R^2\). This analysis was used to
quantify the effect of increasing the amount of simulated training data
and to compare the predictive behavior of the two neural-network
architectures.

Second, the neural-network approaches were compared with Bayesian and
exact Gaussian maximum-likelihood estimation (MLE) under a common
simulation setting. The statistical methods provide model-based
benchmarks against which the predictive accuracy of the FNN and LSTM can
be assessed for estimation of the Hurst parameter and the resulting
fractional option prices.

Computational performance was also considered. For the neural-network
methods, training cost and post-training prediction time were distinguished,
since model training is performed once whereas subsequent prediction can
be carried out rapidly. For the Bayesian and MLE approaches, estimation
must be performed for each return sequence. The comparison therefore
examines both statistical accuracy and computational efficiency.

All simulations, statistical estimation, neural-network training, and
performance evaluations were implemented in R version 4.5.1. The
neural-network models were fitted using the \texttt{keras3} package
(version 1.5.1) with TensorFlow version 2.20.0 as the computational
backend. All computations were performed on a 13-inch MacBook Air
(2025) equipped with an Apple M4 processor and 24 GB of memory, running
macOS Sequoia 15.7.4. Computational times reported in this paper
therefore correspond to this specific software and hardware environment
and are intended primarily for within-study comparison rather than as
hardware-independent benchmarks.

Together, these analyses assess whether the neural-network models can
provide competitive approximation to the model-based statistical
procedures while substantially reducing the computational cost associated
with repeated estimation and option pricing.

\subsection{Simulation Results}
\label{sec:simulation_results}

This subsection presents the results of the simulation study
and evaluates the predictive performance of the feedforward
neural network (FNN) and long short-term memory (LSTM) models.
The results are examined in terms of their ability to predict
European call prices, put prices, and the Hurst parameter.
Prediction accuracy is assessed using the selected performance
metrics, with particular attention to the effect of sample
size and the differences between the two architectures.

\subsubsection{Feedforward Neural Network Results}
\label{subsec:fnn_results}

The predictive performance of the feedforward neural network (FNN) was
evaluated across total simulated sample sizes of
\(N=5{,}000\), \(25{,}000\), \(50{,}000\), \(100{,}000\),
\(200{,}000\), and \(500{,}000\). Performance was assessed separately
for the call price, put price, and Hurst parameter using the test-set
root mean squared error (RMSE), mean absolute error (MAE), correlation,
and coefficient of determination (\(R^2\)). In addition to point
prediction, empirical residual-based prediction intervals were examined to assess
predictive uncertainty.

\paragraph{Test-Set Predictive Performance.}

Table~\ref{tab:fnn_test_performance} reports the test-set predictive
performance of the FNN across the six total simulated sample sizes.
Increasing the amount of simulated data generally improves predictive
accuracy for all three outputs.

\begin{table}[H]
\centering
\caption{Test-set prediction performance of the FNN model across total simulated sample sizes.}
\label{tab:fnn_test_performance}
\begin{tabular}{clcccc}
\toprule
Sample Size & Quantity & RMSE & MAE & Correlation & $R^2$ \\
\midrule

\multirow{3}{*}{5,000}
& Call Price      & 2.205 & 1.748 & 0.842 & 0.700 \\
& Put Price       & 2.206 & 1.747 & 0.843 & 0.700 \\
& Hurst Parameter & 0.070 & 0.056 & 0.848 & 0.712 \\
\midrule

\multirow{3}{*}{25,000}
& Call Price      & 1.824 & 1.419 & 0.896 & 0.788 \\
& Put Price       & 1.827 & 1.421 & 0.896 & 0.787 \\
& Hurst Parameter & 0.056 & 0.045 & 0.906 & 0.813 \\
\midrule

\multirow{3}{*}{50,000}
& Call Price      & 1.769 & 1.357 & 0.906 & 0.800 \\
& Put Price       & 1.770 & 1.358 & 0.906 & 0.800 \\
& Hurst Parameter & 0.051 & 0.040 & 0.925 & 0.848 \\
\midrule

\multirow{3}{*}{100,000}
& Call Price      & 1.703 & 1.319 & 0.915 & 0.813 \\
& Put Price       & 1.703 & 1.318 & 0.915 & 0.813 \\
& Hurst Parameter & 0.047 & 0.037 & 0.934 & 0.869 \\
\midrule

\multirow{3}{*}{200,000}
& Call Price      & 1.720 & 1.305 & 0.922 & 0.809 \\
& Put Price       & 1.720 & 1.305 & 0.922 & 0.809 \\
& Hurst Parameter & 0.043 & 0.034 & 0.944 & 0.888 \\
\midrule

\multirow{3}{*}{500,000}
& Call Price      & \textbf{1.623} & \textbf{1.258} & \textbf{0.927} & \textbf{0.831} \\
& Put Price       & \textbf{1.623} & \textbf{1.258} & \textbf{0.927} & \textbf{0.831} \\
& Hurst Parameter & \textbf{0.043} & \textbf{0.034} & \textbf{0.947} & \textbf{0.893} \\
\bottomrule
\end{tabular}
\end{table}

For the call price, the RMSE decreases from \(2.205\) at
\(N=5{,}000\) to \(1.623\) at \(N=500{,}000\), while the MAE
decreases from \(1.748\) to \(1.258\). The put-price results are
nearly identical, with the RMSE decreasing from \(2.206\) to
\(1.623\) and the MAE decreasing from \(1.747\) to \(1.258\).
The improvement is also reflected in the correlation and \(R^2\).
For the call price, correlation increases from \(0.842\) to
\(0.927\), while \(R^2\) increases from \(0.700\) to \(0.831\).

The FNN also exhibits substantial improvement in predicting the Hurst
parameter. Its RMSE decreases from \(0.0699\) at \(N=5{,}000\) to
\(0.0425\) at \(N=500{,}000\), while its MAE decreases from
\(0.0556\) to \(0.0339\). At the same time, correlation increases
from \(0.8480\) to \(0.9472\), and \(R^2\) increases from \(0.7121\)
to \(0.8927\). These results indicate that the FNN becomes increasingly
effective at extracting information about the Hurst parameter directly
from the simulated return sequences as the amount of training data
increases.

Although the overall trend is favorable, the improvement is not
strictly monotonic for every metric. In particular, the call- and
put-price RMSE values increase slightly between \(N=100{,}000\) and
\(N=200{,}000\), from approximately \(1.703\) to \(1.720\), before
declining to their minimum values at \(N=500{,}000\). In contrast,
the corresponding MAE values continue to decrease over this range.
Thus, the small RMSE fluctuation at \(N=200{,}000\) does not alter
the overall improvement observed as the simulated sample size
increases.

\begin{figure}[htbp]
    \centering

    \begin{subfigure}[t]{0.48\textwidth}
        \centering
        \includegraphics[width=\textwidth]
        {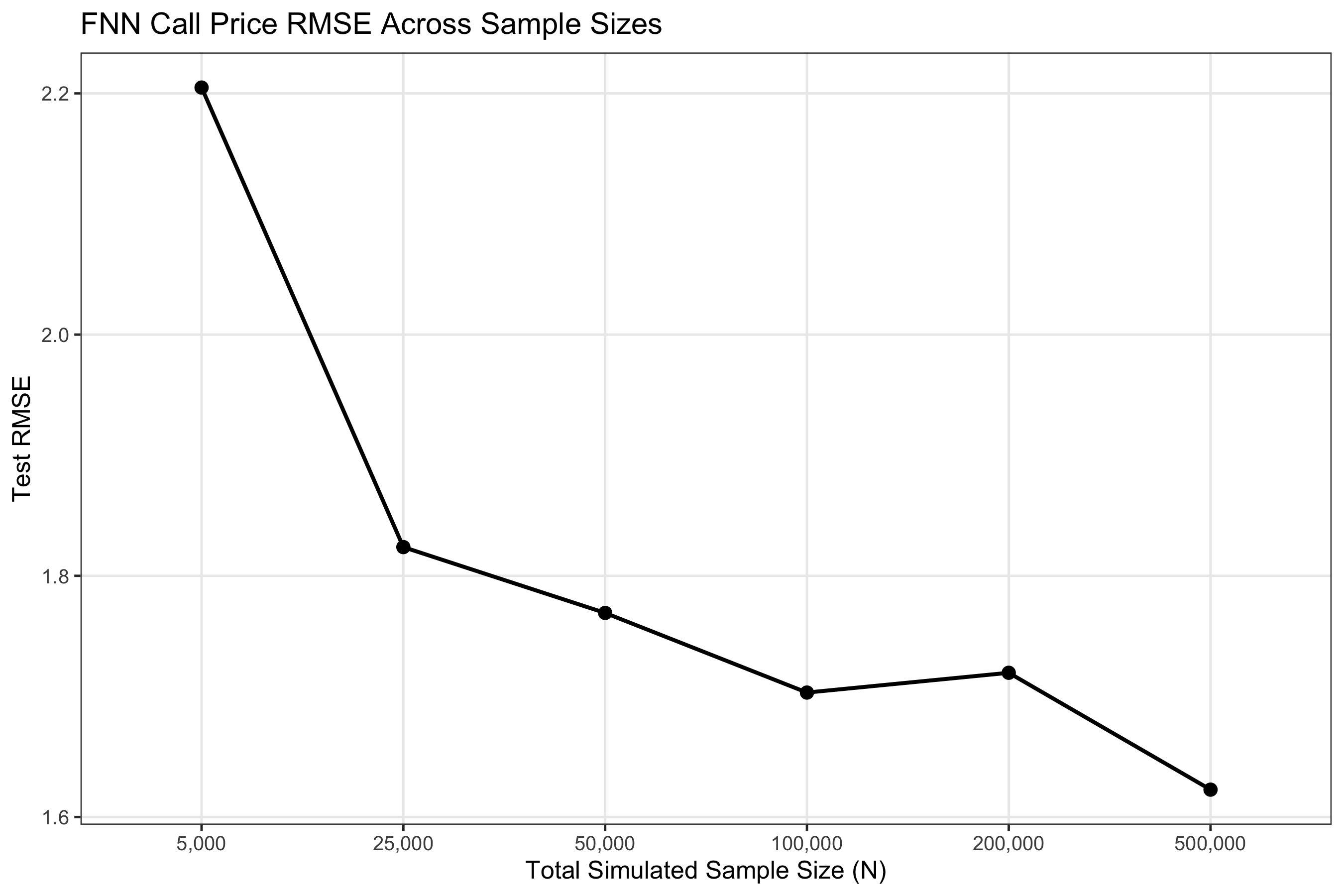}
        \caption{Call-price RMSE.}
    \end{subfigure}
    \hfill
    \begin{subfigure}[t]{0.48\textwidth}
        \centering
        \includegraphics[width=\textwidth]
        {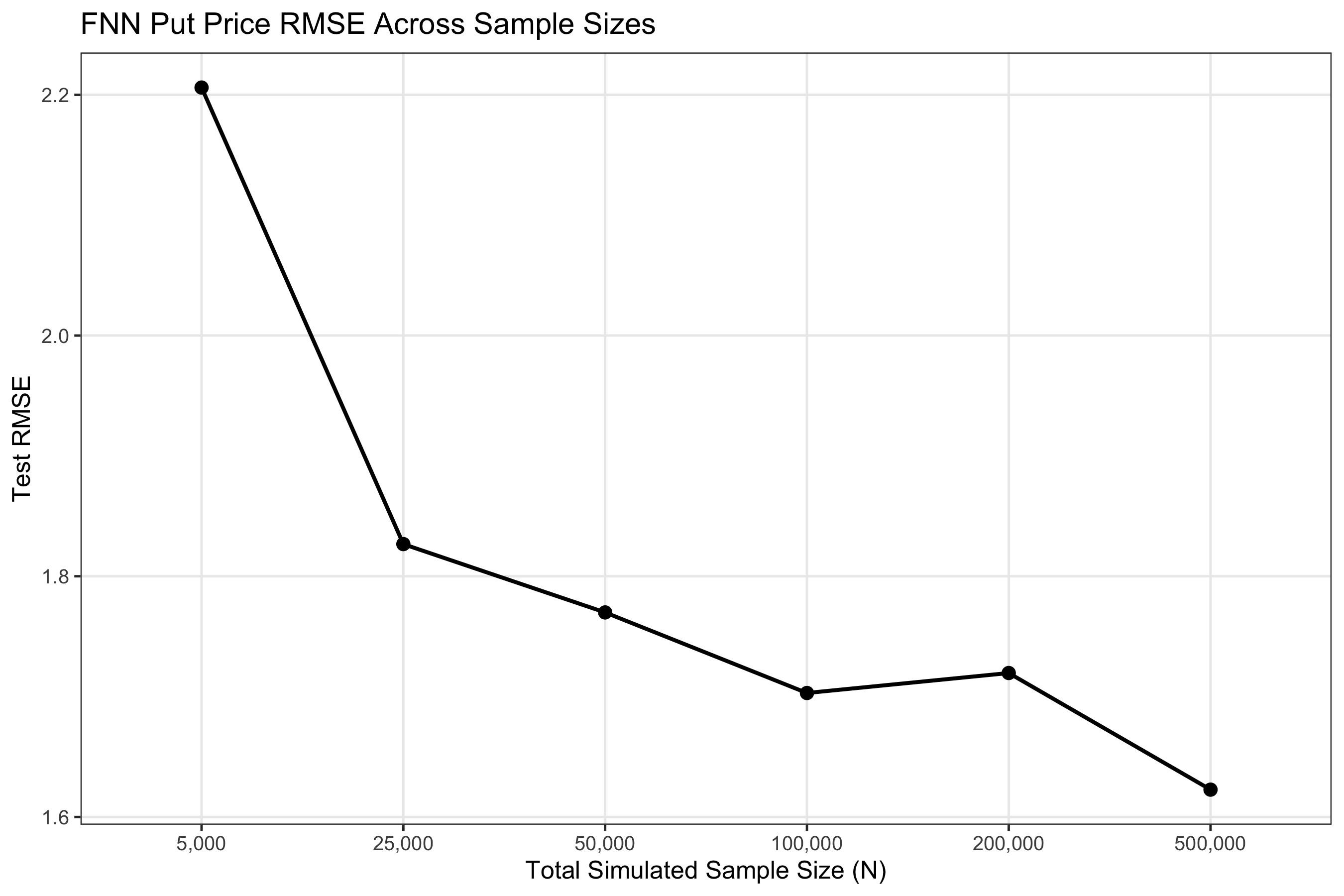}
        \caption{Put-price RMSE.}
    \end{subfigure}

    \vspace{0.3cm}

    \begin{subfigure}[t]{0.48\textwidth}
        \centering
        \includegraphics[width=\textwidth]
        {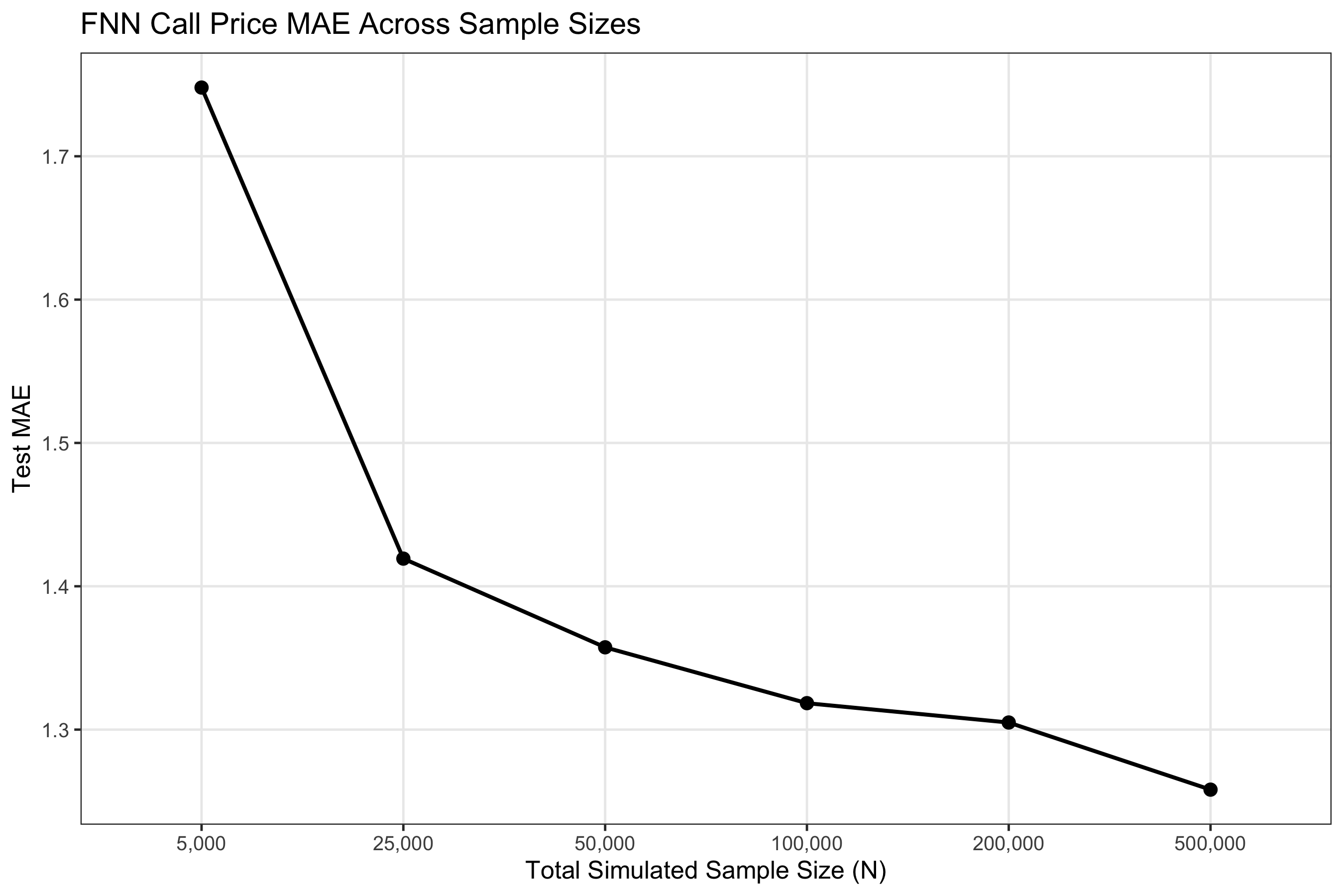}
        \caption{Call-price MAE.}
    \end{subfigure}
    \hfill
    \begin{subfigure}[t]{0.48\textwidth}
        \centering
        \includegraphics[width=\textwidth]
        {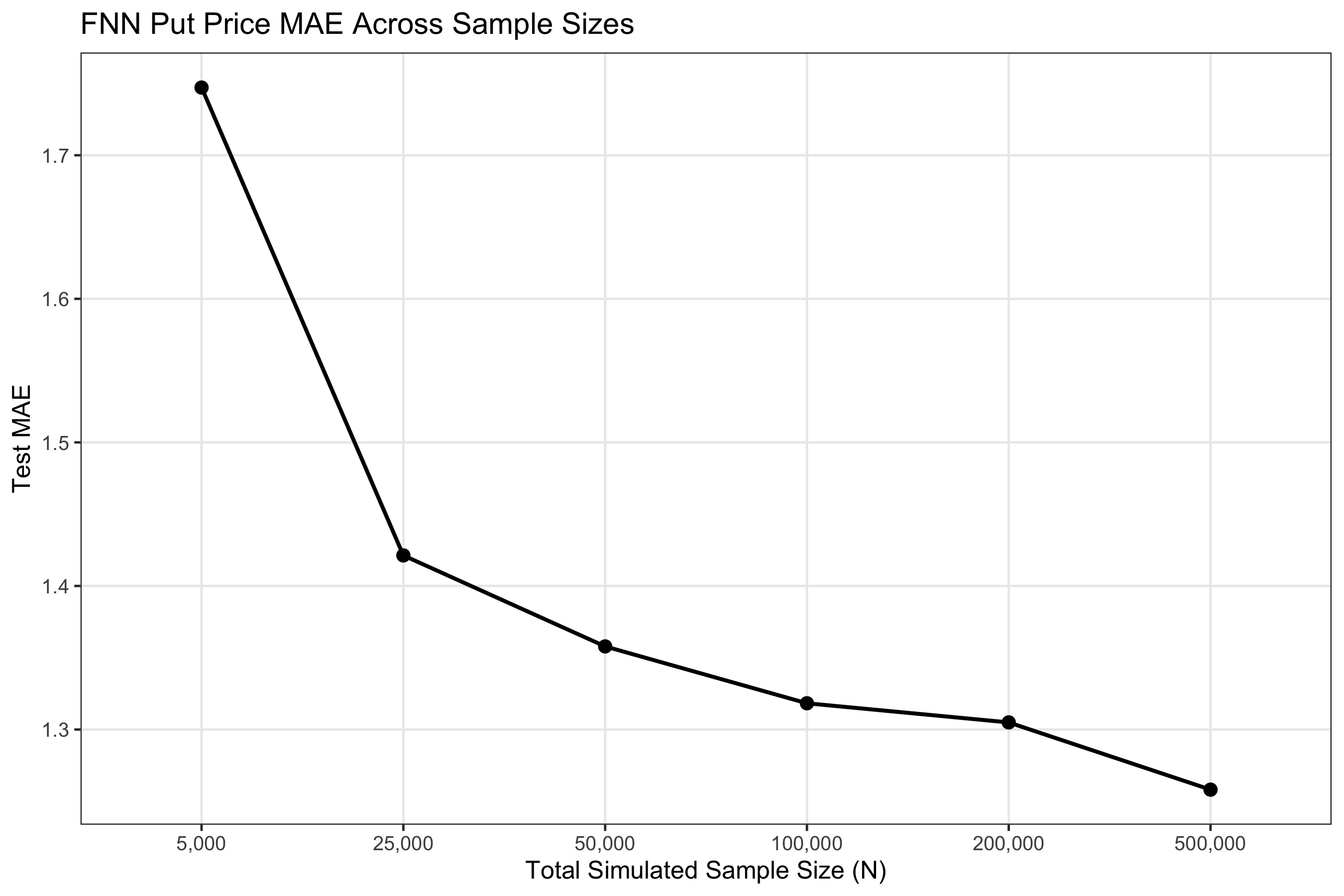}
        \caption{Put-price MAE.}
    \end{subfigure}

\caption{FNN test-set prediction errors for European call and
put prices across total simulated sample sizes. The upper
panels show root mean squared error (RMSE), while the lower
panels show mean absolute error (MAE). Each plotted point
is calculated using 20\% of the corresponding simulated
dataset as test observations, representing 1,000, 5,000,
10,000, 20,000, 40,000, and 100,000 observations for
total sample sizes of 5,000, 25,000, 50,000, 100,000,
200,000, and 500,000, respectively.}
\label{fig:fnn_error_sample_size}
\end{figure}

Figure~\ref{fig:fnn_error_sample_size} provides a visual summary of
the option-price errors. The largest improvement occurs between the
smaller sample sizes, while subsequent reductions are more gradual.

\begin{figure}[htbp]
    \centering

    \begin{subfigure}[t]{0.48\textwidth}
        \centering
        \includegraphics[width=\textwidth]
        {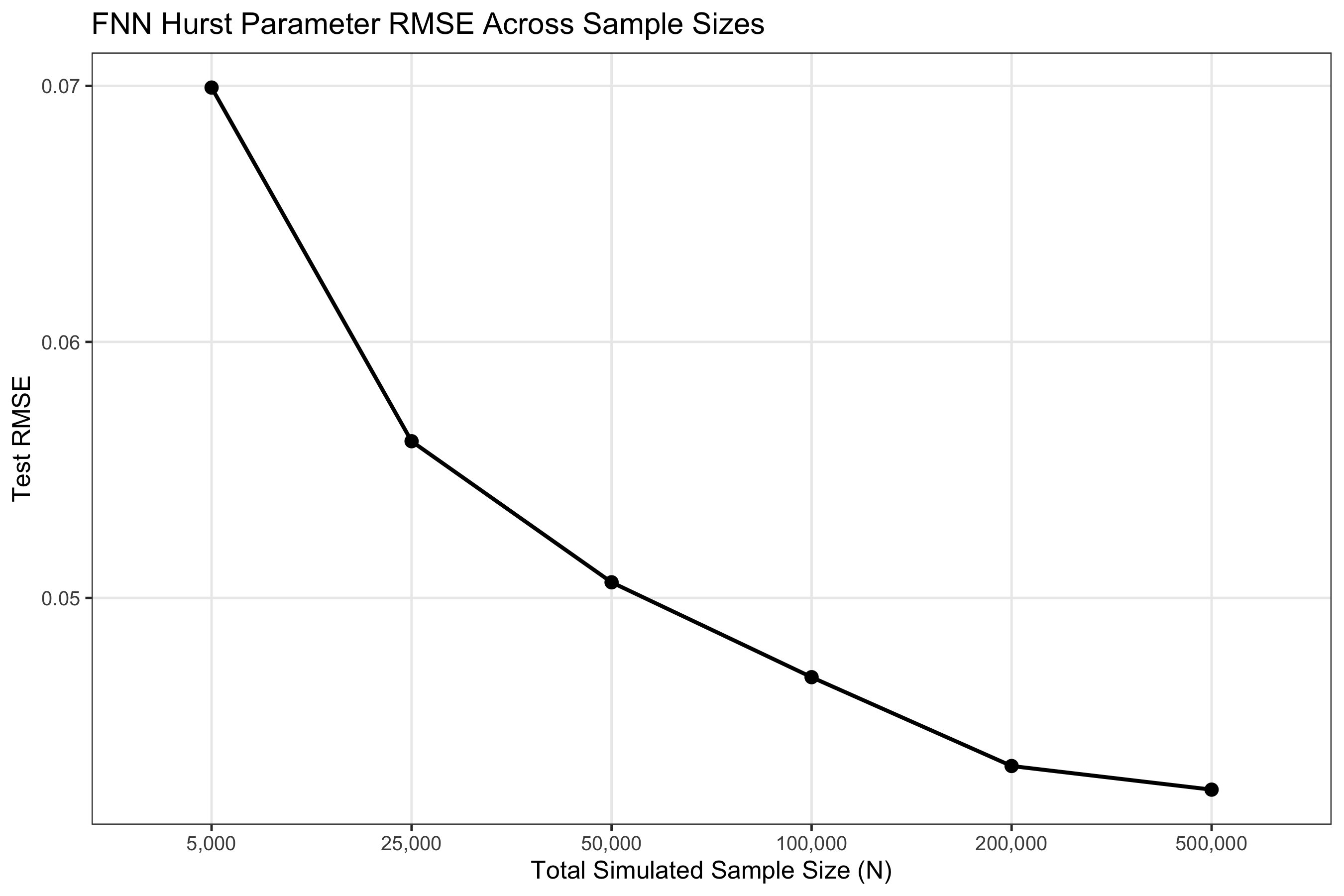}
        \caption{Hurst-parameter RMSE.}
    \end{subfigure}
    \hfill
    \begin{subfigure}[t]{0.48\textwidth}
        \centering
        \includegraphics[width=\textwidth]
        {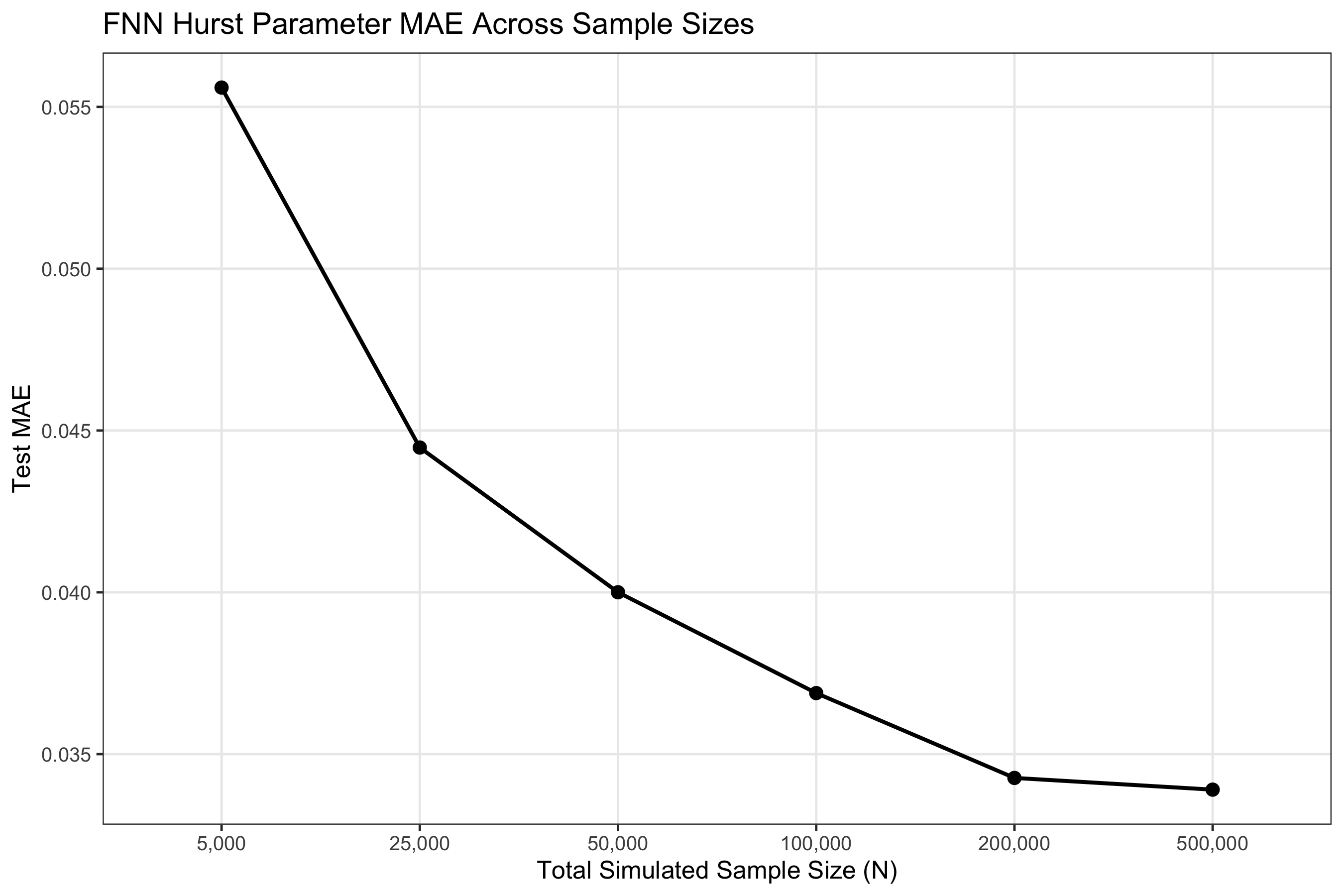}
        \caption{Hurst-parameter MAE.}
    \end{subfigure}

    
\caption{FNN test-set prediction errors for the Hurst parameter
across total simulated sample sizes. The left panel shows
root mean squared error (RMSE), and the right panel shows
mean absolute error (MAE). Each plotted point is based on
1,000, 5,000, 10,000, 20,000, 40,000, and 100,000 test
observations, corresponding to total simulated sample sizes
of 5,000, 25,000, 50,000, 100,000, 200,000, and 500,000,
respectively.}
\label{fig:fnn_H_sample_size}

\end{figure}

Figure~\ref{fig:fnn_H_sample_size} presents the corresponding error
patterns for the Hurst parameter. Unlike the small fluctuation observed
in option-price RMSE, both the RMSE and MAE for \(H\) decrease
consistently across the six sample sizes. The reductions become smaller
at the largest sample sizes, suggesting diminishing marginal gains from
additional simulated training observations.

\paragraph{Empirical Residual-Based Prediction-Interval Performance.}

In addition to point-prediction accuracy, empirical residual-based
\(95\%\) prediction intervals were constructed for each neural-network
output using the empirical distribution of the validation-set residuals.
Specifically, residuals were defined as
\(e_i=y_i-\widehat{y}_i\), and the \(2.5\%\) and \(97.5\%\) empirical
quantiles of the validation residuals were added to the corresponding
point prediction to obtain the lower and upper interval limits. The
resulting intervals were subsequently evaluated on the independent test
set. These intervals provide an empirical characterization of predictive
uncertainty under the simulated validation and test distributions; they
are not Bayesian credible intervals and do not imply a formal
distribution-free coverage guarantee.

Because option prices are nonnegative, lower prediction-interval bounds
for the call and put prices were truncated at zero whenever the
residual-based lower bound was negative. Consequently, although the
underlying residual-quantile width is fixed for a given output, the
realized option-price interval width can be smaller for observations
whose lower bounds are truncated. The resulting average interval width
may therefore vary slightly across evaluation samples, particularly
for put options with predicted values close to zero.

Table~\ref{tab:fnn_interval_performance} reports both empirical
test-set coverage and average interval width. Reporting these
quantities jointly is important because satisfactory coverage is most
informative when it is achieved without unnecessarily wide prediction
intervals.

\begin{table}[H]
\centering
\caption{Empirical coverage and average width of the FNN \(95\%\)
prediction intervals across total simulated sample sizes.}
\label{tab:fnn_interval_performance}
\begin{tabular}{clcc}
\toprule
Sample Size & Quantity & Coverage (\%) & Average Width \\
\midrule

\multirow{3}{*}{5,000}
& Call Price      & 94.800 & 8.178 \\
& Put Price       & 94.600 & 7.861 \\
& Hurst Parameter & 94.000 & 0.241 \\
\midrule

\multirow{3}{*}{25,000}
& Call Price      & 95.320 & 6.981 \\
& Put Price       & 95.340 & 6.743 \\
& Hurst Parameter & 94.820 & 0.206 \\
\midrule

\multirow{3}{*}{50,000}
& Call Price      & 94.380 & 6.637 \\
& Put Price       & 94.380 & 6.429 \\
& Hurst Parameter & 94.890 & 0.188 \\
\midrule

\multirow{3}{*}{100,000}
& Call Price      & 94.630 & 6.252 \\
& Put Price       & 94.660 & 6.096 \\
& Hurst Parameter & 94.675 & 0.173 \\
\midrule

\multirow{3}{*}{200,000}
& Call Price      & 94.978 & 6.188 \\
& Put Price       & 94.978 & 6.073 \\
& Hurst Parameter & 94.855 & 0.163 \\
\midrule

\multirow{3}{*}{500,000}
& Call Price      & 94.844 & \textbf{5.891} \\
& Put Price       & 94.844 & \textbf{5.800} \\
& Hurst Parameter & 94.985 & \textbf{0.159} \\
\bottomrule
\end{tabular}
\end{table}

Coverage remains close to the nominal \(95\%\) level for all three
outputs across all six sample sizes. For the call price, empirical
coverage ranges from \(94.380\%\) to \(95.320\%\), while put-price
coverage ranges from \(94.380\%\) to \(95.340\%\). Coverage for the
Hurst parameter ranges from \(94.000\%\) to \(94.985\%\). Thus, the
empirical residual-based prediction intervals maintain coverage close to the nominal
level throughout the sequence of experiments.

Importantly, approximately nominal coverage is maintained while the
prediction intervals become progressively narrower as the sample size
increases. The average call-price interval width decreases from
\(8.178\) at \(N=5{,}000\) to \(5.891\) at \(N=500{,}000\),
while the corresponding put-price interval width decreases from
\(7.861\) to \(5.800\). For the Hurst parameter, the average interval
width decreases from \(0.241\) to \(0.159\). Consequently, increasing
the training sample size improves the precision of the empirical
prediction intervals while maintaining coverage close to the nominal
\(95\%\) level.

\begin{figure}[htbp]
    \centering
    \includegraphics[width=0.80\textwidth]
    {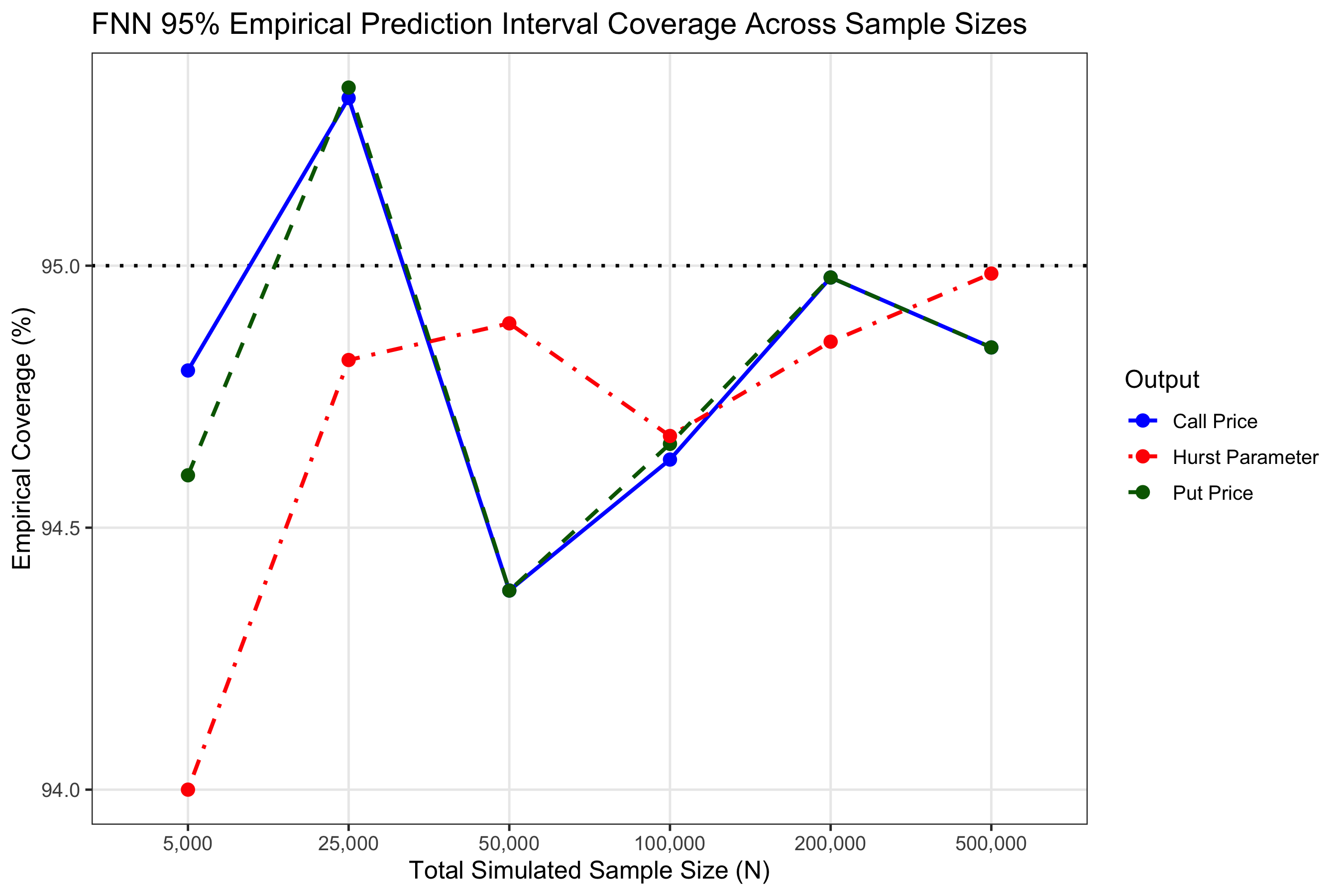}
    \caption{Empirical coverage of the FNN \(95\%\) prediction
    intervals for the call price, put price, and Hurst parameter across
    total simulated sample sizes. The horizontal dotted line indicates
    the nominal \(95\%\) coverage level.}
    \label{fig:fnn_coverage}
\end{figure}

Figure~\ref{fig:fnn_coverage} displays the empirical coverage
probabilities across the six sample sizes. The horizontal dotted line
represents the nominal \(95\%\) coverage level.

\paragraph{Detailed Predictive Performance at \(N=500{,}000\).}

Because the model trained using \(N=500{,}000\) achieved the strongest
overall test-set performance among the FNN specifications considered,
including the lowest option-price RMSE and the lowest RMSE and MAE for
the Hurst parameter, this specification was selected for more detailed
diagnostic evaluation.

\begin{figure}[htbp]
    \centering

    \begin{subfigure}[t]{0.32\textwidth}
        \centering
        \includegraphics[width=\textwidth]
        {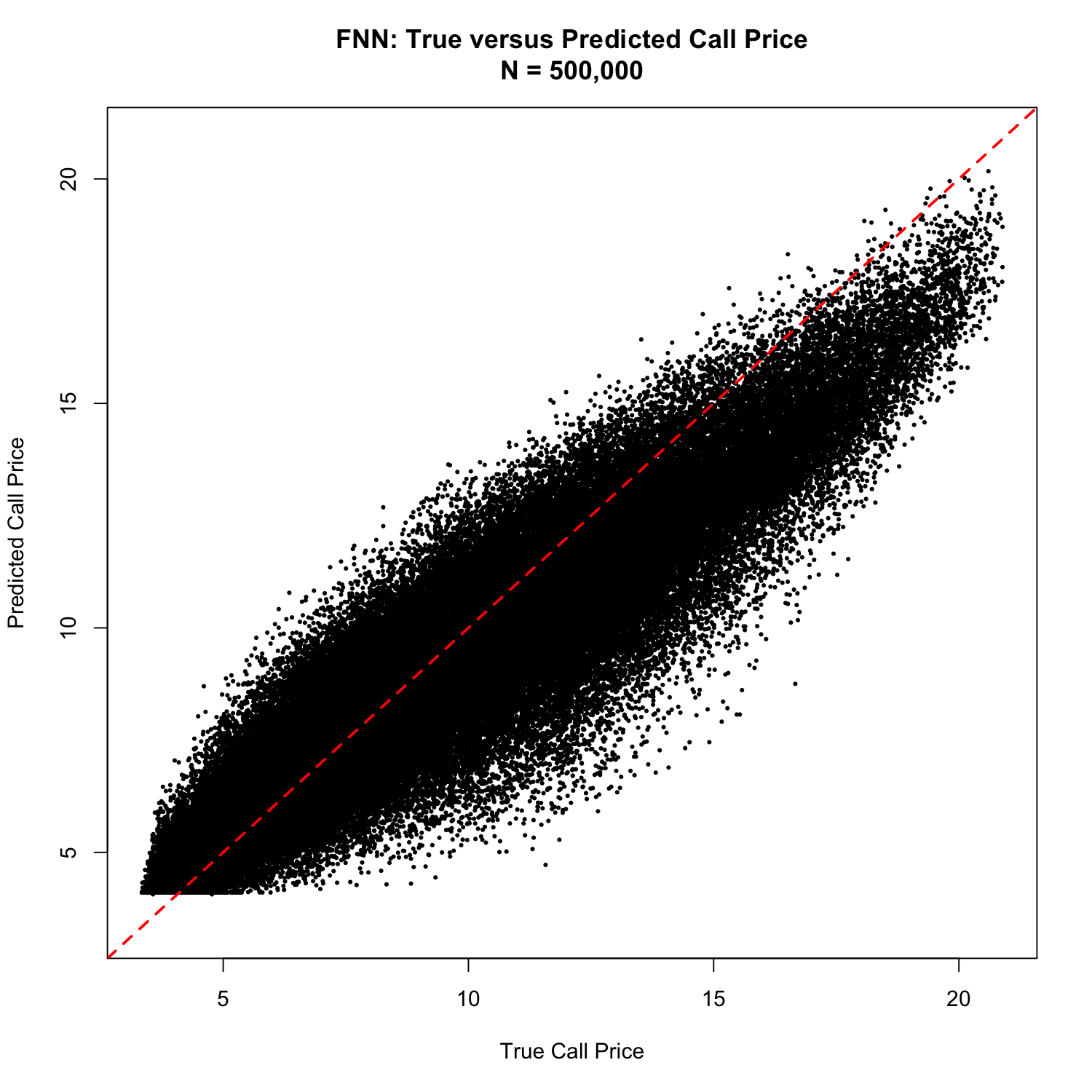}
        \caption{Call price.}
    \end{subfigure}
    \hfill
    \begin{subfigure}[t]{0.32\textwidth}
        \centering
        \includegraphics[width=\textwidth]
        {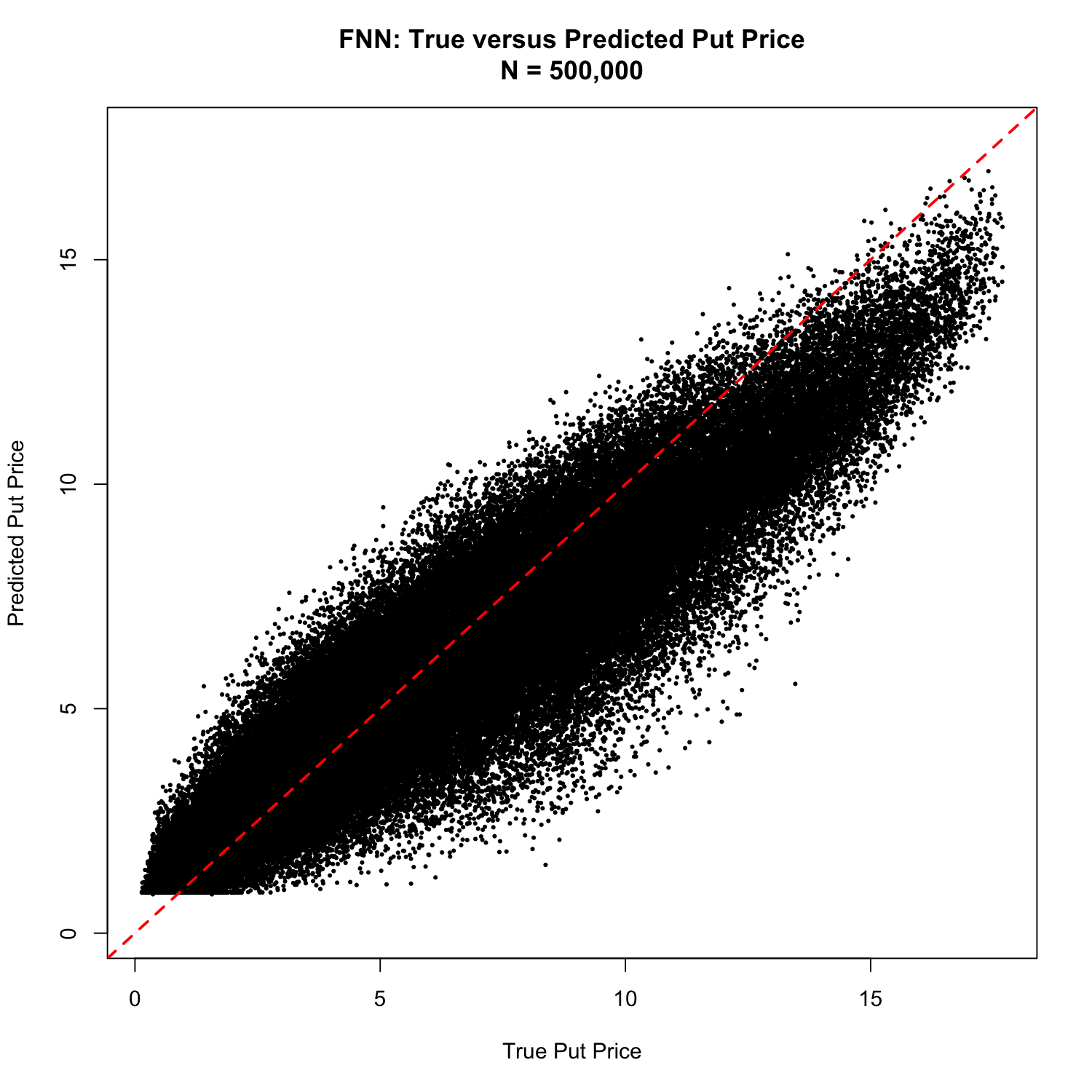}
        \caption{Put price.}
    \end{subfigure}
    \hfill
    \begin{subfigure}[t]{0.32\textwidth}
        \centering
        \includegraphics[width=\textwidth]
        {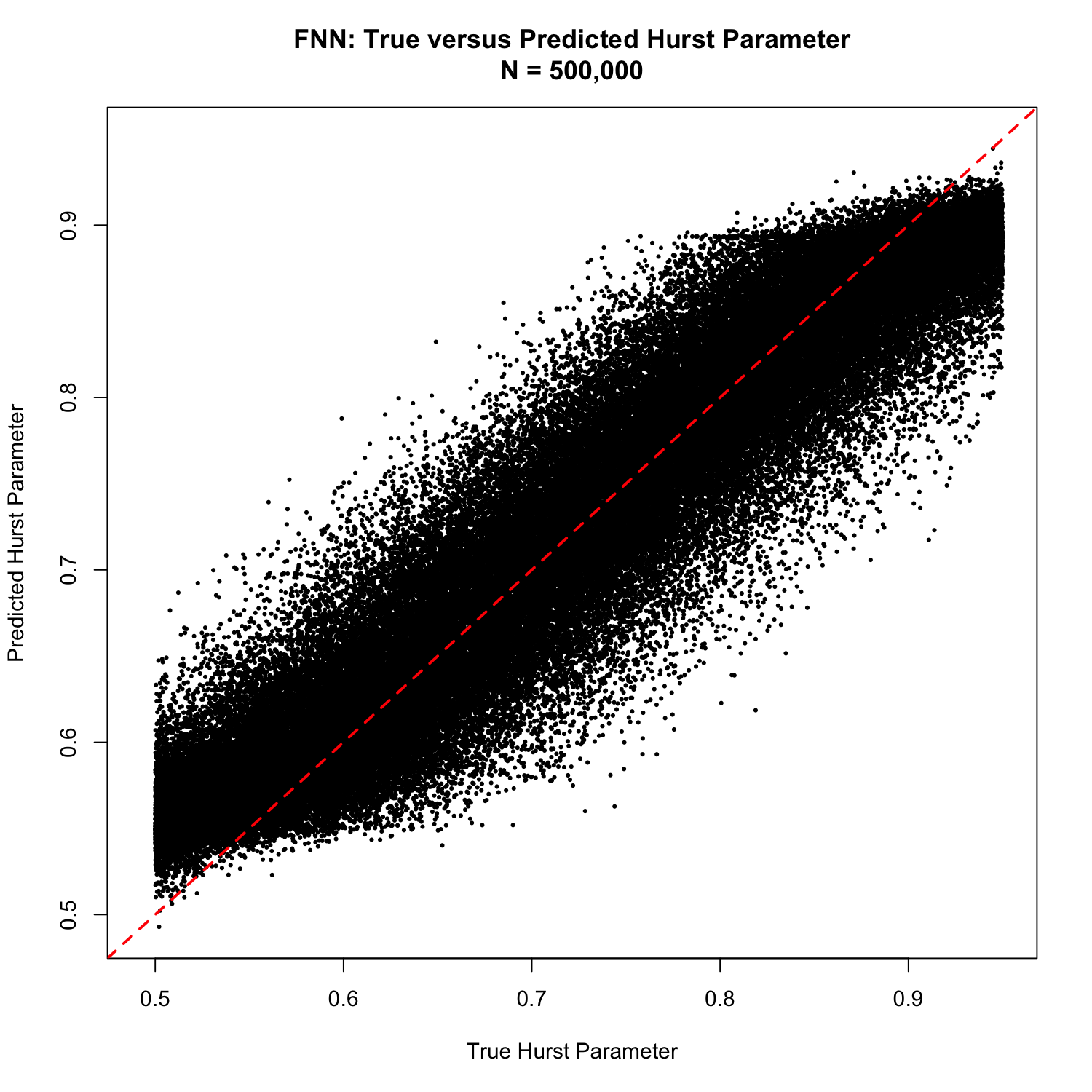}
        \caption{Hurst parameter.}
    \end{subfigure}

    \caption{True versus predicted test-set values from the FNN trained
    using a total simulated sample size of \(N=500{,}000\). The dashed
    \(45^\circ\) line represents perfect agreement between the true and
    predicted values.}
    \label{fig:fnn_true_pred_500k}
\end{figure}

Figure~\ref{fig:fnn_true_pred_500k} compares the true simulated values
with the corresponding FNN predictions for the call price, put price,
and Hurst parameter. The dashed \(45^\circ\) line represents perfect
agreement between the true and predicted values.

The predicted option prices exhibit a strong positive relationship with
their corresponding simulated values. The call- and put-price plots
display nearly identical patterns, consistent with their similar
aggregate error measures. Some dispersion around the \(45^\circ\)
reference line remains, particularly toward the extremes of the
option-price range.

The Hurst-parameter predictions also track the simulated values closely.
However, some compression toward the center of the parameter range is
visible. Predictions near the lower and upper boundaries tend to be
drawn inward relative to the \(45^\circ\) line, indicating greater
difficulty in reproducing values near the boundaries of the simulated
range.

\paragraph{Training and Validation Behavior.}
Table~\ref{tab:fnn_training} summarizes the training and validation
behavior of the FNN across the six experiments. Training was permitted
to continue for a maximum of \(250\) epochs; however, the actual number
of completed epochs was determined by the early-stopping criterion.
The table therefore reports the number of completed epochs, the selected
best epoch, the best validation loss, the best validation MAE, and the
total training time.

\begin{table}[H]
\centering
\caption{FNN training and validation behavior across total simulated sample sizes.}
\label{tab:fnn_training}
\begin{tabular}{rrrrrrr}
\toprule
Sample Size &
Training Size &
Epochs &
Best Epoch &
Validation Loss &
Validation MAE &
Time (s) \\
\midrule
5,000   & 3,200   & 104 & 79 & 0.267 & 0.411 & 17.630 \\
25,000  & 16,000  & 61  & 36 & 0.199 & 0.351 & 14.080 \\
50,000  & 32,000  & 77  & 52 & 0.178 & 0.326 & 23.760 \\
100,000 & 64,000  & 48  & 23 & 0.165 & 0.313 & 22.410 \\
200,000 & 128,000 & 73  & 48 & 0.162 & 0.307 & 59.380 \\
500,000 & 320,000 & 37  & 12 & \textbf{0.146} & \textbf{0.297} & 64.640 \\
\bottomrule
\end{tabular}
\end{table}

None of the FNN experiments reached the maximum of \(250\) epochs.
The number of completed epochs ranged from \(37\) to \(104\), while
the selected best epoch ranged from \(12\) to \(79\). The best
validation loss decreased from \(0.267\) at \(N=5{,}000\) to
\(0.146\) at \(N=500{,}000\), while the corresponding best validation
MAE decreased from \(0.411\) to \(0.297\). Thus, the improvement
observed on the independent test set was accompanied by improved
validation performance as the amount of simulated data increased.

\begin{figure}[htbp]
    \centering

    \begin{subfigure}[t]{0.48\textwidth}
        \centering
        \includegraphics[width=\textwidth]
        {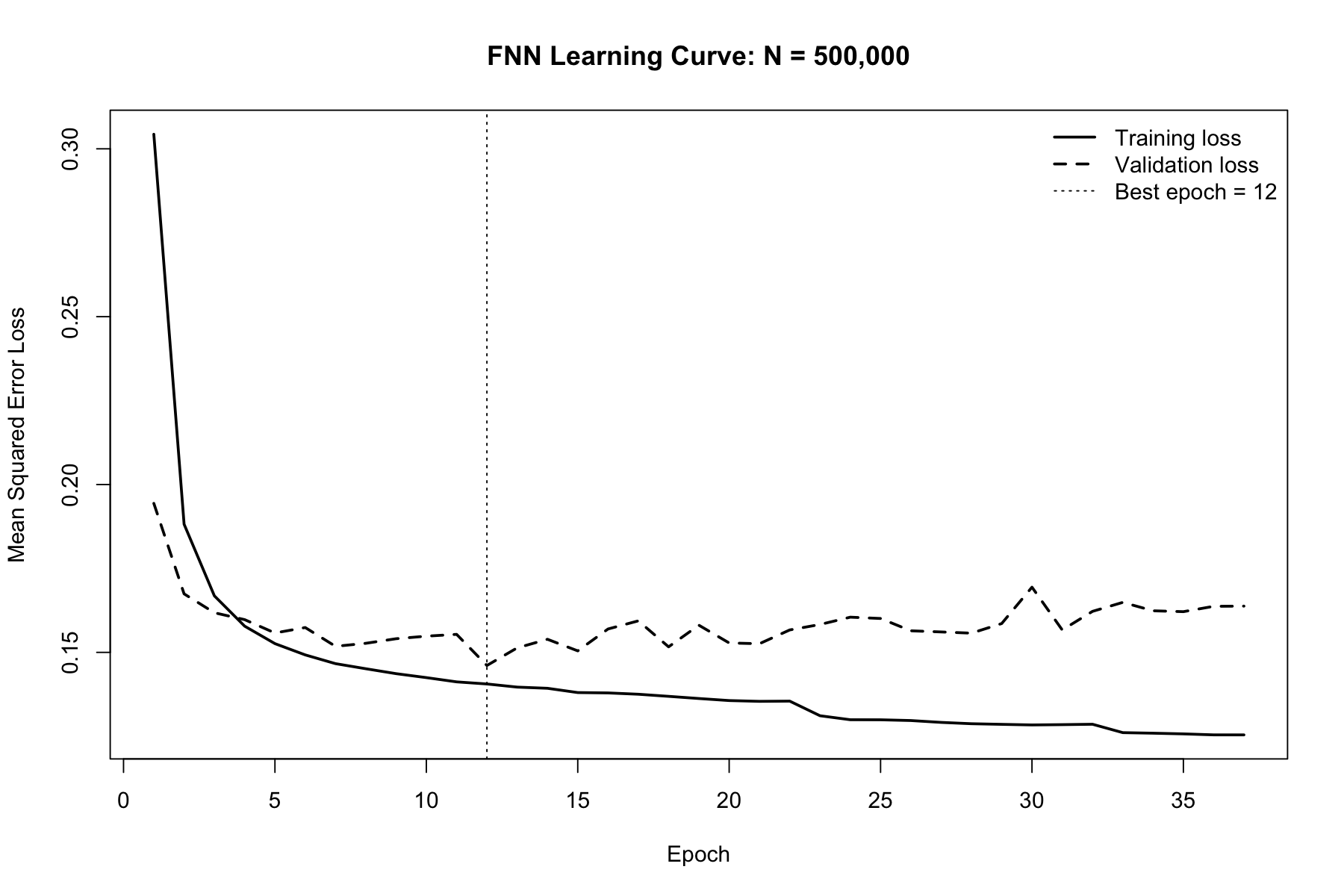}
        \caption{Training and validation loss.}
    \end{subfigure}
    \hfill
    \begin{subfigure}[t]{0.48\textwidth}
        \centering
        \includegraphics[width=\textwidth]
        {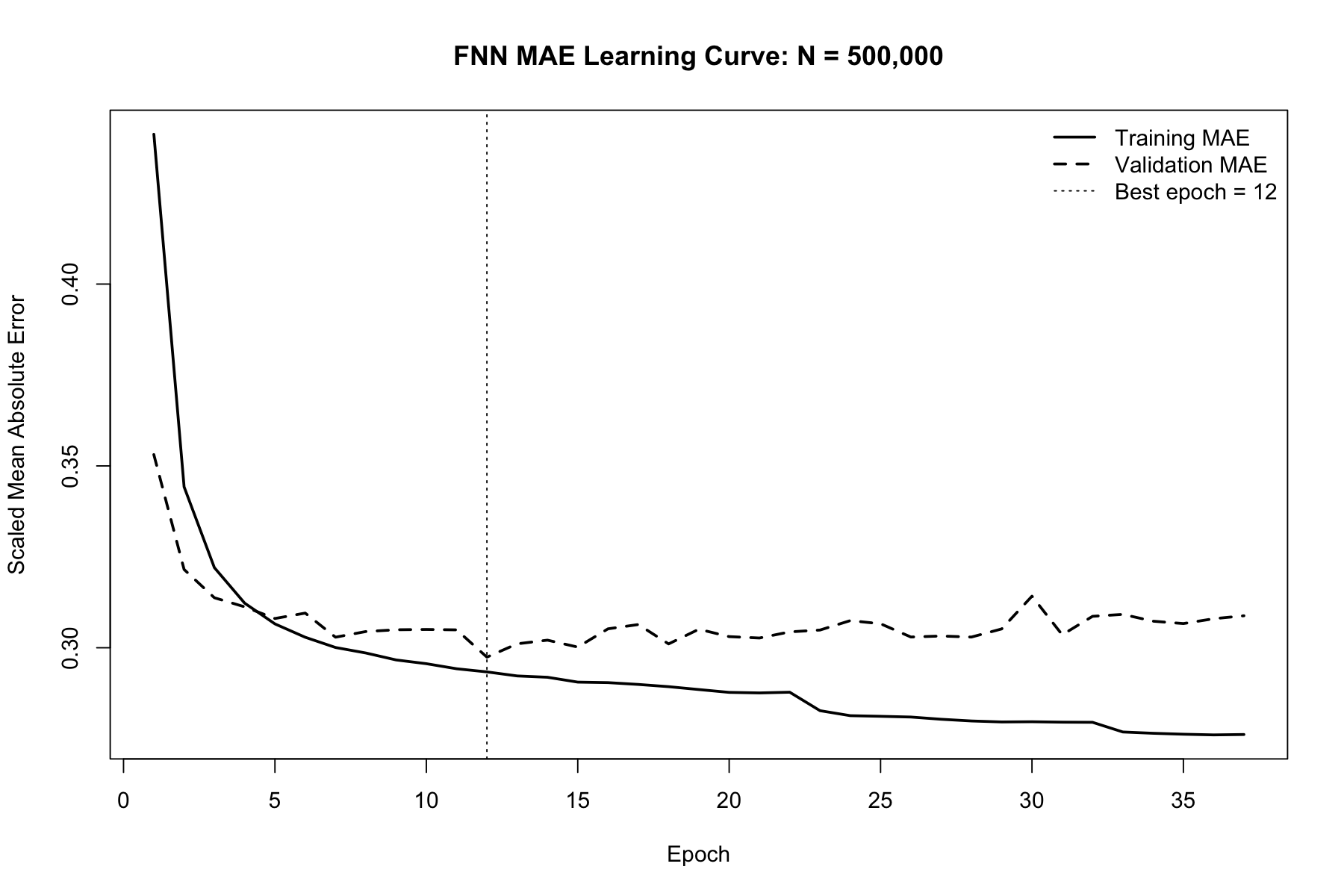}
        \caption{Training and validation MAE.}
    \end{subfigure}

    \caption{FNN learning curves for the \(N=500{,}000\) experiment.
    The vertical dotted line identifies the epoch corresponding to the
    minimum validation loss.}
    \label{fig:fnn_learning}
\end{figure}

For the largest experiment, \(N=500{,}000\), training terminated after
\(37\) epochs, with the minimum validation loss attained at epoch \(12\).
The learning curves in Figure~\ref{fig:fnn_learning} show that both
training and validation errors decrease rapidly during the initial
epochs. After the selected epoch, the training error continues to
decrease while validation performance no longer improves consistently.
This behavior illustrates the role of early stopping in retaining the
model corresponding to the best validation performance rather than the
model from the final training epoch.

\paragraph{Residual Diagnostics.}

\begin{figure}[htbp]
    \centering

    \begin{subfigure}[t]{0.32\textwidth}
        \centering
        \includegraphics[width=\textwidth]
        {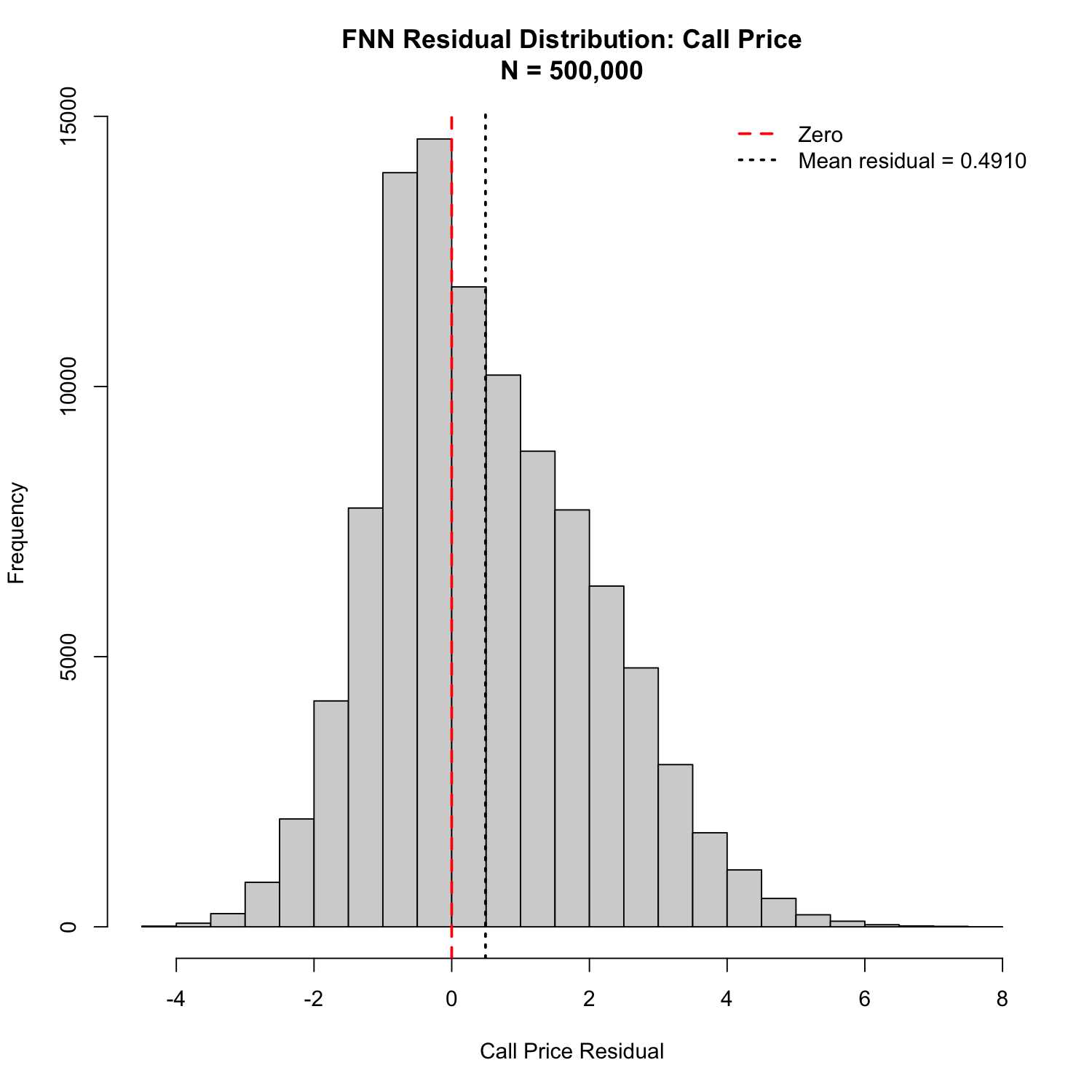}
        \caption{Call-price residual distribution.}
    \end{subfigure}
    \hfill
    \begin{subfigure}[t]{0.32\textwidth}
        \centering
        \includegraphics[width=\textwidth]
        {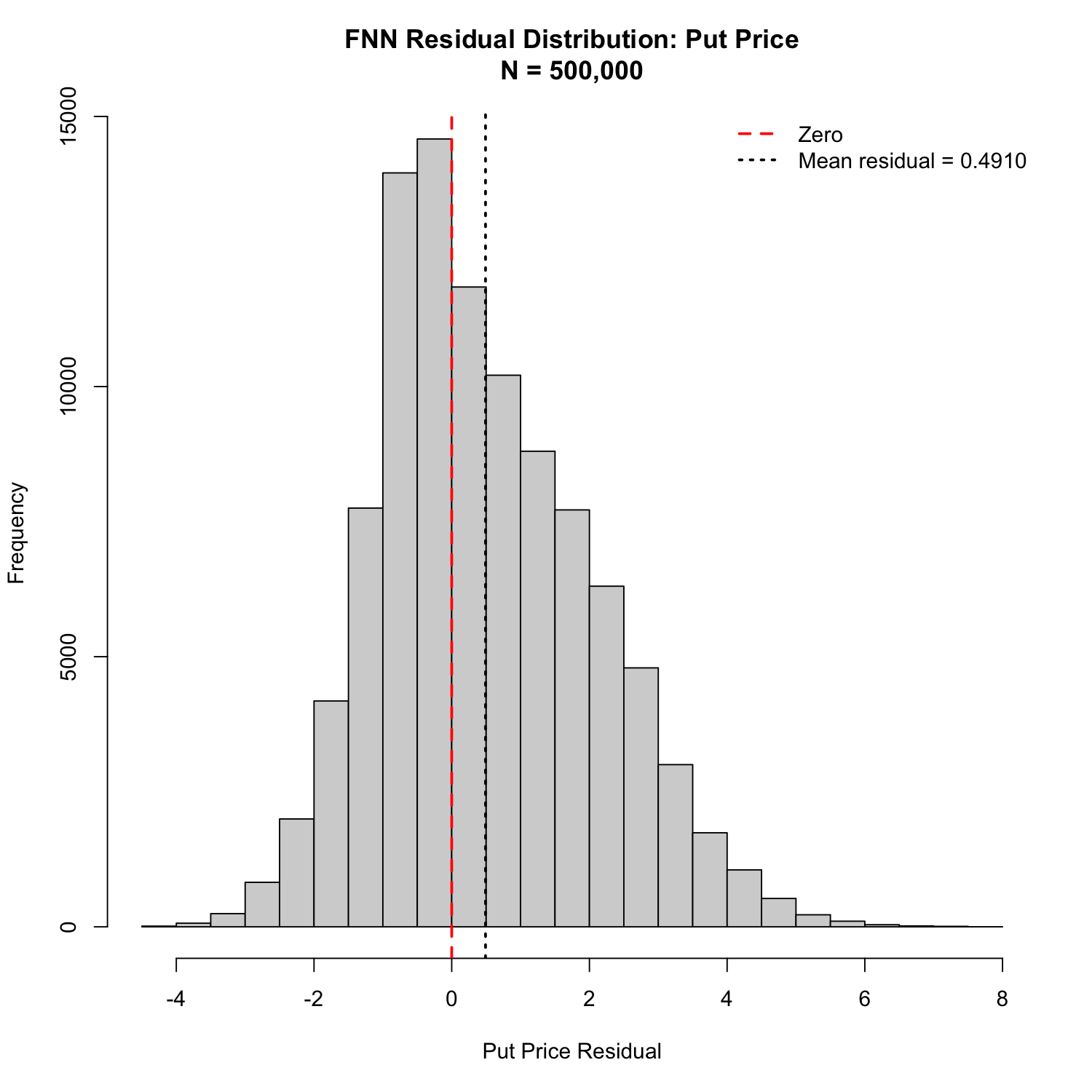}
        \caption{Put-price residual distribution.}
    \end{subfigure}
    \hfill
    \begin{subfigure}[t]{0.32\textwidth}
        \centering
        \includegraphics[width=\textwidth]
        {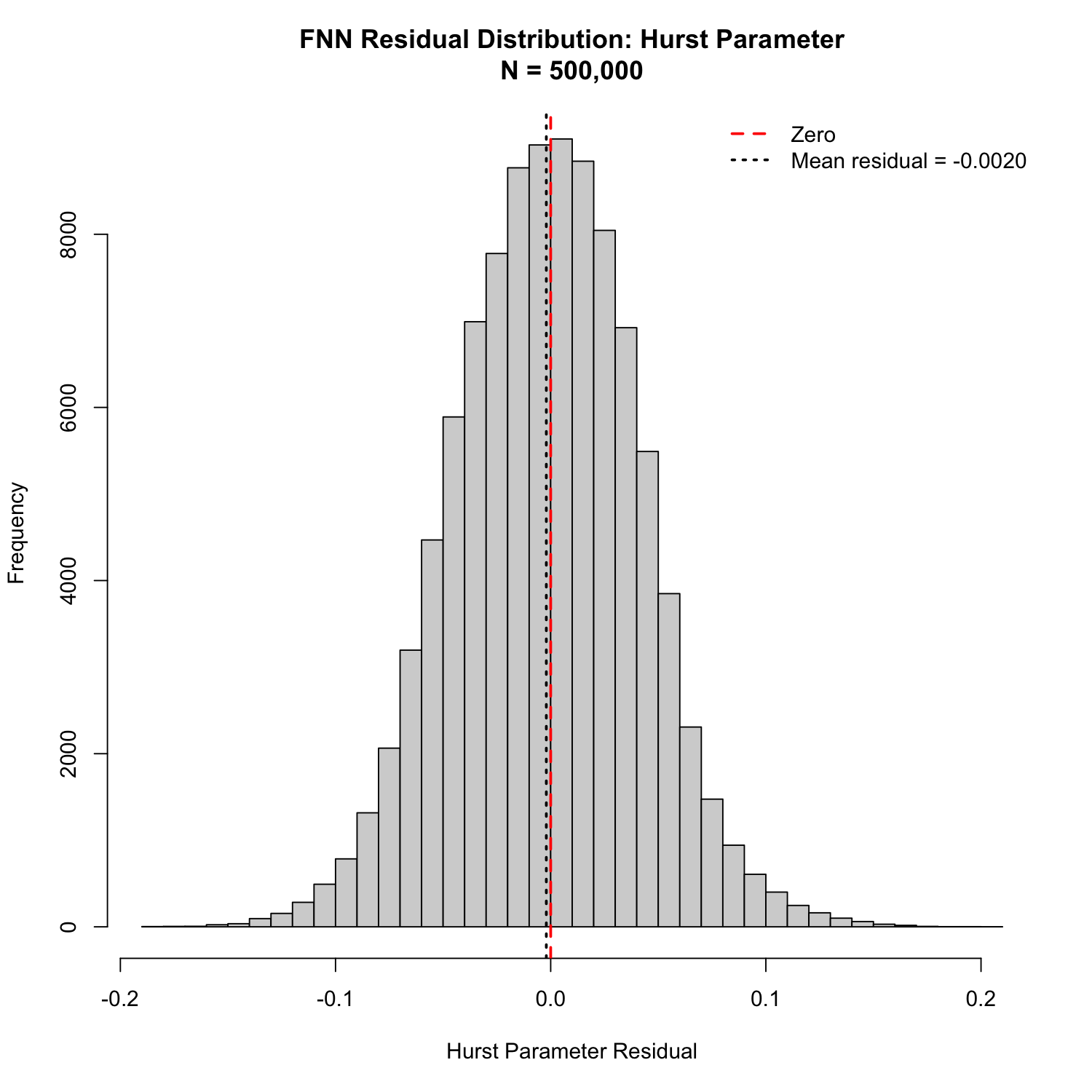}
        \caption{Hurst parameter residual distribution.}
    \end{subfigure}

    \vspace{0.3cm}

    \begin{subfigure}[t]{0.32\textwidth}
        \centering
        \includegraphics[width=\textwidth]
        {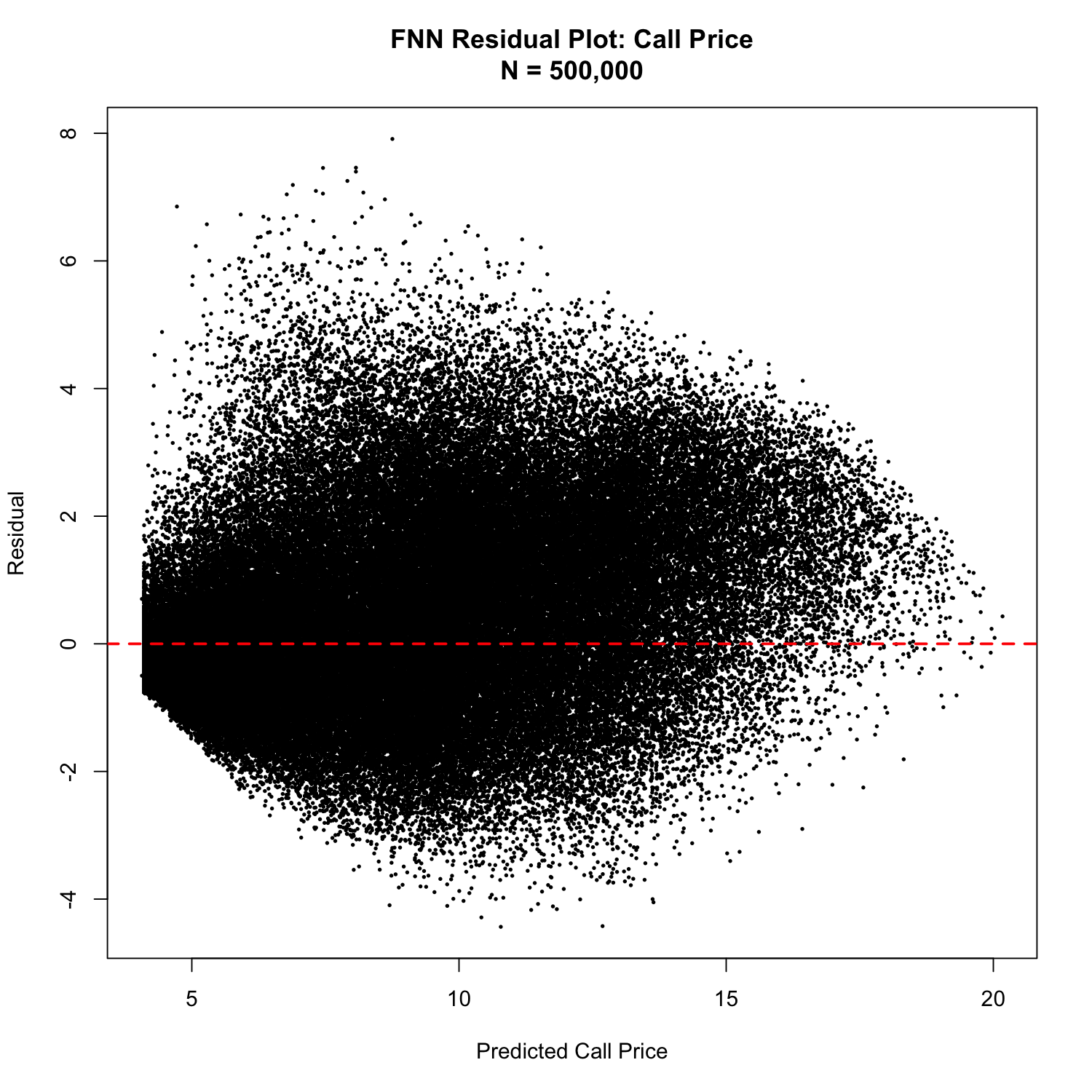}
        \caption{Call residuals versus predicted values.}
    \end{subfigure}
    \hfill
    \begin{subfigure}[t]{0.32\textwidth}
        \centering
        \includegraphics[width=\textwidth]
        {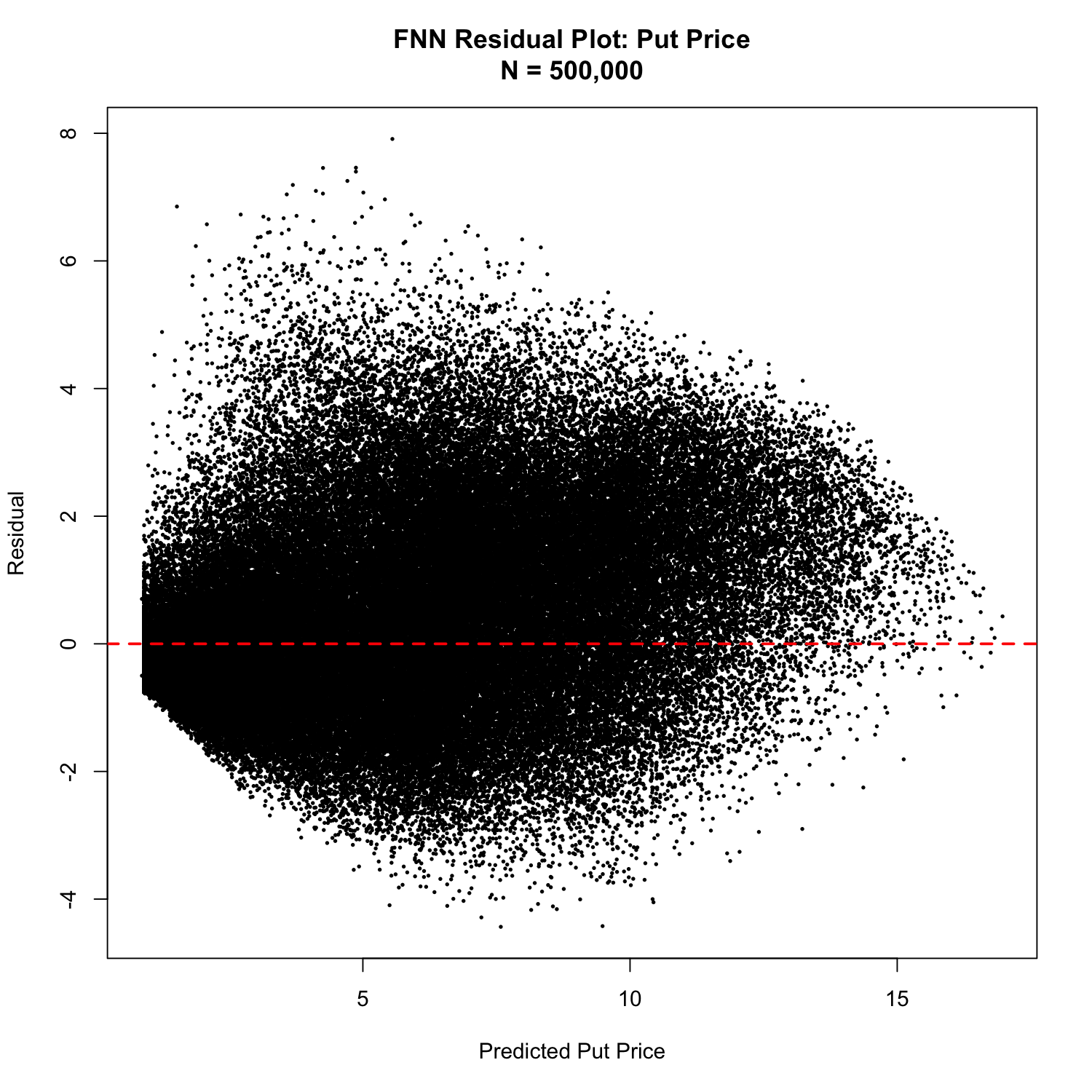}
        \caption{Put residuals versus predicted values.}
    \end{subfigure}
    \hfill
    \begin{subfigure}[t]{0.32\textwidth}
        \centering
        \includegraphics[width=\textwidth]
        {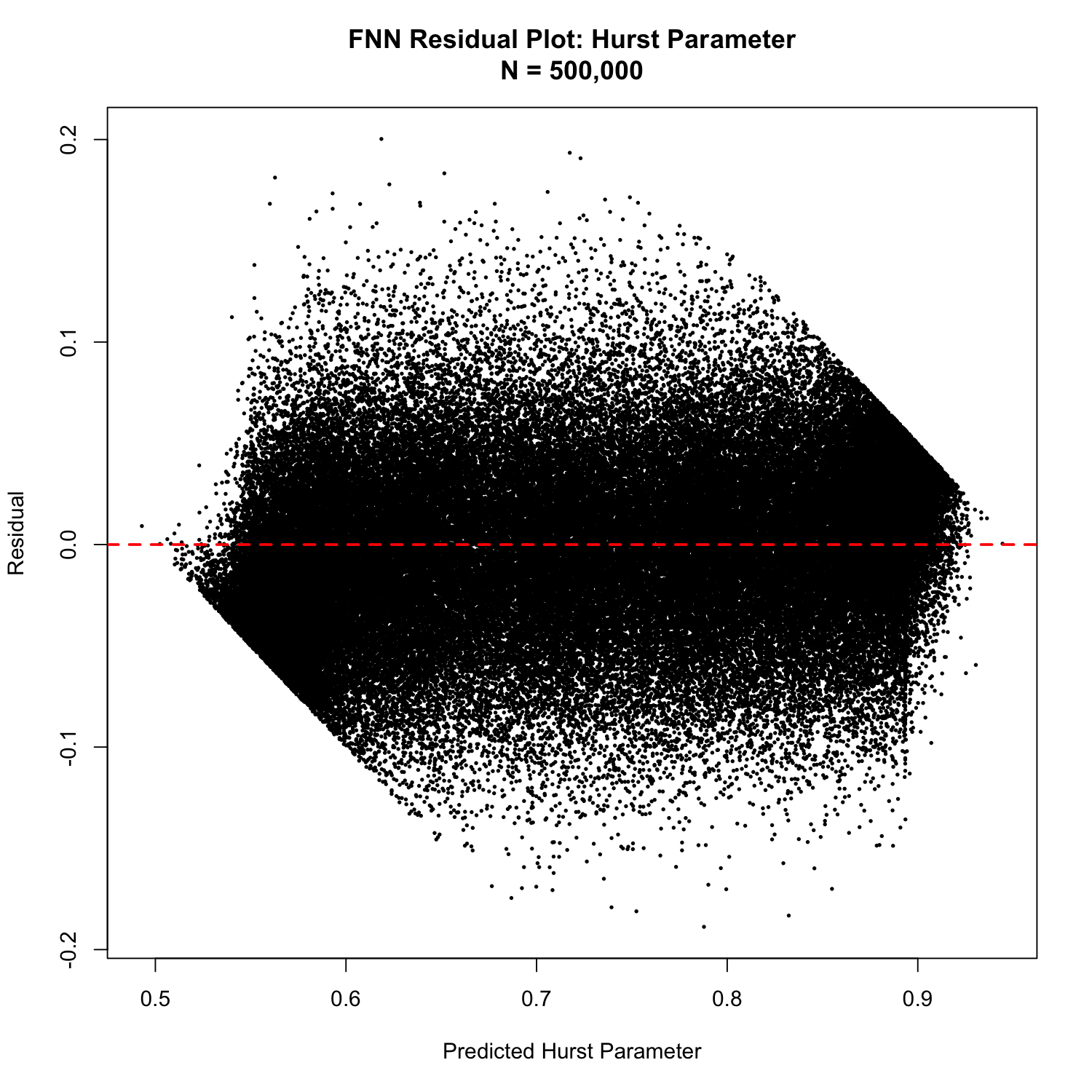}
        \caption{Hurst parameter residuals versus predicted values.}
    \end{subfigure}

    \caption{Residual diagnostics for the FNN trained using a total
    simulated sample size of \(N=500{,}000\). The upper panels show
    the residual distributions, while the lower panels show residuals
    versus the corresponding predicted values. Residuals are defined
    as the true simulated value minus the FNN prediction.}
    \label{fig:fnn_residual_diagnostics}
\end{figure}

Residual diagnostics were examined for the \(N=500{,}000\) FNN using
the independent test set, with residuals defined as the true simulated
value minus the corresponding FNN prediction. The residual
distributions and residual-versus-predicted plots are presented in
Figure~\ref{fig:fnn_residual_diagnostics}.

For the call and put prices, the residual distributions are similar,
consistent with the nearly identical test-set performance observed for
the two option-price outputs. Both distributions exhibit a small
positive shift, with a mean residual of approximately \(0.491\).
Because the residual is defined as the true value minus the predicted
value, the positive mean indicates a modest average tendency of the FNN
to underpredict option prices. The residual-versus-predicted plots also
show that the dispersion of the residuals varies across the range of
predicted option prices. Thus, although the overall prediction accuracy
is strong, the remaining option-price errors are not completely
homogeneous across the prediction range.

For the Hurst parameter, the residual distribution is approximately
centered at zero, with a mean residual of approximately \(-0.002\),
indicating little overall systematic bias. However, the
residual-versus-predicted plot reveals a structured pattern near the
boundaries of the predicted range. This behavior is consistent with
the compression toward the center observed in
Figure~\ref{fig:fnn_true_pred_500k}: predictions near the lower and
upper portions of the simulated Hurst-parameter range are more
difficult for the FNN to reproduce accurately. Therefore, while the
average residual for \(H\) is close to zero, the residual diagnostics
indicate that prediction errors depend to some extent on the location
within the parameter range.

\paragraph{Illustration of empirical residual-based prediction intervals.}

\begin{figure}[htbp]
    \centering

    \begin{subfigure}[t]{0.32\textwidth}
        \centering
        \includegraphics[width=\textwidth]
        {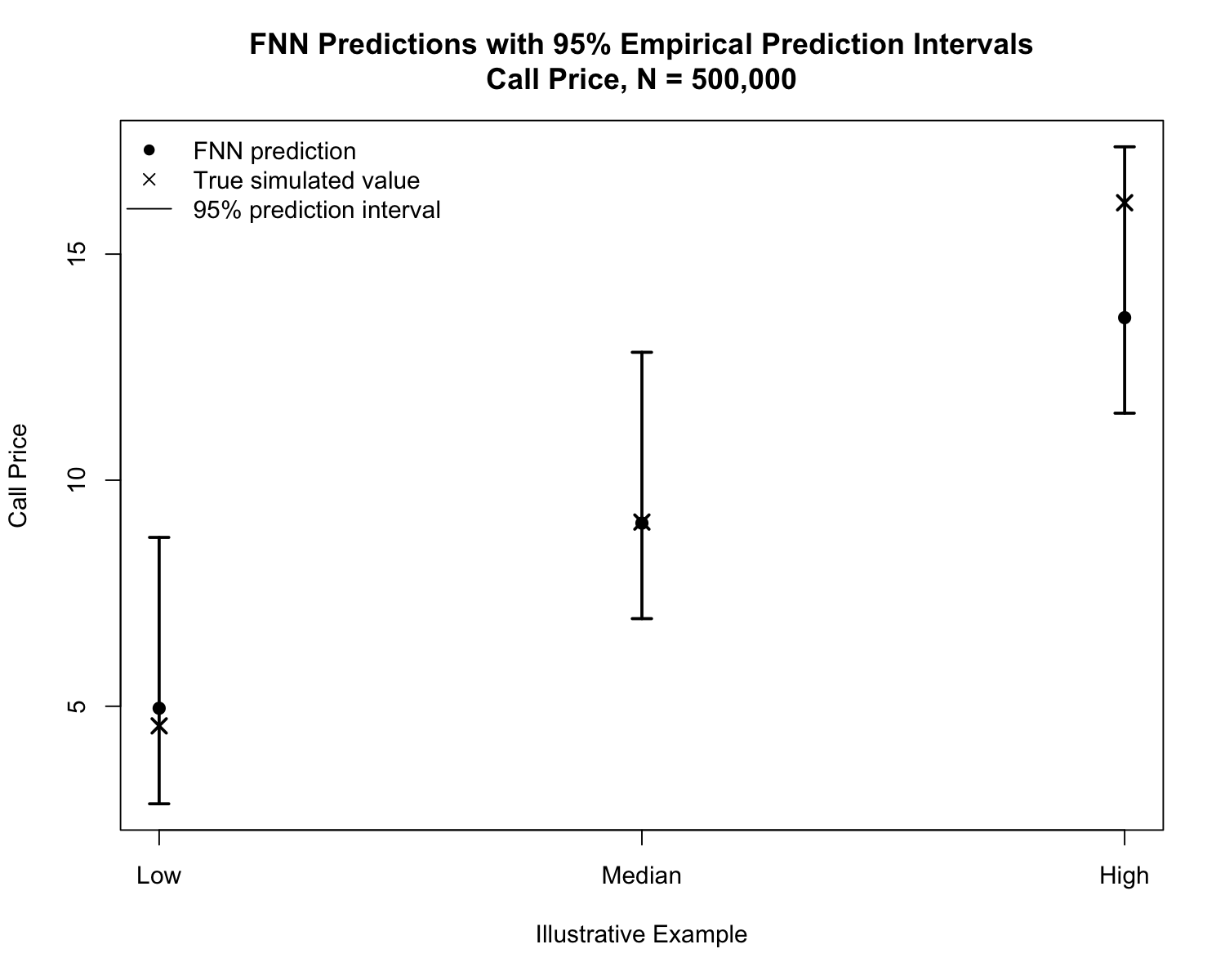}
        \caption{Call price.}
    \end{subfigure}
    \hfill
    \begin{subfigure}[t]{0.32\textwidth}
        \centering
        \includegraphics[width=\textwidth]
        {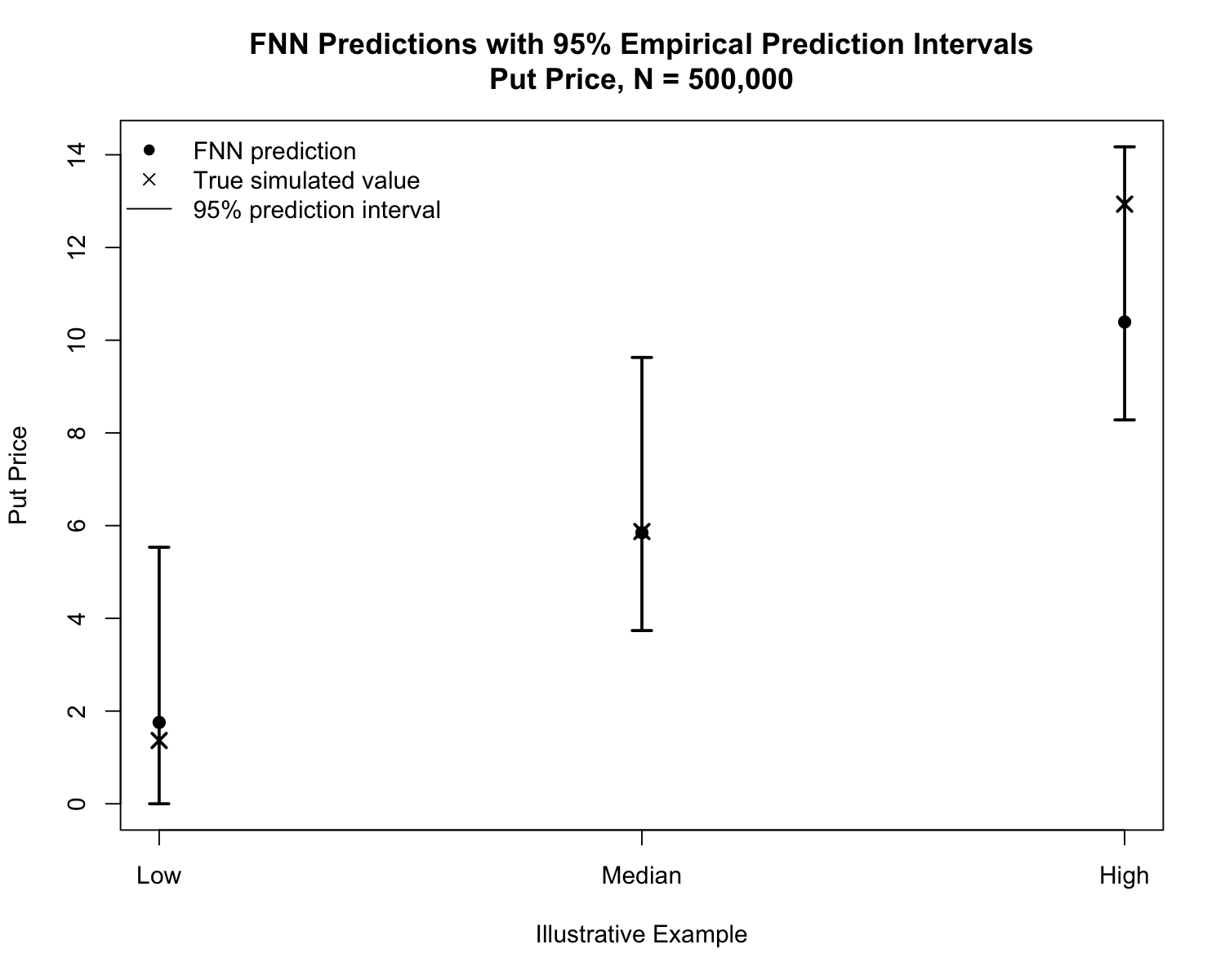}
        \caption{Put price.}
    \end{subfigure}
    \hfill
    \begin{subfigure}[t]{0.32\textwidth}
        \centering
        \includegraphics[width=\textwidth]
        {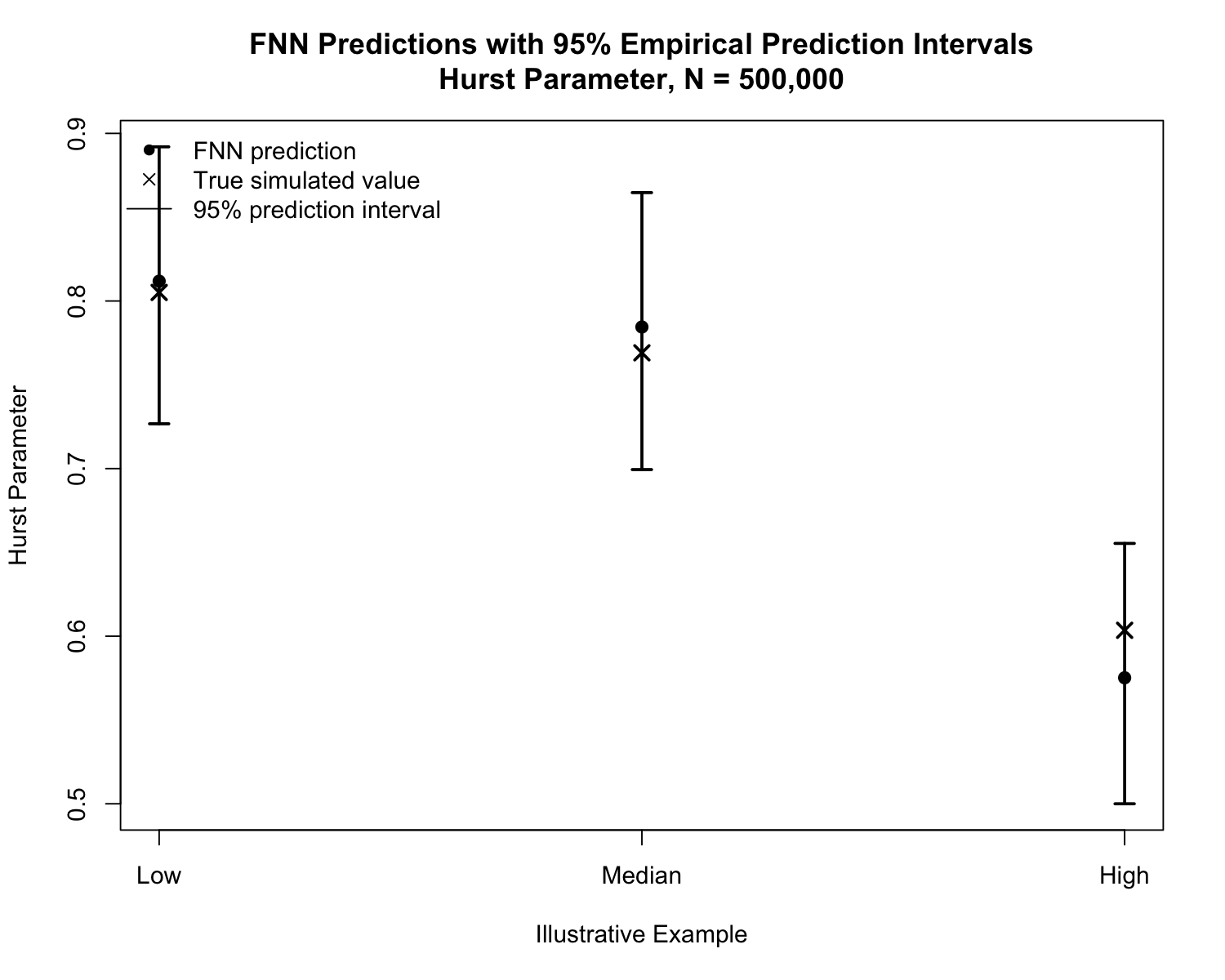}
        \caption{Hurst parameter.}
    \end{subfigure}

    \caption{Illustrative FNN predictions and empirical \(95\%\)
    prediction intervals for representative low, median, and high
    predictions from the \(N=500{,}000\) experiment.}
    \label{fig:fnn_PI_examples}
\end{figure}

Figure~\ref{fig:fnn_PI_examples} provides illustrative examples of the
empirical \(95\%\) prediction intervals obtained from the
\(N=500{,}000\) FNN. For each output, representative observations from
the lower, middle, and upper portions of the predicted-value
distribution are displayed. The FNN point prediction, true simulated
value, and corresponding empirical prediction interval are shown
simultaneously.

These examples illustrate the predictive uncertainty at different
locations of the response distribution. In each displayed case, the
true simulated value lies within the corresponding prediction interval,
consistent with the approximately nominal aggregate coverage reported
in Table~\ref{tab:fnn_interval_performance} and
Figure~\ref{fig:fnn_coverage}.

Overall, the FNN results demonstrate a substantial benefit from
increasing the amount of simulated training data. From
\(N=5{,}000\) to \(N=500{,}000\), prediction errors decrease,
correlations and \(R^2\) values increase, and the empirical prediction
intervals become narrower while maintaining coverage close to the
nominal \(95\%\) level. The validation results exhibit a consistent
improvement as well, with both the best validation loss and validation
MAE attaining their lowest values in the \(N=500{,}000\) experiment.
Among the FNN specifications considered, the \(N=500{,}000\) model
therefore provides the strongest overall predictive performance and
is retained for the subsequent model comparisons.

\subsubsection{Long Short-Term Memory Results}
\label{subsec:lstm_results}

The predictive performance of the long short-term memory (LSTM) model
was evaluated across the same total simulated sample sizes used for the
FNN:
\(N=5{,}000\), \(25{,}000\), \(50{,}000\), \(100{,}000\),
\(200{,}000\), and \(500{,}000\). Performance was assessed separately
for the call price, put price, and Hurst parameter using the test-set
root mean squared error (RMSE), mean absolute error (MAE), correlation,
and coefficient of determination (\(R^2\)). In addition to point
prediction, empirical residual-based prediction intervals were examined to assess
predictive uncertainty.

\paragraph{Test-Set Predictive Performance.}

Table~\ref{tab:lstm_test_performance} reports the test-set predictive
performance of the LSTM across the six total simulated sample sizes.
Increasing the amount of simulated data generally improves predictive
accuracy for all three outputs, although the magnitude of improvement
becomes smaller at the larger sample sizes.

\begin{table}[H]
\centering
\caption{Test-set prediction performance of the LSTM model across total simulated sample sizes.}
\label{tab:lstm_test_performance}
\begin{tabular}{clcccc}
\toprule
Sample Size & Quantity & RMSE & MAE & Correlation & $R^2$ \\
\midrule

\multirow{3}{*}{5,000}
& Call Price      & 1.385 & 1.084 & 0.940 & 0.882 \\
& Put Price       & 1.385 & 1.085 & 0.940 & 0.882 \\
& Hurst Parameter & 0.039 & 0.030 & 0.955 & 0.911 \\
\midrule

\multirow{3}{*}{25,000}
& Call Price      & 1.320 & 1.029 & 0.943 & 0.889 \\
& Put Price       & 1.320 & 1.029 & 0.943 & 0.889 \\
& Hurst Parameter & 0.035 & 0.027 & 0.965 & 0.928 \\
\midrule

\multirow{3}{*}{50,000}
& Call Price      & 1.317 & 1.021 & 0.943 & 0.889 \\
& Put Price       & 1.317 & 1.021 & 0.943 & 0.889 \\
& Hurst Parameter & 0.034 & 0.027 & 0.966 & 0.931 \\
\midrule

\multirow{3}{*}{100,000}
& Call Price      & 1.304 & 1.013 & 0.944 & 0.891 \\
& Put Price       & 1.304 & 1.013 & 0.944 & 0.891 \\
& Hurst Parameter & 0.034 & 0.027 & 0.966 & 0.933 \\
\midrule

\multirow{3}{*}{200,000}
& Call Price      & \textbf{1.297} & \textbf{1.000} & 0.944 & 0.892 \\
& Put Price       & \textbf{1.297} & \textbf{1.000} & 0.944 & 0.892 \\
& Hurst Parameter & 0.033 & 0.026 & 0.967 & 0.933 \\
\midrule

\multirow{3}{*}{500,000}
& Call Price      & 1.298 & 1.006 & \textbf{0.945} & \textbf{0.892} \\
& Put Price       & 1.298 & 1.006 & \textbf{0.945} & \textbf{0.892} \\
& Hurst Parameter & \textbf{0.033} & \textbf{0.026}
& \textbf{0.967} & \textbf{0.934} \\
\bottomrule
\end{tabular}
\end{table}

For the call price, the RMSE decreases from \(1.385\) at
\(N=5{,}000\) to \(1.297\) at \(N=200{,}000\), while the MAE
decreases from \(1.084\) to \(1.000\). The put-price results are
nearly identical. The improvement is also reflected in the correlation
and \(R^2\). For the call price, correlation increases from \(0.940\)
at \(N=5{,}000\) to \(0.945\) at \(N=500{,}000\), while \(R^2\)
increases from \(0.882\) to \(0.892\).

The LSTM also exhibits strong performance in predicting the Hurst
parameter. Its RMSE decreases from \(0.039\) at \(N=5{,}000\) to
approximately \(0.033\) at \(N=500{,}000\), while its MAE decreases
from \(0.030\) to \(0.026\). At the same time, correlation increases
from \(0.955\) to \(0.967\), and \(R^2\) increases from \(0.911\) to
\(0.934\).

Although the overall trend is favorable, improvement is not strictly
monotonic for every option-price metric. In particular, the call- and
put-price RMSE values attain their minimum at \(N=200{,}000\), at
approximately \(1.297\), and then increase only slightly to \(1.298\)
at \(N=500{,}000\). The corresponding MAE values change from
\(1.000\) to \(1.006\). These small differences indicate that
option-price predictive performance becomes relatively stable at the
largest sample sizes.

\begin{figure}[htbp]
    \centering

    \begin{subfigure}[t]{0.48\textwidth}
        \centering
        \includegraphics[width=\textwidth]
        {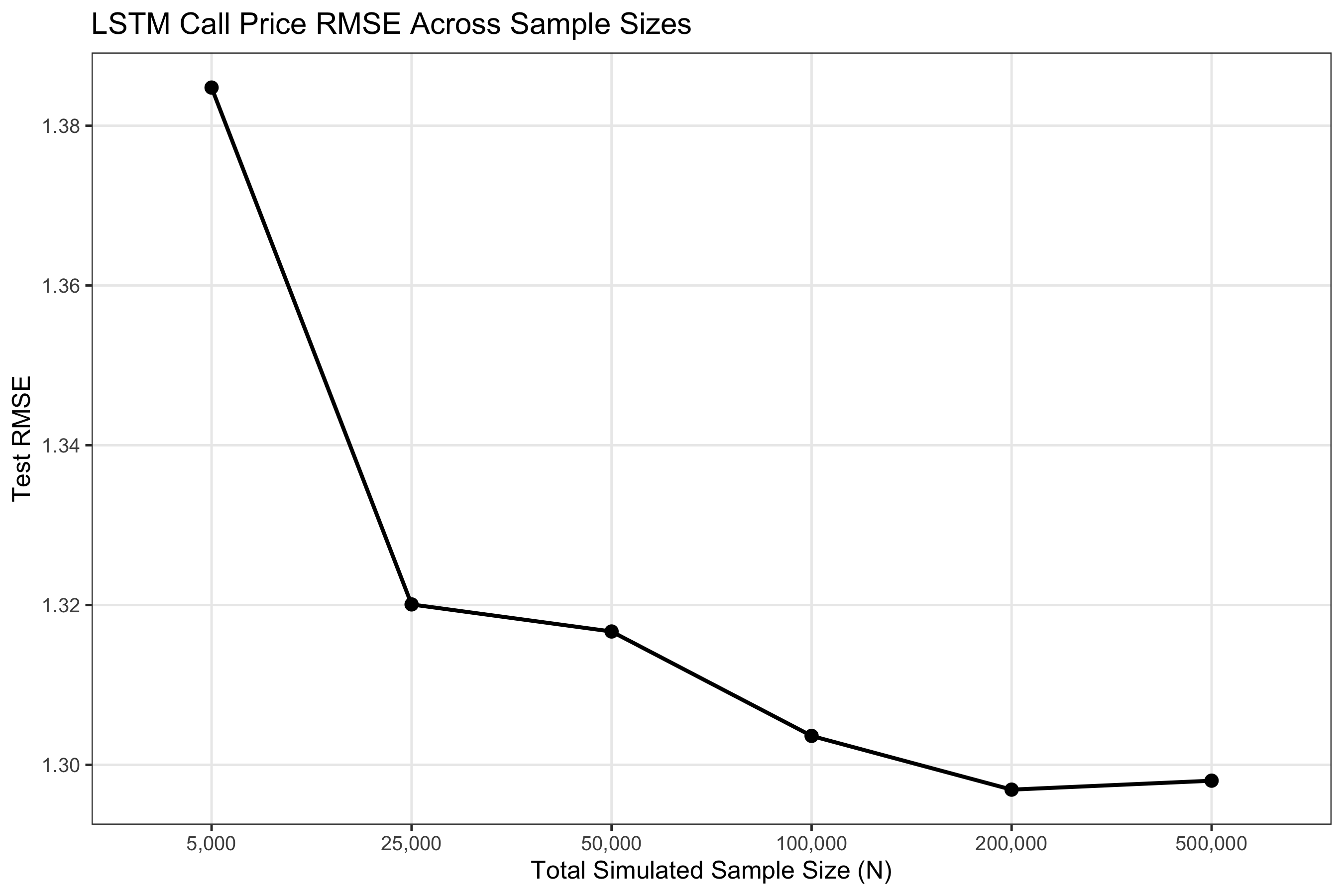}
        \caption{Call-price RMSE.}
    \end{subfigure}
    \hfill
    \begin{subfigure}[t]{0.48\textwidth}
        \centering
        \includegraphics[width=\textwidth]
        {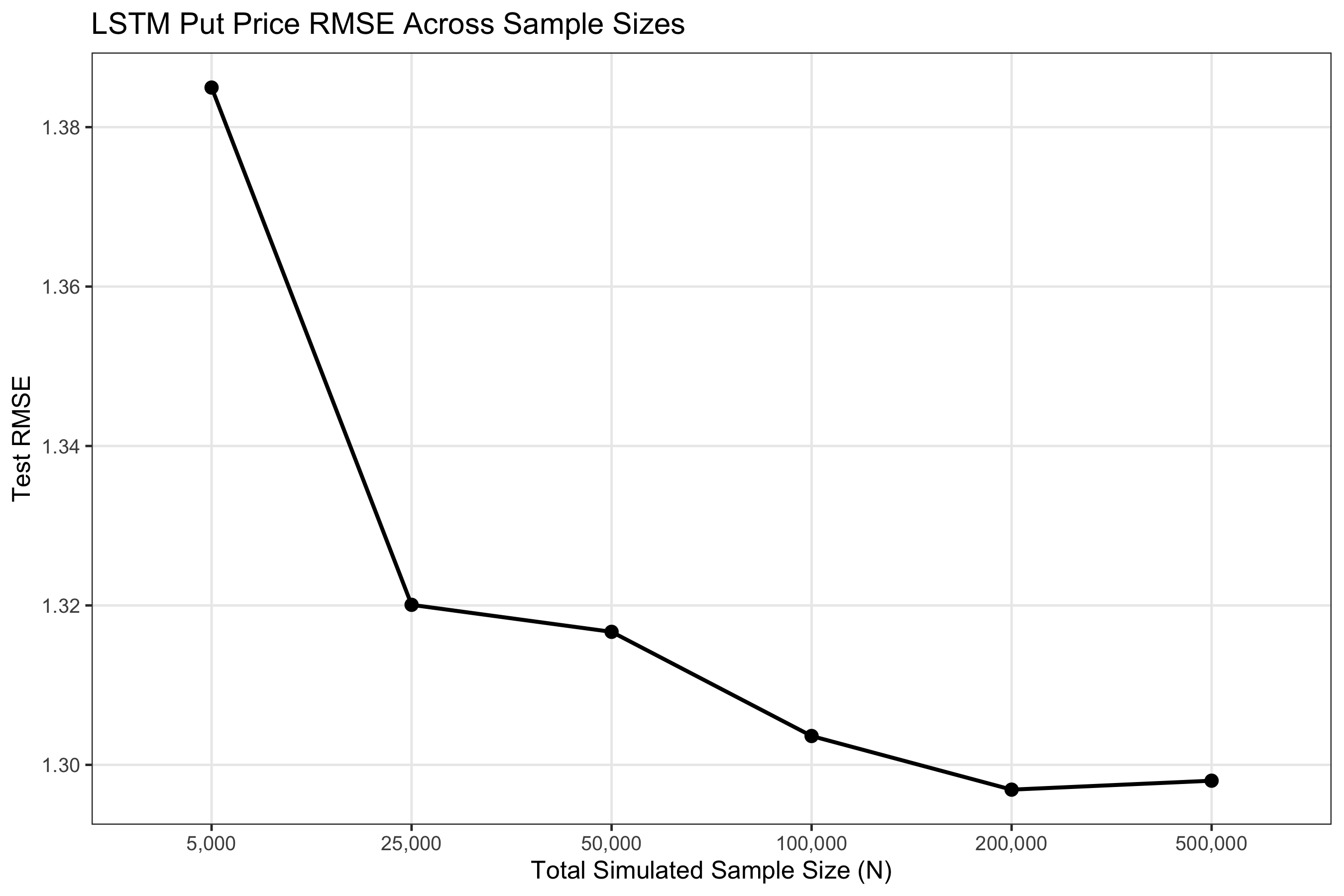}
        \caption{Put-price RMSE.}
    \end{subfigure}

    \vspace{0.3cm}

    \begin{subfigure}[t]{0.48\textwidth}
        \centering
        \includegraphics[width=\textwidth]
        {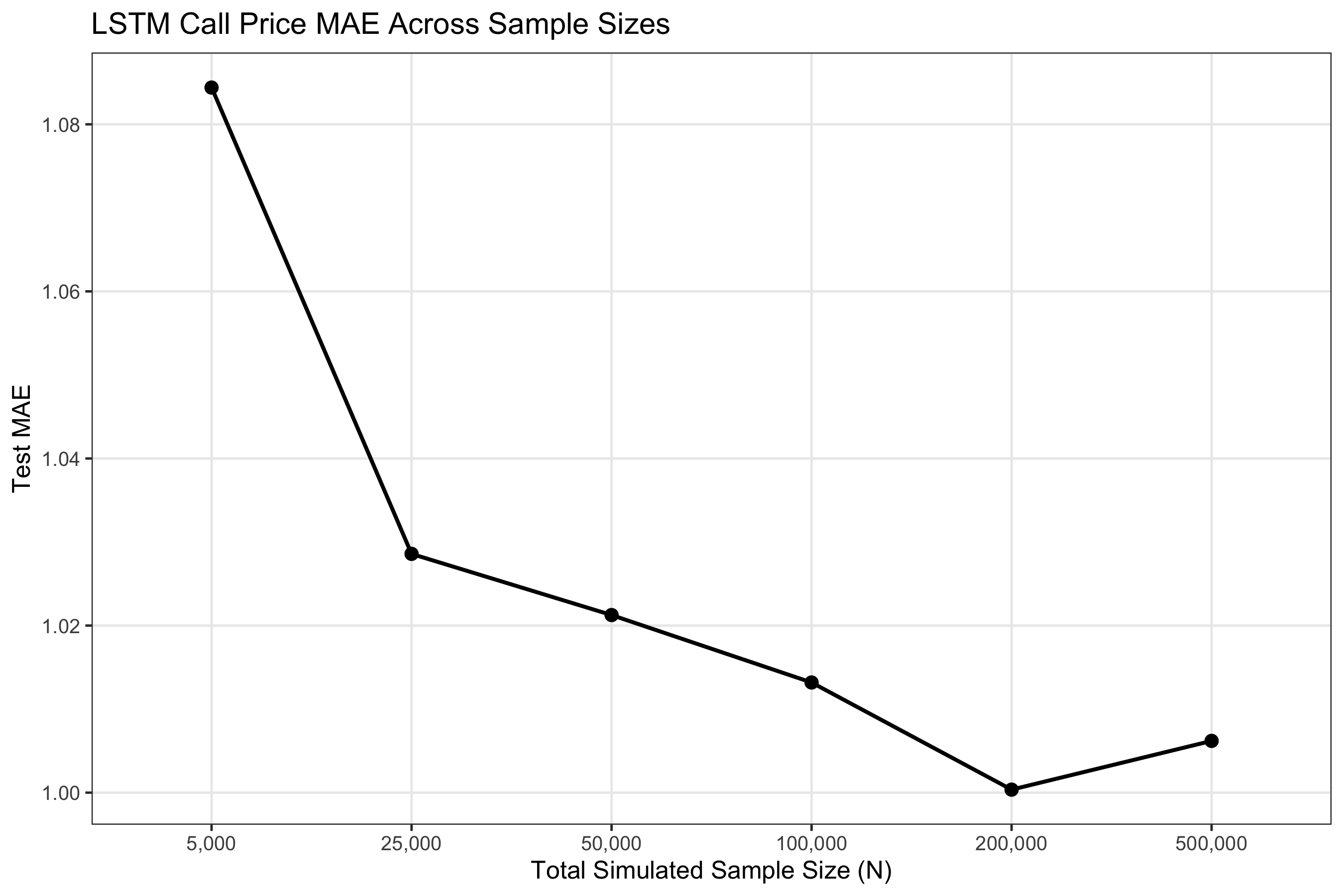}
        \caption{Call-price MAE.}
    \end{subfigure}
    \hfill
    \begin{subfigure}[t]{0.48\textwidth}
        \centering
        \includegraphics[width=\textwidth]
        {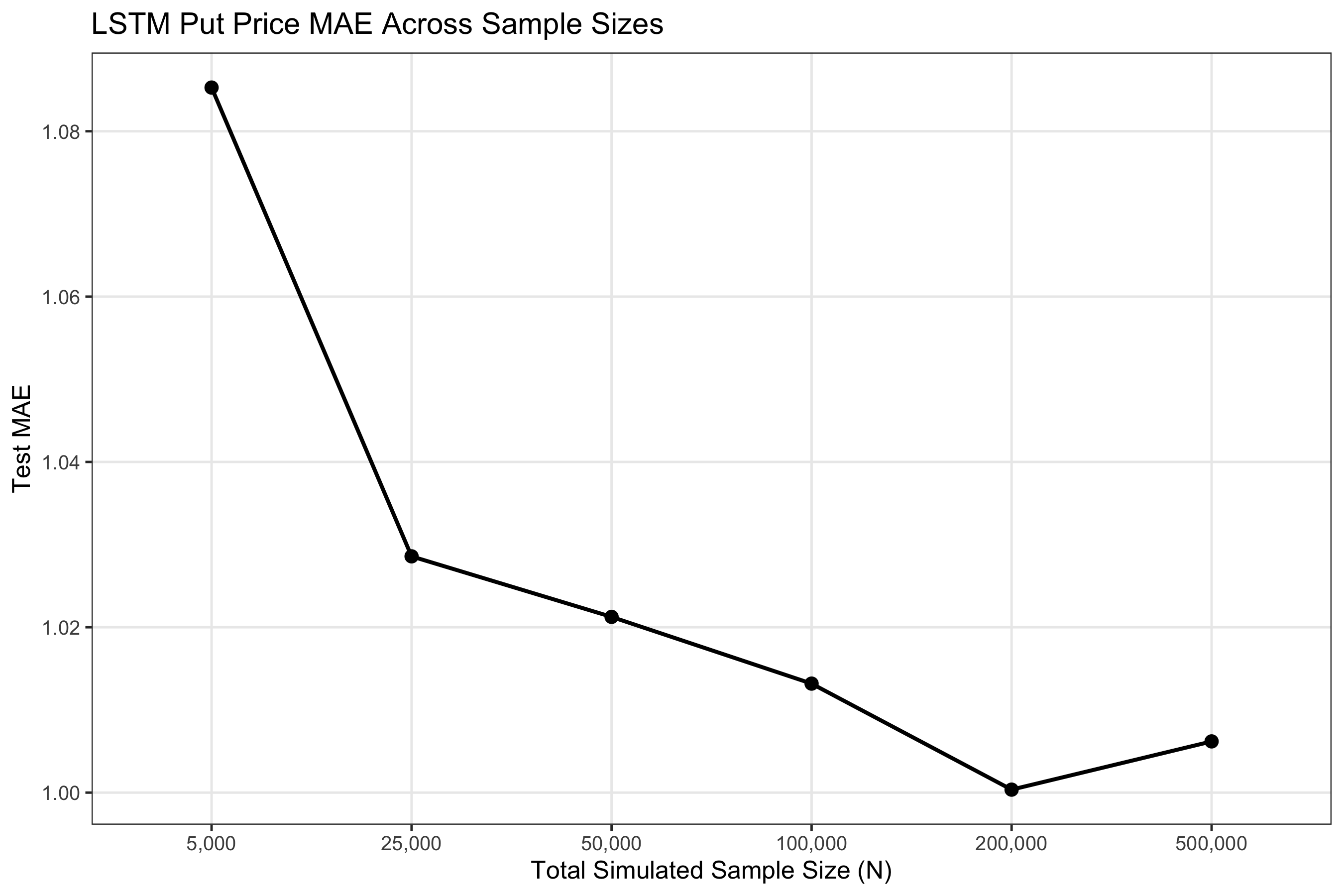}
        \caption{Put-price MAE.}
    \end{subfigure}

\caption{LSTM test-set prediction errors for European call
and put prices across total simulated sample sizes. The
upper panels show RMSE, while the lower panels show MAE.
Each plotted point is based on 1,000, 5,000, 10,000,
20,000, 40,000, and 100,000 simulated test sequences,
corresponding to total sample sizes of 5,000, 25,000,
50,000, 100,000, 200,000, and 500,000, respectively.}

    \label{fig:lstm_error_sample_size}
\end{figure}

Figure~\ref{fig:lstm_error_sample_size} provides a visual summary of
the option-price prediction errors. The largest improvement occurs
between the smaller sample sizes, while subsequent changes become
progressively smaller.

\begin{figure}[htbp]
    \centering

    \begin{subfigure}[t]{0.48\textwidth}
        \centering
        \includegraphics[width=\textwidth]
        {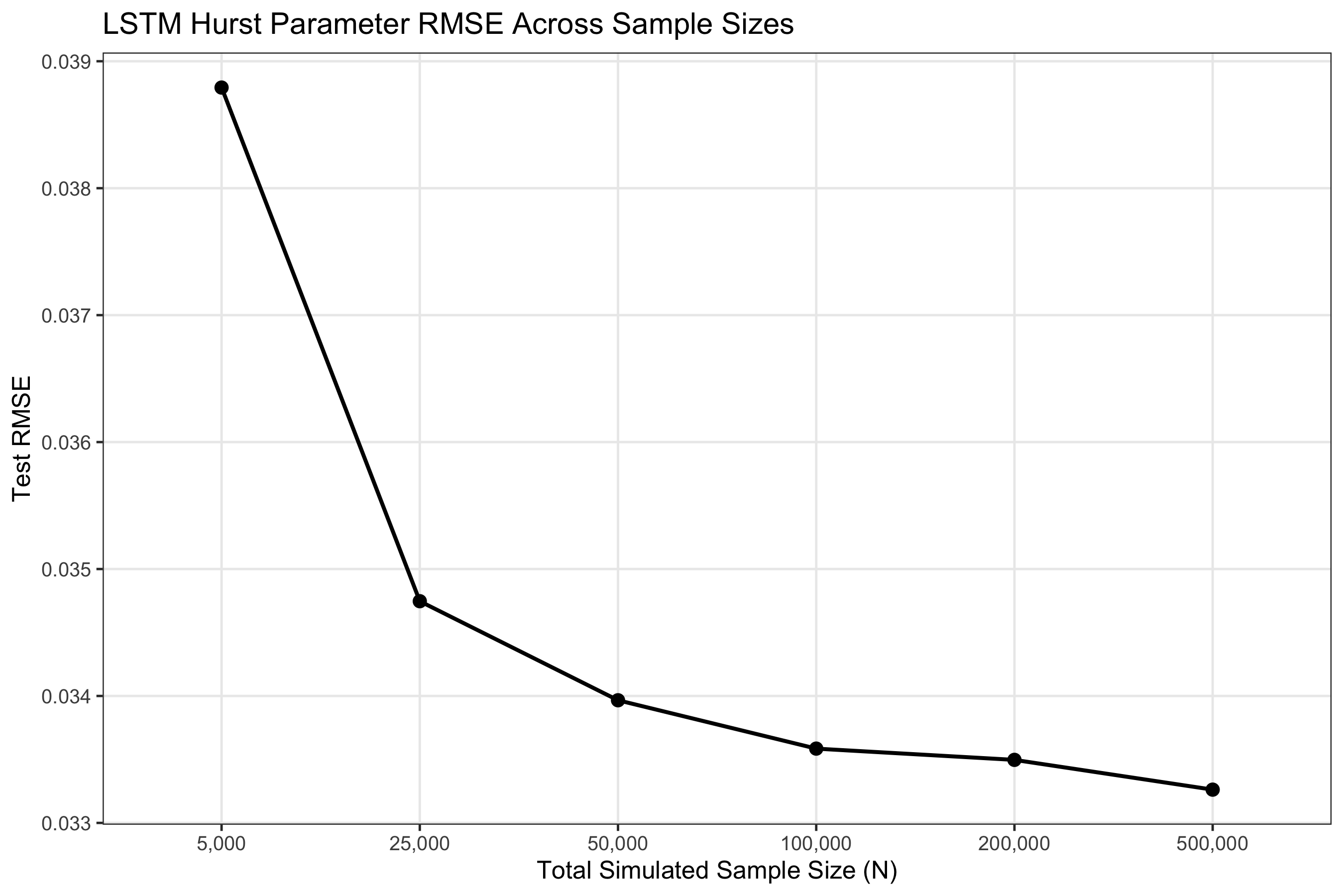}
        \caption{Hurst-parameter RMSE.}
    \end{subfigure}
    \hfill
    \begin{subfigure}[t]{0.48\textwidth}
        \centering
        \includegraphics[width=\textwidth]
        {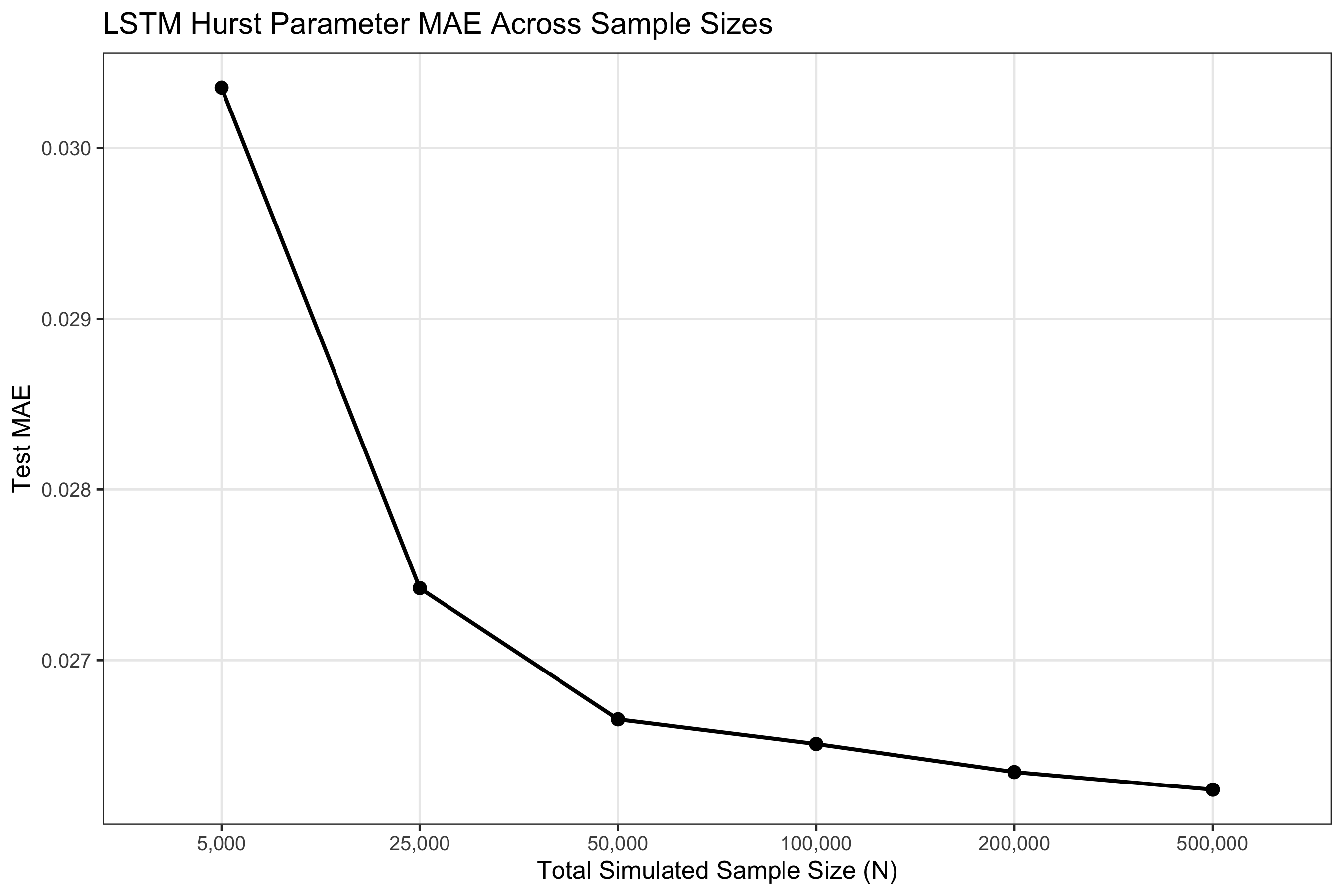}
        \caption{Hurst-parameter MAE.}
    \end{subfigure}

\caption{LSTM test-set prediction errors for the Hurst
parameter across total simulated sample sizes. The left
panel shows RMSE, and the right panel shows MAE. Each
plotted point is based on 1,000, 5,000, 10,000, 20,000,
40,000, and 100,000 simulated test sequences,
corresponding to total sample sizes of 5,000, 25,000,
50,000, 100,000, 200,000, and 500,000, respectively.}

    \label{fig:lstm_H_sample_size}
\end{figure}

Figure~\ref{fig:lstm_H_sample_size} presents the corresponding error
patterns for the Hurst parameter. Both RMSE and MAE decrease
consistently as the simulated sample size increases, indicating
that the LSTM benefits from additional training data when
learning the dependence structure of fractional Gaussian
return sequences. The largest improvement occurs between
the smaller sample sizes, while the reductions become
progressively smaller at larger sample sizes. This suggests
that the LSTM achieves increasingly stable predictive
performance as more training observations become available,
with diminishing gains from further increases in sample size.
Overall, these findings support the effectiveness of the
LSTM architecture for estimating the Hurst parameter and
suggest that relatively accurate predictions can be achieved
without requiring the largest simulated datasets.

\paragraph{Empirical Prediction-Interval Performance.}

In addition to point-prediction accuracy, empirical \(95\%\) prediction
intervals were constructed for each output using the empirical
distribution of the validation-set residuals. Specifically, the
\(2.5\%\) and \(97.5\%\) quantiles of the validation residuals were
added to the corresponding LSTM test-set point predictions to obtain
the lower and upper prediction limits. The resulting intervals were
then evaluated on the independent test set.
Table~\ref{tab:lstm_interval_performance} reports both empirical
test-set coverage and average interval width. Reporting these
quantities jointly is important because satisfactory coverage is most
informative when it is achieved without unnecessarily wide prediction
intervals.

\begin{table}[H]
\centering
\caption{Empirical coverage and average width of the LSTM \(95\%\)
prediction intervals across total simulated sample sizes.}
\label{tab:lstm_interval_performance}
\begin{tabular}{clcc}
\toprule
Sample Size & Quantity & Coverage (\%) & Average Width \\
\midrule

\multirow{3}{*}{5,000}
& Call Price       & 94.900 & 5.524 \\
& Put Price        & 94.900 & 5.367 \\
& Hurst Parameter  & 93.000 & 0.137 \\
\midrule

\multirow{3}{*}{25,000}
& Call Price       & 95.460 & 5.399 \\
& Put Price        & 95.460 & 5.220 \\
& Hurst Parameter  & 94.440 & 0.129 \\
\midrule

\multirow{3}{*}{50,000}
& Call Price       & 95.170 & 5.340 \\
& Put Price        & 95.170 & 5.158 \\
& Hurst Parameter  & 94.410 & 0.126 \\
\midrule

\multirow{3}{*}{100,000}
& Call Price       & 94.820 & 5.189 \\
& Put Price        & 94.820 & 5.028 \\
& Hurst Parameter  & 94.805 & \textbf{0.125} \\
\midrule

\multirow{3}{*}{200,000}
& Call Price       & 94.805 & 5.184 \\
& Put Price        & 94.805 & 5.021 \\
& Hurst Parameter  & 94.920 & 0.127 \\
\midrule

\multirow{3}{*}{500,000}
& Call Price       & 94.870 & \textbf{5.177} \\
& Put Price        & 94.870 & \textbf{5.011} \\
& Hurst Parameter  & 94.909 & 0.126 \\
\bottomrule
\end{tabular}
\end{table}

Coverage for the call- and put-price prediction intervals remains
close to the nominal \(95\%\) level across all six sample sizes,
ranging from \(94.805\%\) to \(95.460\%\). The corresponding average
interval widths generally decrease as the amount of simulated data
increases. For the call price, the average width decreases from
\(5.524\) at \(N=5{,}000\) to \(5.177\) at \(N=500{,}000\), while
the put-price interval width decreases from \(5.367\) to \(5.011\).
Thus, the option-price intervals become narrower at larger sample
sizes while maintaining coverage close to the nominal level.

For the Hurst parameter, empirical coverage is lower at the smallest
sample size, with a value of \(93.000\%\) at \(N=5{,}000\). Coverage
increases to \(94.440\%\) at \(N=25{,}000\) and remains close to the
nominal \(95\%\) level for the larger sample sizes, reaching
\(94.909\%\) at \(N=500{,}000\). The average Hurst-parameter interval
width decreases from \(0.137\) at \(N=5{,}000\) to \(0.125\) at
\(N=100{,}000\). At the two largest sample sizes, the width remains
essentially stable, with values of \(0.127\) and \(0.126\) for
\(N=200{,}000\) and \(N=500{,}000\), respectively. This pattern
suggests that most of the gain in interval precision for the Hurst
parameter is achieved by approximately \(N=100{,}000\).

\begin{figure}[H]
\centering
\includegraphics[width=0.80\textwidth]
{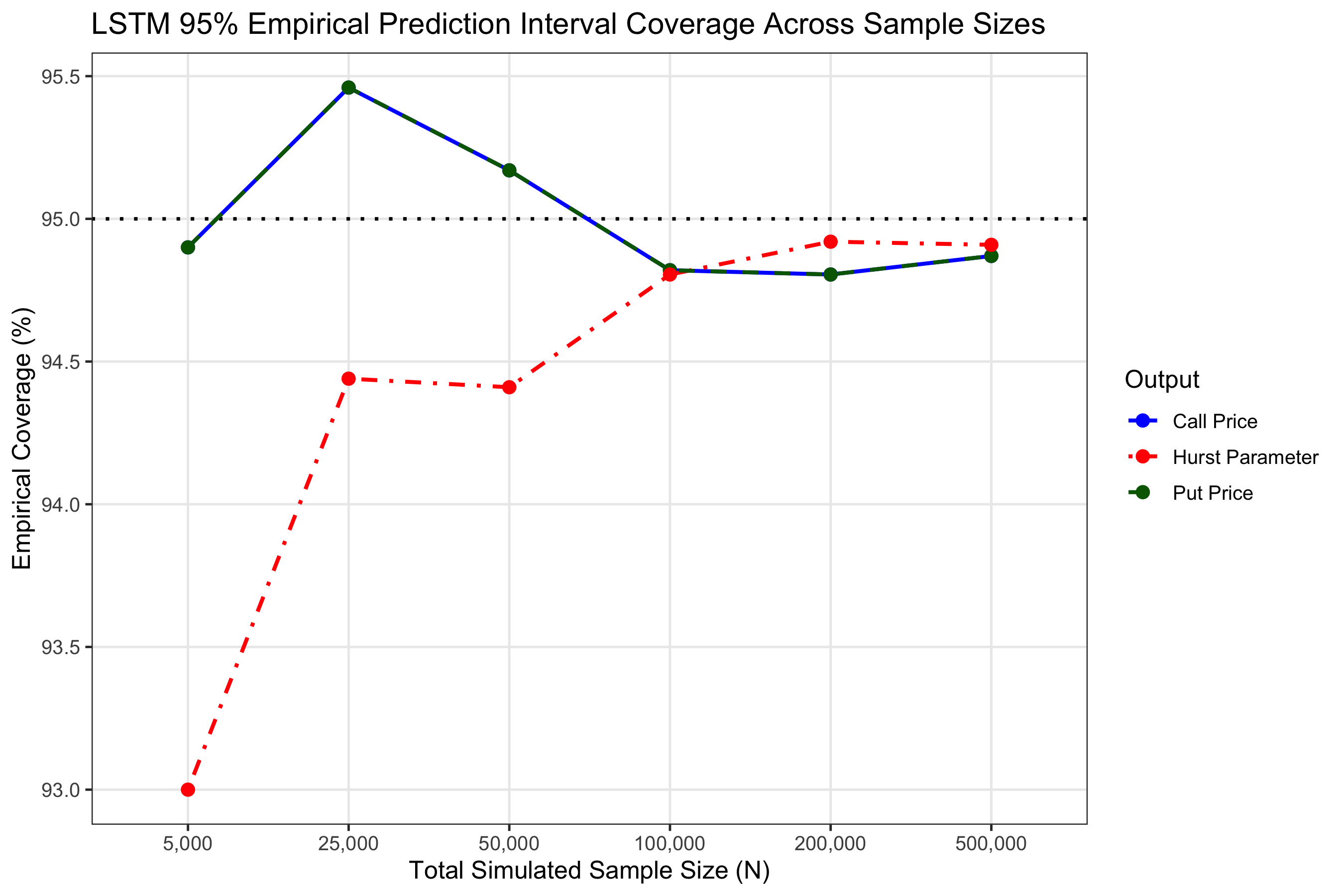}
\caption{Empirical coverage of the LSTM \(95\%\) prediction intervals
for the call price, put price, and Hurst parameter across total
simulated sample sizes. The horizontal dotted line indicates the
nominal \(95\%\) coverage level.}
\label{fig:lstm_coverage}
\end{figure}

Figure~\ref{fig:lstm_coverage} provides a graphical comparison of the
empirical coverage probabilities across sample sizes. The horizontal
dotted line represents the nominal \(95\%\) coverage level. The call-
and put-price coverage curves nearly overlap throughout the six
experiments and remain close to the nominal level. In contrast, the
Hurst-parameter coverage begins below the nominal level at the
smallest sample size and moves progressively closer to \(95\%\) as
the sample size increases.

Overall, the LSTM prediction intervals provide approximately nominal
coverage for the option-price outputs across all sample sizes, while
the Hurst-parameter intervals approach nominal coverage as the sample
size increases. At the same time, the reduction in average interval
width indicates improved predictive precision, with the gains becoming
comparatively small at the largest sample sizes.

\paragraph{Detailed Predictive Performance at \(N=500{,}000\).}

Because the LSTM model exhibited strong and relatively stable predictive
performance at the largest simulated sample sizes, the model trained using
\(N=500{,}000\) was selected for more detailed diagnostic evaluation.
Although the minimum call- and put-price RMSE and MAE values were attained
at \(N=200{,}000\), the differences between the \(N=200{,}000\) and
\(N=500{,}000\) experiments were negligible. In addition, the
\(N=500{,}000\) model achieved the highest correlation and \(R^2\) values
for the option-price outputs and the strongest overall performance for
prediction of the Hurst parameter. The largest-sample model was therefore
retained as the final LSTM specification for detailed diagnostic analysis
and subsequent model comparisons.

\begin{figure}[htbp]
    \centering

    \begin{subfigure}[t]{0.32\textwidth}
        \centering
        \includegraphics[width=\textwidth]
        {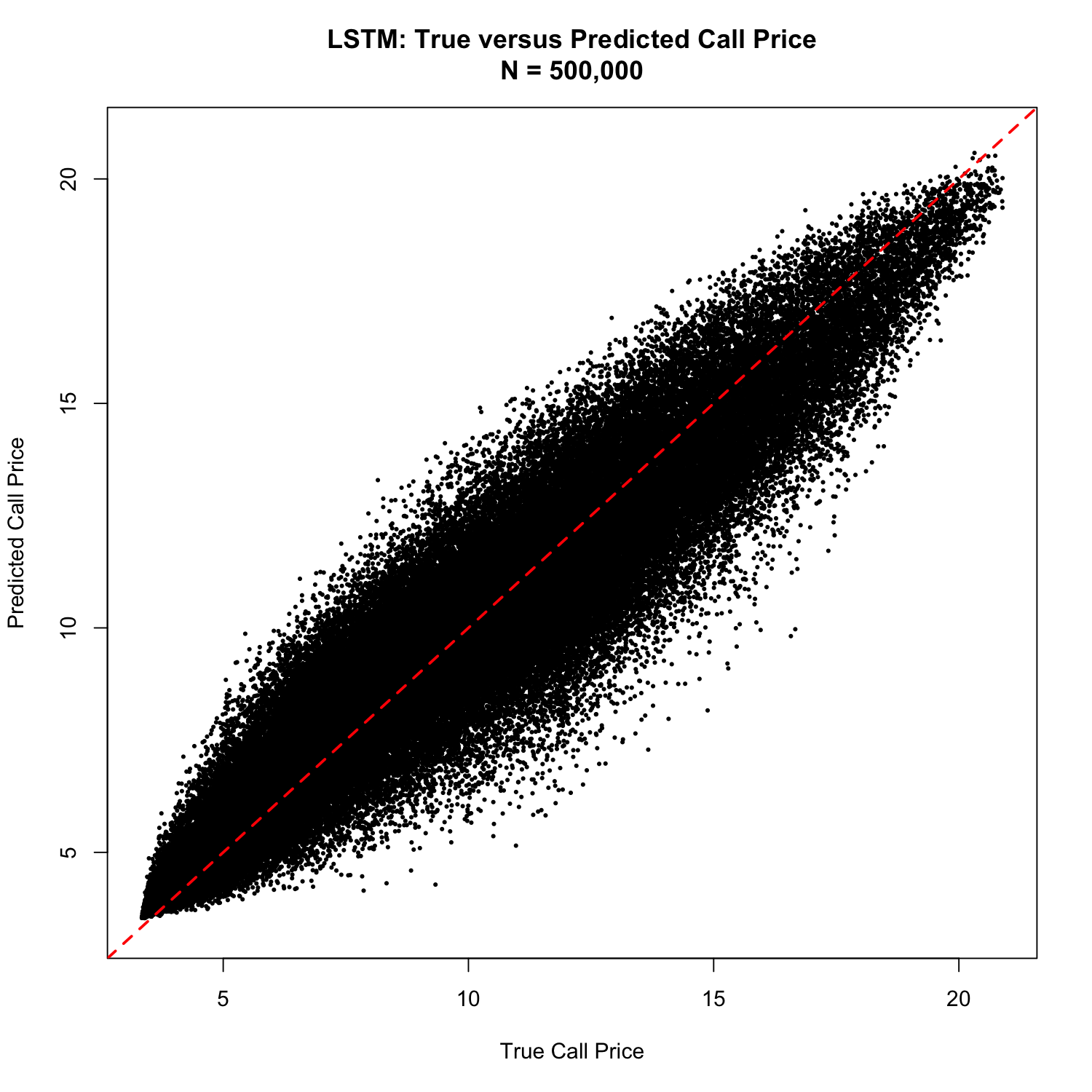}
        \caption{Call price.}
    \end{subfigure}
    \hfill
    \begin{subfigure}[t]{0.32\textwidth}
        \centering
        \includegraphics[width=\textwidth]
        {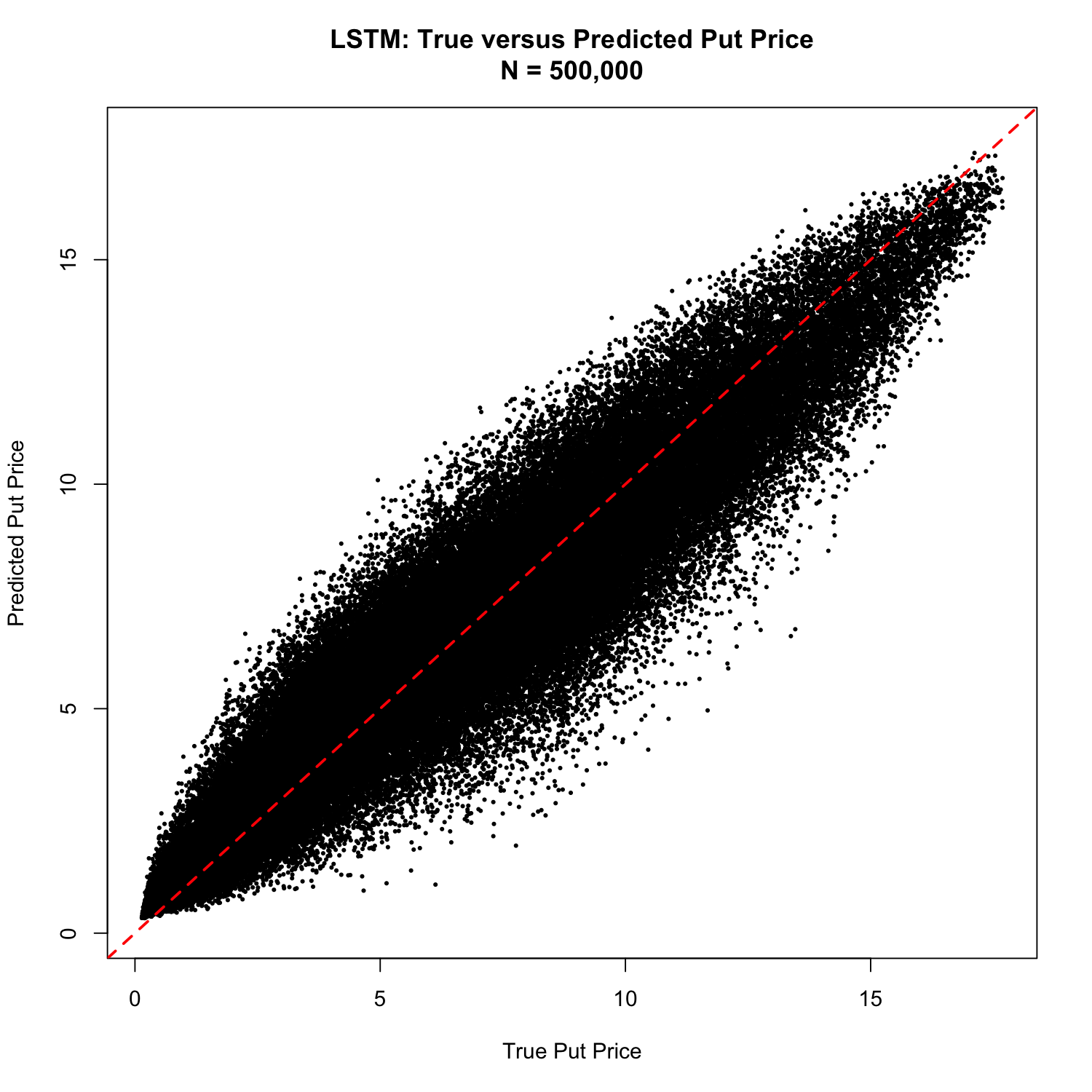}
        \caption{Put price.}
    \end{subfigure}
    \hfill
    \begin{subfigure}[t]{0.32\textwidth}
        \centering
        \includegraphics[width=\textwidth]
        {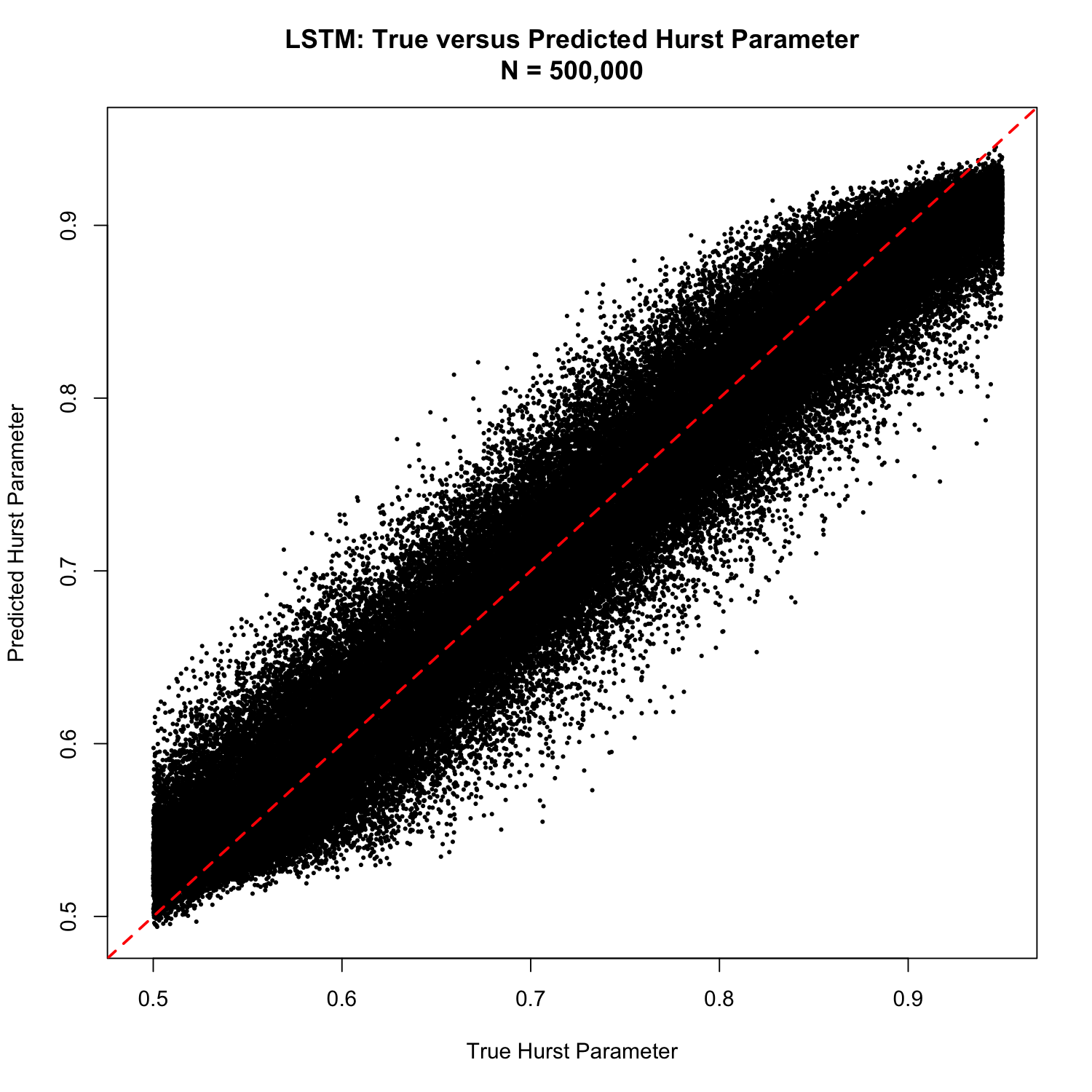}
        \caption{Hurst parameter.}
    \end{subfigure}

    \caption{True versus predicted test-set values from the LSTM trained
    using a total simulated sample size of \(N=500{,}000\). The dashed
    \(45^\circ\) line represents perfect agreement between the true and
    predicted values.}
    \label{fig:lstm_true_pred_500k}
\end{figure}

Figure~\ref{fig:lstm_true_pred_500k} compares the true simulated values
with the corresponding LSTM predictions for the call price, put price,
and Hurst parameter. The dashed \(45^\circ\) line represents perfect
agreement between the true and predicted values.

The predicted call and put prices exhibit strong agreement with their
corresponding simulated values. This visual agreement is consistent with
the test-set results, for which both option-price outputs achieve an RMSE
of approximately \(1.298\), an MAE of approximately \(1.006\), a
correlation of \(0.945\), and an \(R^2\) of \(0.892\) at
\(N=500{,}000\). The call- and put-price plots are expected to display
nearly identical patterns because the two outputs have essentially
identical aggregate prediction measures.

The Hurst-parameter predictions also show strong agreement with the true
simulated values. At \(N=500{,}000\), the LSTM achieves an RMSE of
approximately \(0.033\), an MAE of \(0.026\), a correlation of \(0.967\),
and an \(R^2\) of \(0.934\). These results indicate that the recurrent
architecture is able to extract substantial information about the
long-range dependence structure directly from the ordered return
sequences. Any remaining departures from the \(45^\circ\) line,
particularly near the boundaries of the simulated Hurst-parameter range,
can be examined more closely through the residual diagnostics presented
below.

\paragraph{Training and Validation Behavior.}

Table~\ref{tab:lstm_training} summarizes the training and validation
behavior of the LSTM across the six experiments. Training was permitted
to continue for a maximum of \(250\) epochs; however, the actual number
of completed epochs was determined by the early-stopping criterion.
The table therefore reports the completed epochs, selected best epoch,
best validation loss, best validation MAE, and total training time.

\begin{table}[H]
\centering
\caption{LSTM training and validation behavior across total simulated sample sizes.}
\label{tab:lstm_training}
\begin{tabular}{rrrrrrr}
\toprule
Sample Size &
Training Size &
Epochs &
Best Epoch &
Validation Loss &
Validation MAE &
Time (s) \\
\midrule
5,000   & 3,200   & 250 & 225 & 0.110 & 0.261 & 354.486 \\
25,000  & 16,000  & 245 & 220 & 0.096 & 0.241 & 1869.443 \\
50,000  & 32,000  & 179 & 154 & 0.094 & 0.237 & 2598.277 \\
100,000 & 64,000  & 154 & 129 & 0.093 & 0.236 & 4807.087 \\
200,000 & 128,000 & 133 & 108 & \textbf{0.092} &
\textbf{0.234} & 8264.781 \\
500,000 & 320,000 & 111 & 86 & \textbf{0.092} &
\textbf{0.234} & \(17{,}250.000^{*}\) \\
\bottomrule
\end{tabular}

\begin{flushleft}
\footnotesize
\(^{*}\)Estimated active computation time. The original run was
interrupted, so the recorded wall-clock elapsed time was not used.
The estimate was obtained from the observed computational time during
uninterrupted portions of the experiment and the corresponding
training workload.
\end{flushleft}
\end{table}

The number of completed epochs generally decreases as the sample size
increases. The \(N=5{,}000\) experiment reached the maximum of
\(250\) epochs, whereas the \(N=500{,}000\) experiment completed
\(111\) epochs, with the best model selected at epoch \(86\).
This behavior indicates that, with larger training sets, the
early-stopping criterion identified the optimal validation performance
after fewer passes through the complete training data.

Validation performance also improves as the amount of simulated data
increases. The best validation loss decreases from \(0.110\) at
\(N=5{,}000\) to approximately \(0.092\) at \(N=500{,}000\), while
the corresponding best validation MAE decreases from \(0.261\) to
approximately \(0.234\). The magnitude of the improvement becomes
smaller at the larger sample sizes, suggesting that validation
performance begins to stabilize as the amount of training data
increases.

Training time generally increased with sample size, reflecting the
greater computational burden associated with fitting the LSTM to
larger training sets. The recorded training time increased from
approximately \(354\) seconds at \(N=5{,}000\) to \(8{,}265\) seconds
at \(N=200{,}000\). For \(N=500{,}000\), the active training time was
estimated to be approximately \(17{,}250\) seconds (\(4.79\) hours)
because the original run was interrupted. This estimate is therefore
reported separately from the directly recorded training times.

\paragraph{Residual Diagnostics.}

\begin{figure}[htbp]
    \centering

    \begin{subfigure}[t]{0.32\textwidth}
        \centering
        \includegraphics[width=\textwidth]
        {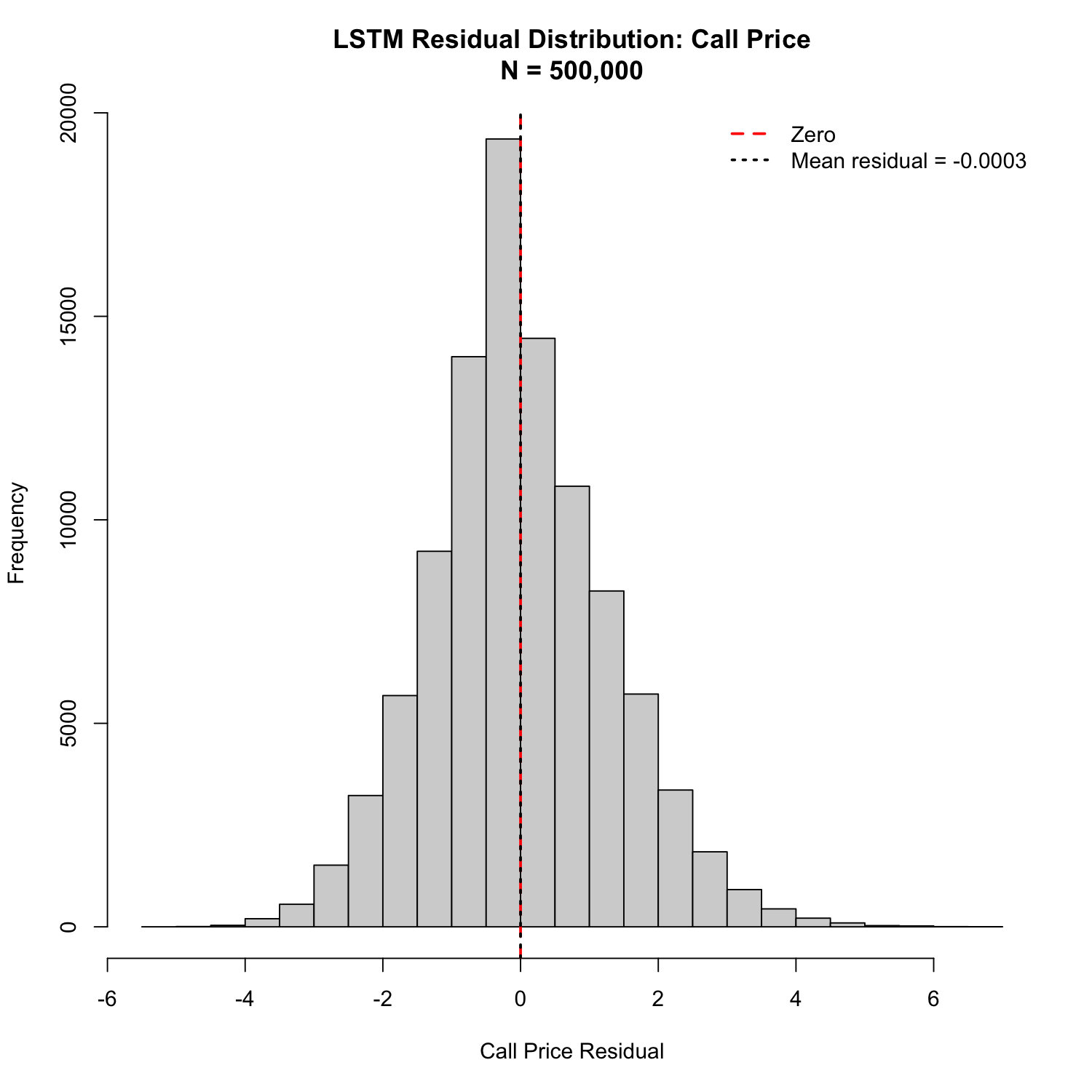}
        \caption{Call-price residual distribution.}
    \end{subfigure}
    \hfill
    \begin{subfigure}[t]{0.32\textwidth}
        \centering
        \includegraphics[width=\textwidth]
        {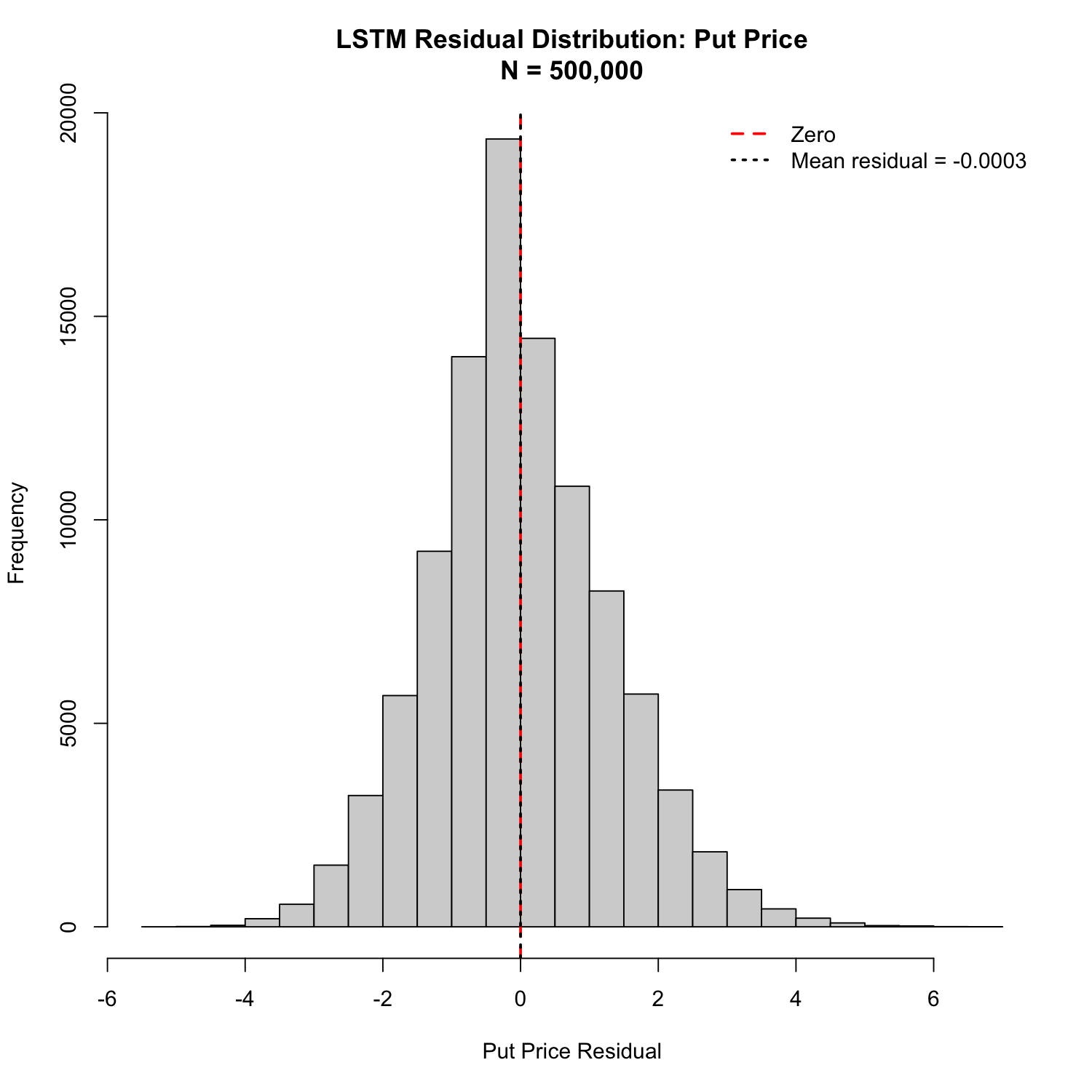}
        \caption{Put-price residual distribution.}
    \end{subfigure}
    \hfill
    \begin{subfigure}[t]{0.32\textwidth}
        \centering
        \includegraphics[width=\textwidth]
        {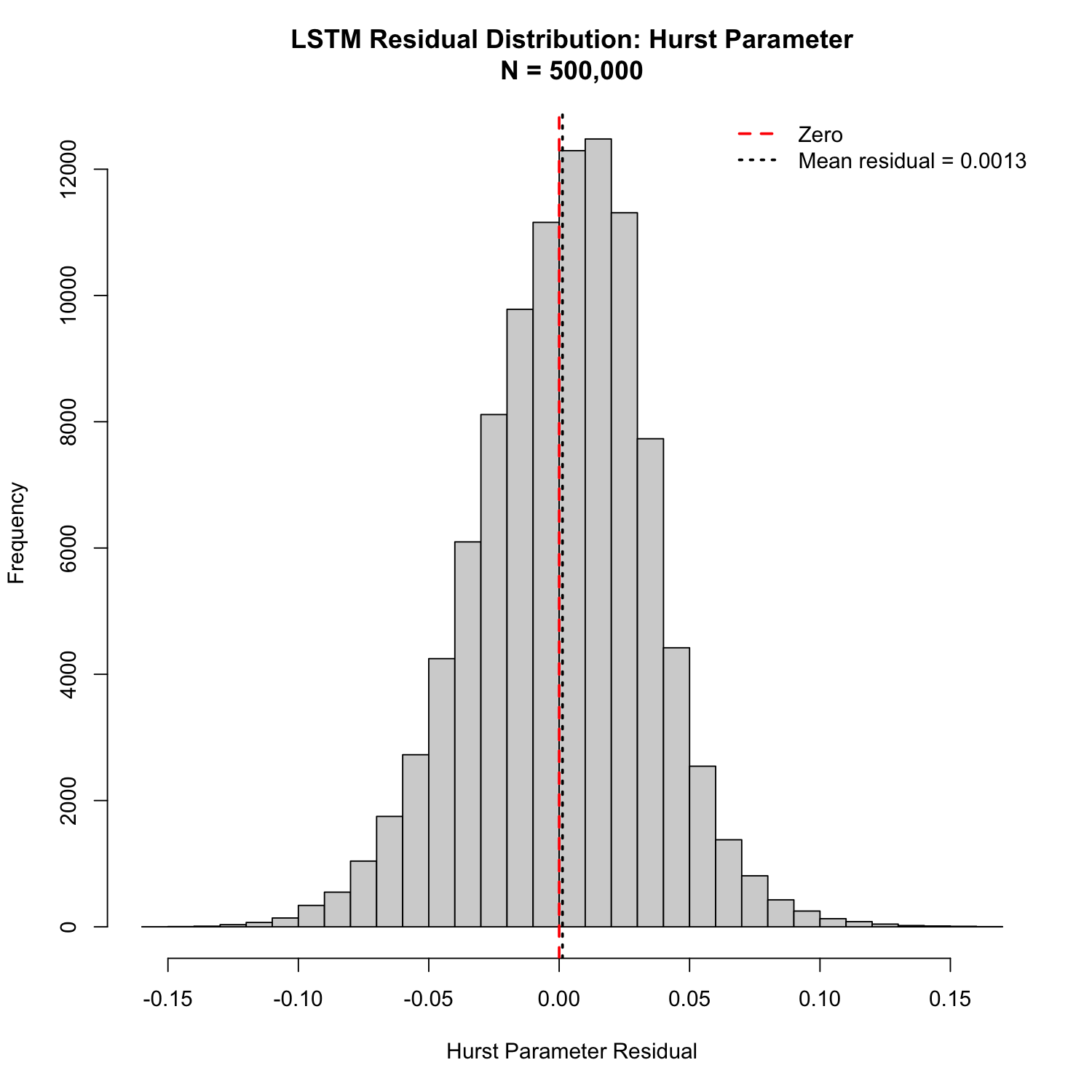}
        \caption{Hurst parameter residual distribution.}
    \end{subfigure}

    \vspace{0.3cm}

    \begin{subfigure}[t]{0.32\textwidth}
        \centering
        \includegraphics[width=\textwidth]
        {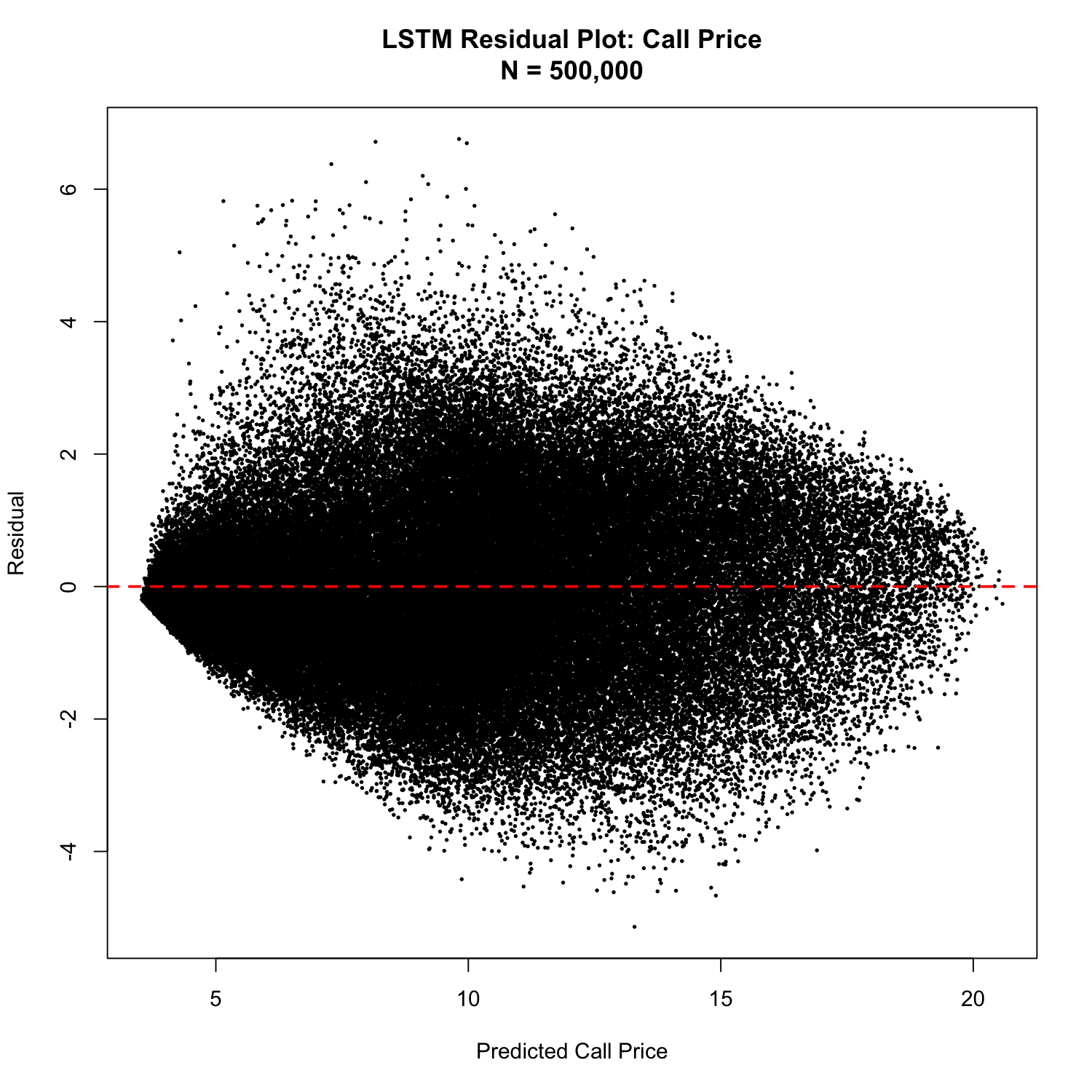}
        \caption{Call residuals versus predicted values.}
    \end{subfigure}
    \hfill
    \begin{subfigure}[t]{0.32\textwidth}
        \centering
        \includegraphics[width=\textwidth]
        {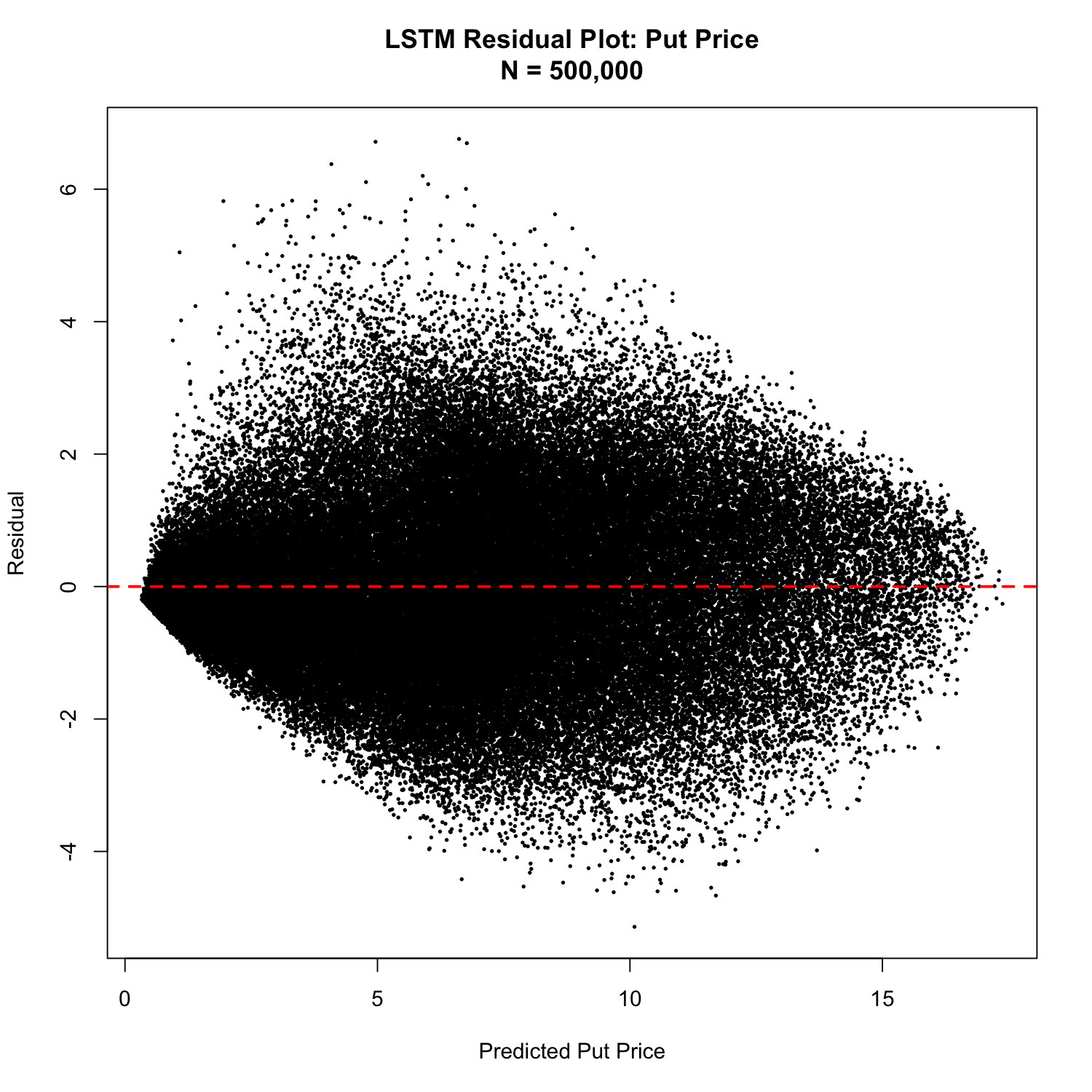}
        \caption{Put residuals versus predicted values.}
    \end{subfigure}
    \hfill
    \begin{subfigure}[t]{0.32\textwidth}
        \centering
        \includegraphics[width=\textwidth]
        {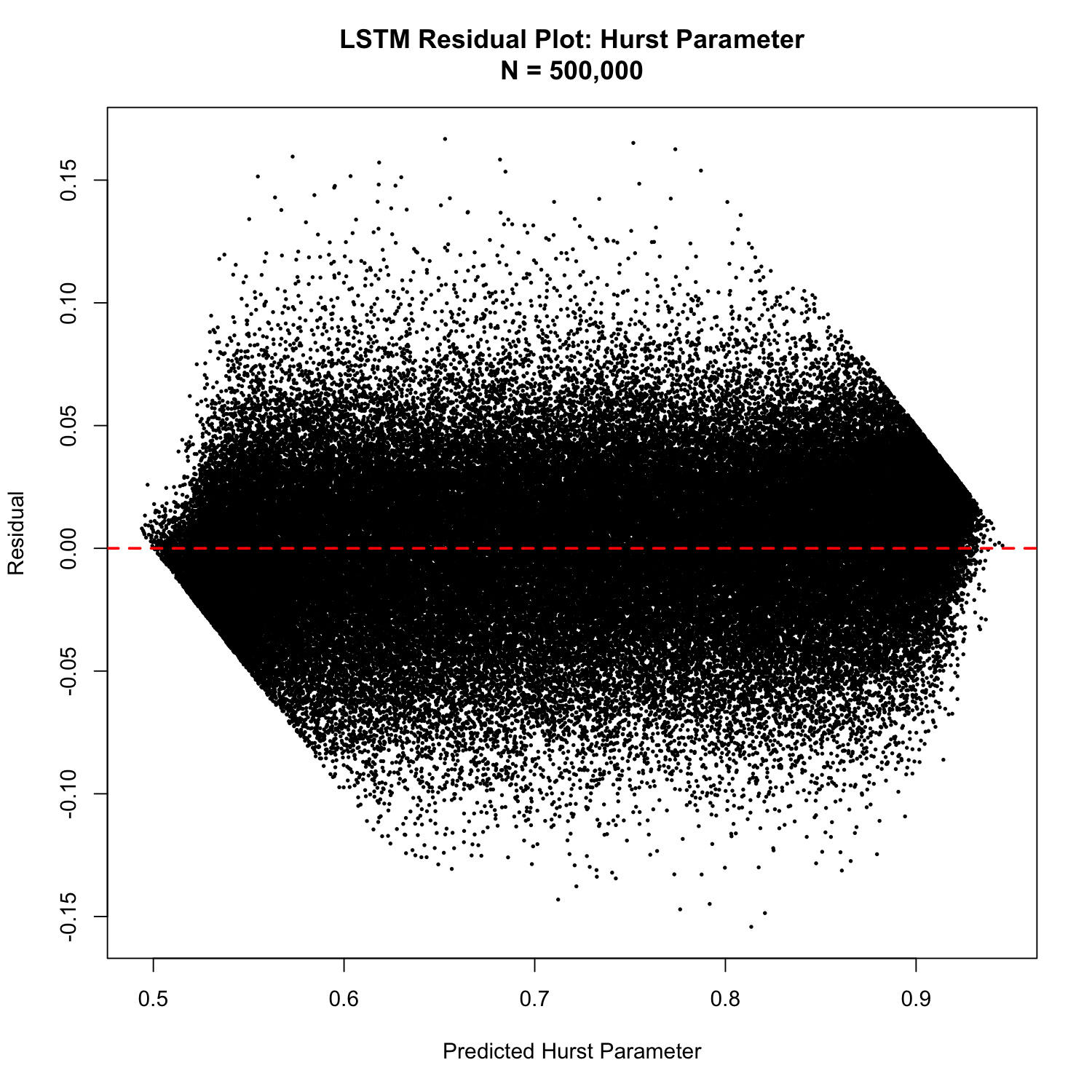}
        \caption{Hurst-parameter residuals versus predicted values.}
    \end{subfigure}

    \caption{Residual diagnostics for the LSTM trained using a total
    simulated sample size of \(N=500{,}000\). The upper panels show
    the residual distributions, while the lower panels show residuals
    versus the corresponding predicted values. Residuals are defined
    as the true simulated value minus the LSTM prediction.}
    \label{fig:lstm_residual_diagnostics}
\end{figure}

Residual diagnostics were examined for the \(N=500{,}000\) LSTM
using the independent test set, with residuals defined as the true
simulated value minus the corresponding LSTM prediction. The residual
distributions and residual-versus-predicted plots are presented in
Figure~\ref{fig:lstm_residual_diagnostics}.

For the call and put prices, the residual distributions are nearly
identical and centered very close to zero, with mean residuals of
approximately \(-0.0003\) for both outputs. Thus, there is little
evidence of systematic average overprediction or underprediction.
The residual-versus-predicted plots nevertheless show that the
dispersion of the prediction errors varies across the predicted-price
range, indicating that the remaining error is not completely
homogeneous.

For the Hurst parameter, the residual distribution is also centered
close to zero, with a mean residual of approximately \(0.001\),
indicating negligible average bias. However, the
residual-versus-predicted plot exhibits a structured pattern near the
boundaries of the predicted range. This behavior is consistent with
the compression toward the center observed in the true-versus-predicted
plot and indicates that prediction accuracy varies somewhat across the
simulated Hurst-parameter range.

\paragraph{Illustration of empirical residual-based prediction intervals.}

\begin{figure}[htbp]
    \centering

    \begin{subfigure}[t]{0.32\textwidth}
        \centering
        \includegraphics[width=\textwidth]
        {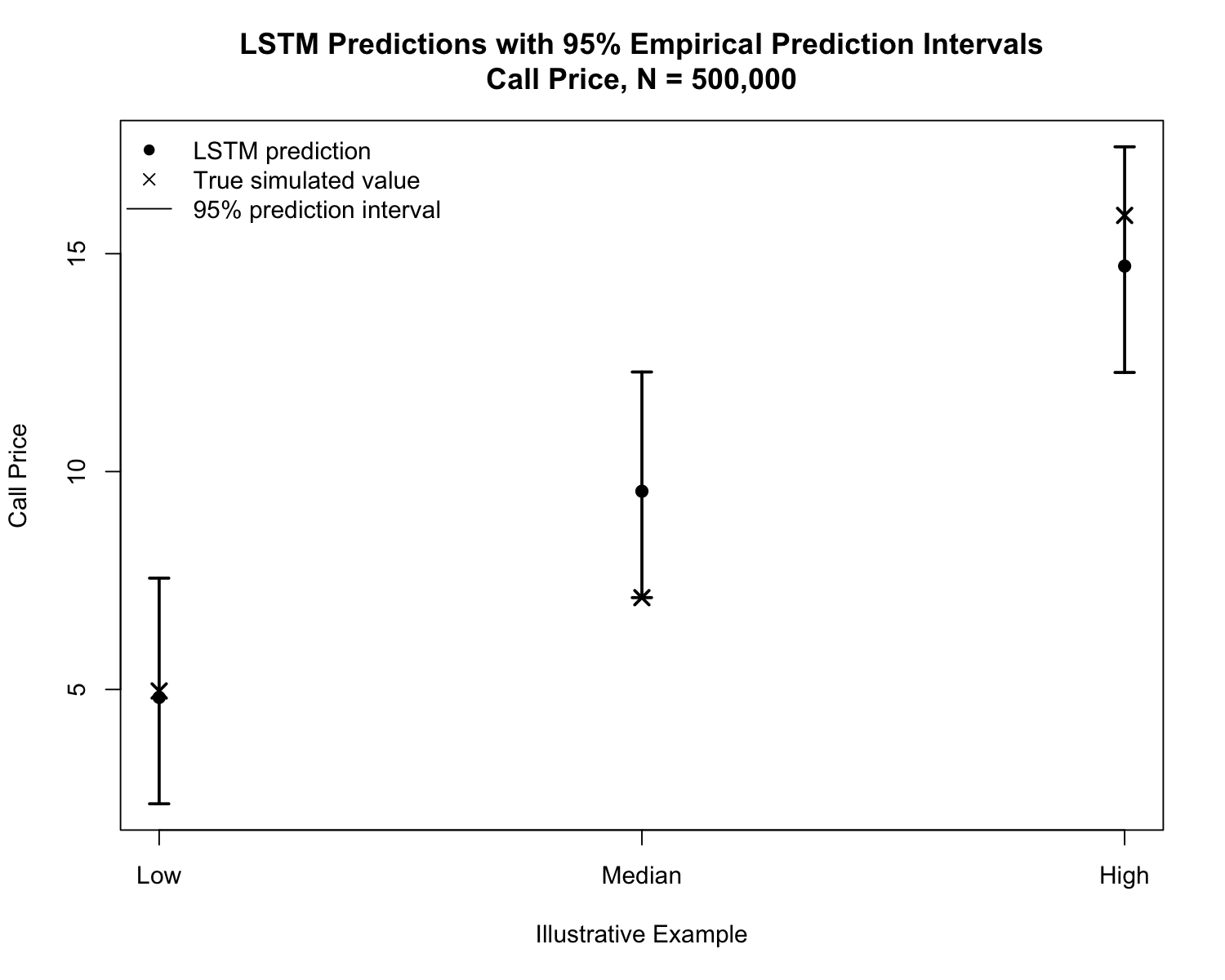}
        \caption{Call price.}
    \end{subfigure}
    \hfill
    \begin{subfigure}[t]{0.32\textwidth}
        \centering
        \includegraphics[width=\textwidth]
        {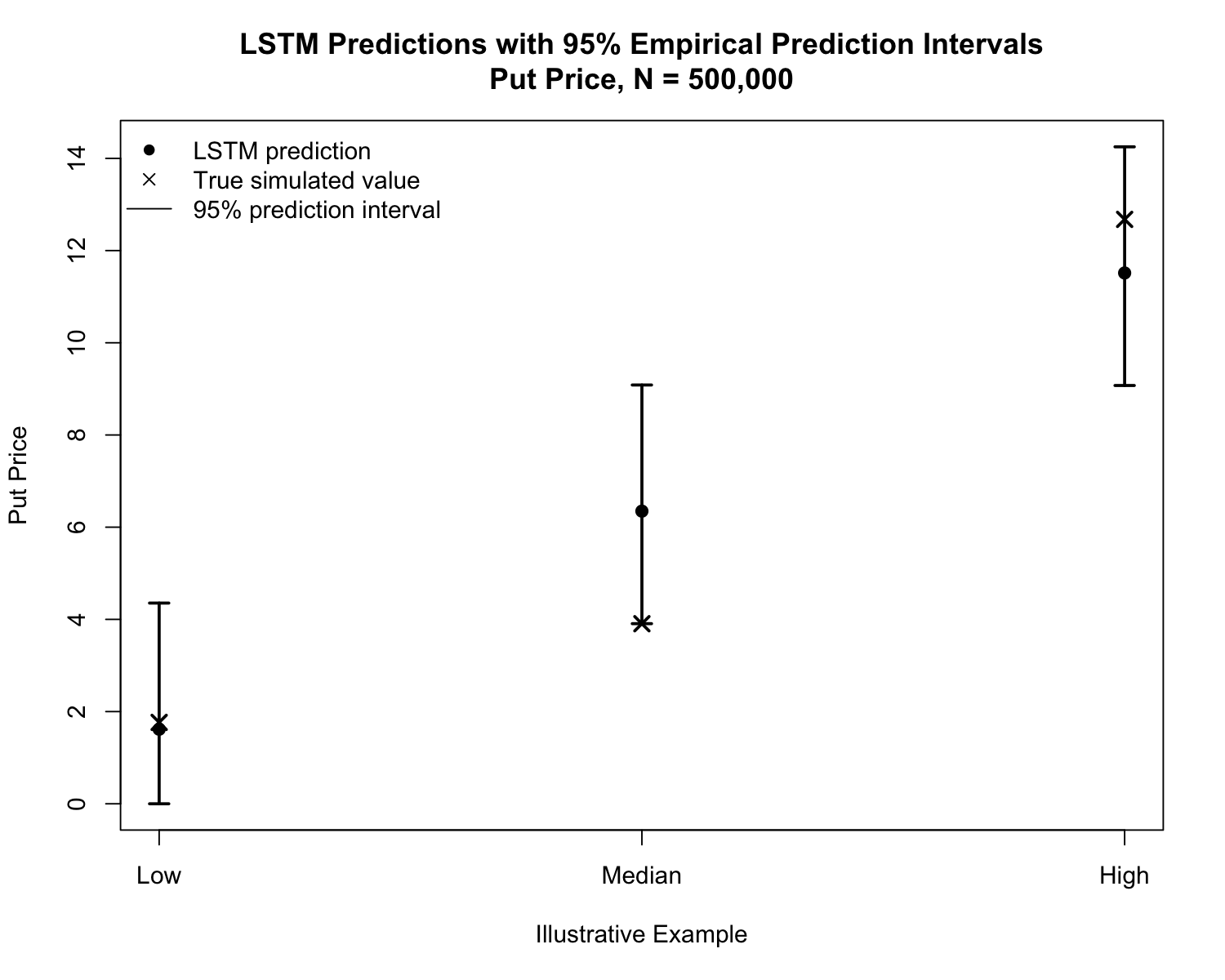}
        \caption{Put price.}
    \end{subfigure}
    \hfill
    \begin{subfigure}[t]{0.32\textwidth}
        \centering
        \includegraphics[width=\textwidth]
        {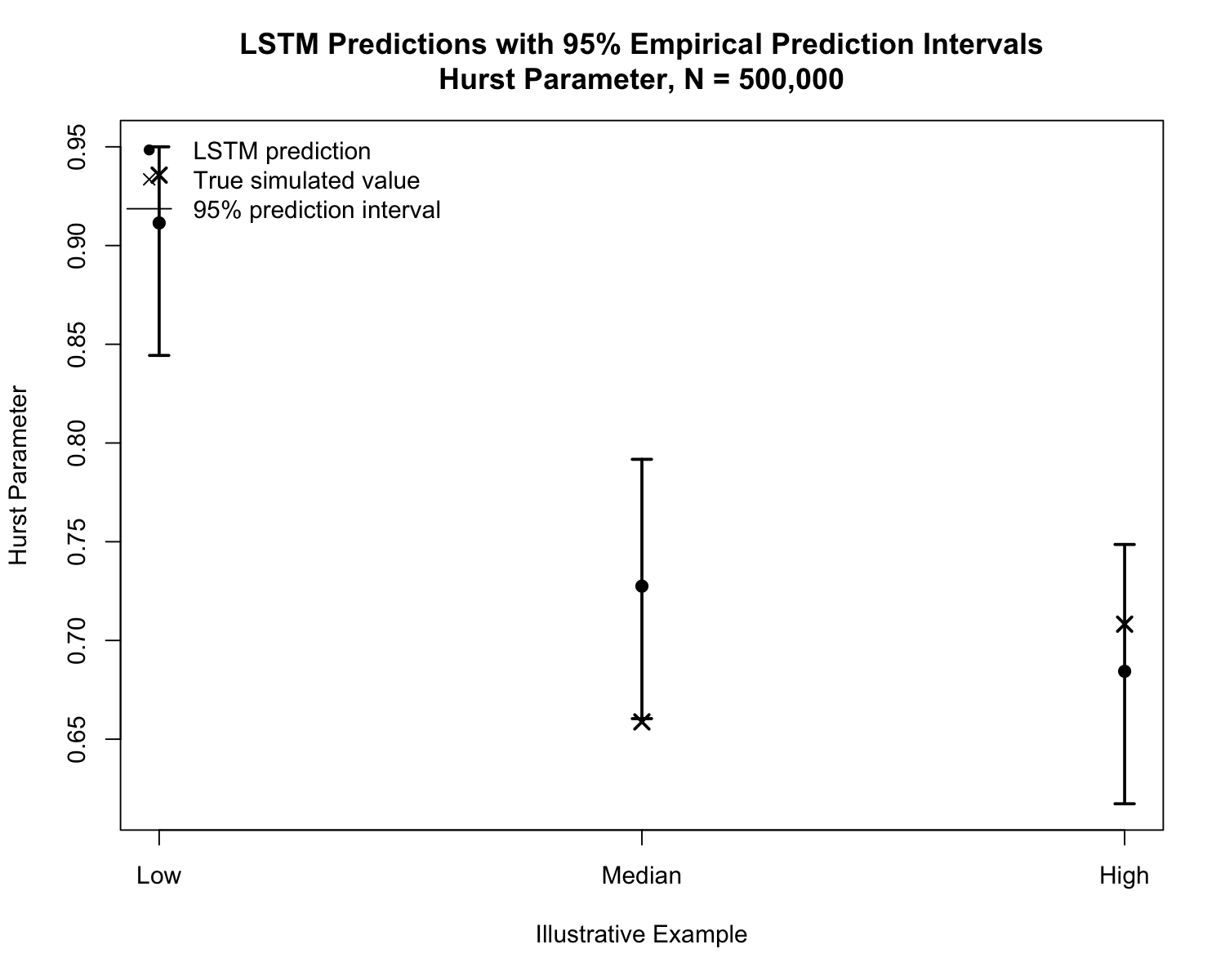}
        \caption{Hurst parameter.}
    \end{subfigure}

    \caption{Illustrative LSTM predictions and empirical \(95\%\)
    prediction intervals for representative low, median, and high
    predictions from the \(N=500{,}000\) experiment.}
    \label{fig:lstm_PI_examples}
\end{figure}

Figure~\ref{fig:lstm_PI_examples} provides illustrative examples of the
empirical \(95\%\) prediction intervals obtained from the
\(N=500{,}000\) LSTM. As in the FNN analysis, representative
observations from the lower, middle, and upper portions of the
predicted-value distribution are displayed for each output. The LSTM
point prediction, true simulated value, and corresponding empirical
prediction interval are shown simultaneously.

These examples complement the aggregate interval results reported in
Table~\ref{tab:lstm_interval_performance}. At \(N=500{,}000\), empirical
coverage is \(94.870\%\) for both the call and put prices and
\(94.909\%\) for the Hurst parameter, all of which are close to the
nominal \(95\%\) level. The corresponding average interval widths are
\(5.177\), \(5.011\), and \(0.126\), respectively. The illustrative
examples therefore provide a direct visualization of the predictive
uncertainty summarized by these aggregate coverage and width measures.

Overall, the LSTM results demonstrate strong predictive performance
across all six simulated sample sizes. Increasing the amount of simulated
data reduces prediction errors and improves the prediction of the Hurst
parameter, although the magnitude of the gains becomes progressively
smaller at the larger sample sizes. For option prices, performance
stabilizes by approximately \(N=200{,}000\), with only negligible
differences between the \(N=200{,}000\) and \(N=500{,}000\)
experiments. For the Hurst parameter, the largest experiment achieves
an RMSE of approximately \(0.033\), an MAE of \(0.026\), a correlation
of \(0.967\), and an \(R^2\) of \(0.934\).

The empirical residual-based prediction intervals also provide useful
information about predictive uncertainty. Option-price coverage remains
reasonably close to the nominal \(95\%\) level across the simulated
sample sizes while the average interval widths generally decrease. For the Hurst parameter, coverage improves from
\(93.000\%\) at \(N=5{,}000\) to \(94.909\%\) at \(N=500{,}000\),
while most of the reduction in interval width is achieved by approximately
\(N=100{,}000\). Taken together, the point-prediction and interval results
suggest that the LSTM extracts the relevant temporal information from
the simulated return sequences efficiently and that its predictive
performance approaches a plateau at the largest sample sizes.

The \(N=500{,}000\) LSTM is therefore retained as the final recurrent
model for the subsequent comparison with the FNN and the model-based
Bayesian and exact Gaussian maximum-likelihood procedures.

\subsection{Comparison of FNN, LSTM, Bayesian, and Exact Gaussian MLE}
\label{subsec:four_method_comparison}

The preceding analyses evaluated the FNN and LSTM architectures across
increasing simulated sample sizes and examined their predictive
performance in detail. To place the neural-network results in a broader
statistical context, this subsection compares the FNN and LSTM with two
model-based inferential approaches: Bayesian estimation and exact Gaussian
maximum likelihood estimation (MLE). The Bayesian benchmark follows the joint estimation framework developed
by Sagor et al.~\cite{Sagor2026}, while the likelihood-based benchmark
uses exact Gaussian maximum likelihood estimation for the fGn return
model. The implementation of both model-based procedures is described
below.

The comparison was conducted using a common set of \(1{,}000\)
simulated return sequences selected from the independent test pool of
the \(N=500{,}000\) experiment. To ensure representation across the
simulated Hurst-parameter range, the test observations were stratified
into ten equal-width intervals over \(H\in[0.50,0.95]\), and \(100\)
sequences were randomly selected without replacement from each interval.
Each selected sequence contains \(252\) return observations and was
evaluated by all four methods. Consequently, the methods are compared
using identical return sequences and true parameter values while
maintaining representation across the full range of persistence
considered in the simulation.

The four methods are compared from three complementary perspectives.
First, their ability to estimate the Hurst parameter is examined using
root mean squared error (RMSE), mean absolute error (MAE), bias,
correlation, and interval performance. Second, their option-price
predictions are compared using the corresponding prediction-error
measures for call and put prices. Finally, computational performance is
examined to distinguish the statistical accuracy of the methods from
their computational cost. This comparison is particularly relevant
because the neural-network models incur substantial computational cost
during training but, once trained, produce predictions through a single
forward pass, whereas the Bayesian and exact Gaussian MLE procedures
perform sequence-specific statistical estimation.

\subsubsection{Bayesian Joint Estimation Procedure}
\label{subsec:bayesian_comparison}

As a model-based inferential benchmark, Bayesian joint estimation was
used to estimate the Hurst parameter, volatility, and mean of each
fractional Gaussian noise return sequence. The procedure follows the
joint Bayesian framework developed in Sagor et al.~\cite{Sagor2026},
with settings adapted to the simulation design of the present paper.
For a return vector
\(\mathbf{d}=(d_1,\ldots,d_n)^\top\), with \(n=252\), the likelihood
was specified as

\[
\mathbf{d}
\mid
\mu_{\mathrm{daily}},
\sigma_{\mathrm{daily}},
H
\sim
N\!\left(
\mu_{\mathrm{daily}}\mathbf{1},
\sigma_{\mathrm{daily}}^2 R(H)
\right),
\]

where \(R(H)\) is the Toeplitz fGn correlation matrix,
\(\mu_{\mathrm{daily}}\) denotes the mean daily return, and
\(\sigma_{\mathrm{daily}}\) is the daily fGn scale. The Hurst parameter
was restricted to the same range used in the simulation experiment,
\(0.50 < H < 0.95\).

The prior distributions used in the four-method comparison were

\[
\mu_{\mathrm{daily}}
\sim
N(0,0.05^2),
\]

\[
\sigma_{\mathrm{daily}}
\sim
\operatorname{LogNormal}
\left(
\log\{\operatorname{sd}(\mathbf{d})\},\,1
\right),
\]

and

\[
H\sim\operatorname{Uniform}(0.50,0.95).
\]

Thus, the prior location for the daily volatility scale was adapted to
the empirical standard deviation of each return sequence, while the
prior for \(H\) was uniform over the Hurst-parameter range represented
in the simulation and neural-network training data.

Posterior sampling was performed using a Metropolis--Hastings MCMC
procedure. A pilot chain of \(2{,}000\) iterations, with the first
\(500\) iterations discarded as burn-in, was used to estimate the
proposal covariance between the transformed Hurst parameter and
log-volatility. This was followed by a longer chain of \(10{,}000\)
iterations, with the first \(3{,}000\) iterations discarded as burn-in.
The long chain included separate updates for the mean, volatility, and
Hurst parameter together with a joint block update for the transformed
Hurst parameter and log-volatility. The block proposal covariance was
estimated from the retained pilot-chain draws to account for posterior
dependence between \(H\) and the volatility scale.

Because the likelihood is expressed in terms of the daily fGn scale,
each retained posterior draw was transformed back to the
simulation-scale volatility according to

\[
\sigma^{(m)}
=
\sigma_{\mathrm{daily}}^{(m)}
252^{H^{(m)}},
\]

where \(m\) indexes the retained posterior draws. Each pair
\(\left(H^{(m)},\sigma^{(m)}\right)\) was then propagated through the
same fractional Black--Scholes pricing function used to generate the
simulation targets, producing posterior distributions for the call and
put option prices.

Posterior medians were used as the Bayesian point estimates of \(H\),
volatility, and the call and put prices. The corresponding \(95\%\)
posterior credible intervals were obtained from the empirical
\(2.5\%\) and \(97.5\%\) quantiles of the retained posterior draws.
Consequently, uncertainty in both the Hurst parameter and volatility is
propagated jointly into the Bayesian option-price distributions.

\subsubsection{Exact Gaussian Maximum Likelihood Estimation}
\label{subsec:exact_mle}

As a classical likelihood-based benchmark, the Hurst parameter and
volatility were estimated using the exact Gaussian likelihood of the
fractional Gaussian noise return sequence \cite{Beran1994}.
The term ``exact'' refers to evaluation of the full finite-sample
Gaussian likelihood based on the fGn covariance matrix, rather than an
approximate frequency-domain likelihood. Let
\(\mathbf{d}=(d_1,\ldots,d_n)^\top\) denote a return vector of length
\(n=252\). Under the assumed fGn model,

$$
\mathbf{d}
\sim
N\!\left(
\mu\mathbf{1},
\sigma_{\mathrm{daily}}^2 R(H)
\right),
$$

where \(R(H)\) is the Toeplitz fGn correlation matrix determined by the
Hurst parameter \(H\), \(\mu\) is the mean daily return, and
\(\sigma_{\mathrm{daily}}\) is the daily fGn scale. The entries of
\(R(H)\) are determined by the fGn autocorrelation function

$$
\rho_H(k)
=
\frac{1}{2}
\left(
|k+1|^{2H}
-
2|k|^{2H}
+
|k-1|^{2H}
\right).
$$

Rather than jointly optimizing over all three parameters, the likelihood
was profiled with respect to \(\mu\) and
\(\sigma_{\mathrm{daily}}^2\). For each candidate value of \(H\), the
generalized least-squares estimate of the mean is

$$
\widehat{\mu}(H)
=
\frac{
\mathbf{1}^{\top}R(H)^{-1}\mathbf{d}
}{
\mathbf{1}^{\top}R(H)^{-1}\mathbf{1}
},
$$

and the corresponding maximum-likelihood estimate of the daily variance is

$$
\widehat{\sigma}_{\mathrm{daily}}^2(H)
=
\frac{1}{n}
\left(
\mathbf{d}-\widehat{\mu}(H)\mathbf{1}
\right)^{\top}
R(H)^{-1}
\left(
\mathbf{d}-\widehat{\mu}(H)\mathbf{1}
\right).
$$

Substitution of these estimates into the Gaussian likelihood gives the
profile negative log-likelihood

$$
\operatorname{NLL}_{p}(H)
=
\frac{1}{2}
\left[
n\log(2\pi)
+
\log|R(H)|
+
n\log\!\left\{
\widehat{\sigma}_{\mathrm{daily}}^2(H)
\right\}
+
n
\right].
$$

The estimate \(\widehat{H}\) was obtained by numerically minimizing the
profile negative log-likelihood over the bounded interval
\(0.50001 \leq H \leq 0.94999\), consistent with the Hurst-parameter
range used in the simulation experiment. The corresponding estimates
\(\widehat{\mu}\) and
\(\widehat{\sigma}_{\mathrm{daily}}\) were then evaluated at
\(\widehat{H}\). For numerical stability and computational efficiency,
the covariance calculations were performed using a Cholesky
decomposition rather than explicit matrix inversion, with a small
diagonal nugget added to the fGn correlation matrix.

For consistency with the simulation design, both the Bayesian and
exact Gaussian MLE procedures restrict the Hurst parameter to the
range used in data generation. These bounds provide the statistical
methods with prior knowledge of the simulated parameter space and
may improve estimation performance, particularly near the boundaries.
Therefore, the comparison should be interpreted within this
controlled simulation setting rather than as evidence of
unrestricted estimation performance.

A \(95\%\) profile-likelihood confidence interval for \(H\) was obtained
using the likelihood-ratio criterion. Specifically, the interval consists
of values of \(H\) satisfying

$$
\operatorname{NLL}_{p}(H)
\leq
\operatorname{NLL}_{p}(\widehat{H})
+
\frac{1}{2}\chi^2_{1,0.95},
$$

with the lower and upper endpoints determined numerically subject to the
same parameter bounds.

When the profile likelihood was minimized at or very near the lower
optimization boundary, the resulting estimate may appear as \(0.500\)
after rounding in the reported tables. Such values therefore represent
boundary estimates close to the Brownian-motion case rather than
unconstrained estimates below \(H=0.5\).

For option pricing, the estimated daily fGn scale was transformed back
to the simulation-scale volatility according to the fractional scaling
relationship

$$
\widehat{\sigma}
=
\widehat{\sigma}_{\mathrm{daily}}\,
252^{\widehat{H}}.
$$

The resulting estimates
\((\widehat{H},\widehat{\sigma})\)
were then substituted into the same fractional Black--Scholes pricing
function used throughout the four-method comparison to obtain the Exact
Gaussian MLE call- and put-price estimates.

To quantify uncertainty in the MLE-based option prices, \(95\%\)
parametric-bootstrap confidence intervals were constructed using
\(B=200\) bootstrap replications for each test sequence. Conditional on
the fitted MLE parameters, each bootstrap return sequence was generated
from

$$
\mathbf{d}^{*}
\sim
N\!\left(
\widehat{\mu}\mathbf{1},
\widehat{\sigma}_{\mathrm{daily}}^{\,2}
R(\widehat{H})
\right).
$$

The Exact Gaussian MLE procedure was then refitted to each bootstrap
sequence, producing
\(\widehat{H}^{*}\),
\(\widehat{\mu}^{*}\), and
\(\widehat{\sigma}_{\mathrm{daily}}^{*}\). The bootstrap daily scale was
transformed using

$$
\widehat{\sigma}^{*}
=
\widehat{\sigma}_{\mathrm{daily}}^{*}\,
252^{\widehat{H}^{*}},
$$

and the corresponding fractional Black--Scholes call and put prices were
recalculated. The empirical \(2.5\%\) and \(97.5\%\) quantiles of the
resulting bootstrap call- and put-price distributions were used as the
lower and upper limits of the respective \(95\%\) parametric-bootstrap
confidence intervals.

The two MLE interval procedures therefore serve different purposes:
profile likelihood is used to quantify uncertainty in the Hurst-parameter
estimate, whereas the parametric bootstrap propagates uncertainty in the
fitted fGn model through the fractional option-pricing calculation. These
intervals are consequently interpreted as frequentist confidence
intervals and are distinct from both the Bayesian posterior credible
intervals and the empirical residual-based prediction intervals used for
the neural-network models.

\subsubsection{Hurst Parameter Estimation}

Table~\ref{tab:four_method_hurst} summarizes Hurst-parameter estimation
performance for the four methods over the common set of \(1{,}000\)
test sequences.

\begin{table}[H]
\centering
\caption{Hurst-parameter estimation performance across the four methods
on the common set of \(1{,}000\) simulated test sequences.}
\label{tab:four_method_hurst}
\begin{tabular}{lrrrrr}
\toprule
Method & RMSE & MAE & Bias & Correlation & \(R^2\) \\
\midrule
FNN
& 0.043 & 0.034 & 0.002 & 0.947 & 0.892 \\

LSTM
& \textbf{0.033} & \textbf{0.026} & -0.001
& \textbf{0.967} & \textbf{0.935} \\

Bayesian
& 0.039 & 0.031 & \textbf{-0.001}
& 0.953 & 0.908 \\

Exact Gaussian MLE
& 0.043 & 0.034 & -0.014
& 0.952 & 0.891 \\
\bottomrule
\end{tabular}
\end{table}

Among the four methods, the LSTM provides the strongest overall
point-estimation performance for the Hurst parameter. Its RMSE of
\(0.033\) is lower than those of the Bayesian method (\(0.039\)),
FNN (\(0.043\)), and exact Gaussian MLE (\(0.043\)). A similar
pattern is observed for the MAE, with the LSTM achieving the smallest
value of \(0.026\), followed by the Bayesian method (\(0.031\)),
exact Gaussian MLE (\(0.034\)), and FNN (\(0.034\)).

The LSTM also exhibits the strongest association between the estimated
and true Hurst parameters, with a correlation of \(0.967\) and an
\(R^2\) of \(0.935\). The Bayesian method follows with a correlation
of \(0.953\) and an \(R^2\) of \(0.908\). The FNN and exact Gaussian
MLE yield nearly identical RMSE values after rounding, although the
exact Gaussian MLE exhibits a more pronounced negative bias of
\(-0.014\). In contrast, the FNN, LSTM, and Bayesian methods exhibit
only small biases, with values of \(0.002\), \(-0.001\), and
\(-0.001\), respectively.

\begin{figure}[H]
\centering
\includegraphics[width=0.92\textwidth]
{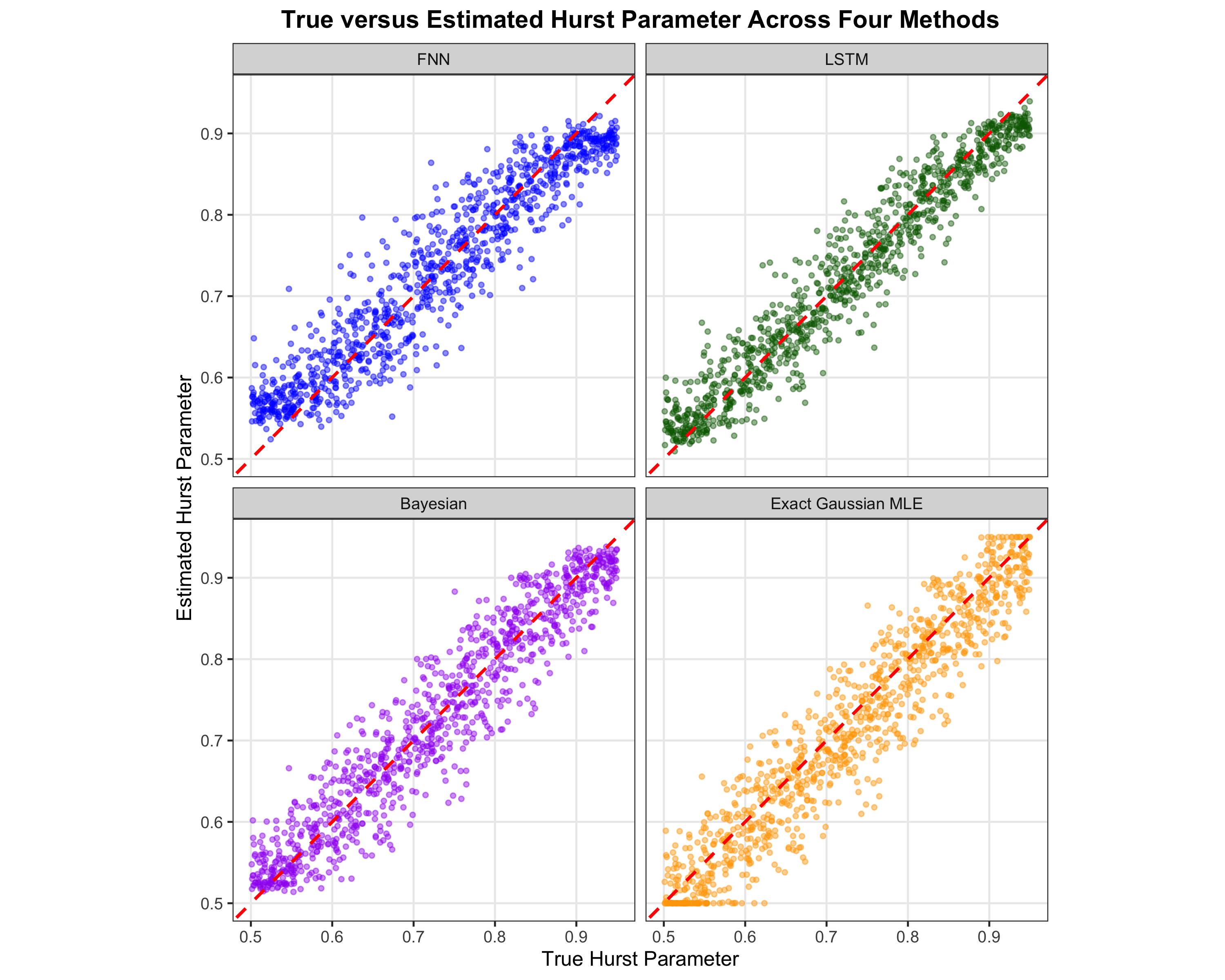}
\caption{True versus estimated Hurst parameters for the FNN, LSTM,
Bayesian, and exact Gaussian MLE methods across the common set of
\(1{,}000\) simulated test sequences. The dashed diagonal line
represents perfect agreement.}
\label{fig:four_method_h_true_estimated}
\end{figure}

Figure~\ref{fig:four_method_h_true_estimated} compares the estimated
Hurst parameters directly with their true simulated values. The dashed
diagonal line represents perfect agreement between the true and
estimated values.

The scatterplots reinforce the numerical results. All four methods
capture the overall relationship between the simulated and estimated
Hurst parameters, but the LSTM estimates are more tightly concentrated
around the diagonal reference line. The FNN exhibits greater dispersion,
particularly across the middle portion of the Hurst-parameter range.
The Bayesian estimates also track the true values closely, although
with somewhat greater dispersion than the LSTM. The exact Gaussian MLE
shows a visible tendency toward lower estimates in portions of the
parameter range, consistent with its negative empirical bias.

\begin{figure}[H]
\centering
\includegraphics[width=0.92\textwidth]
{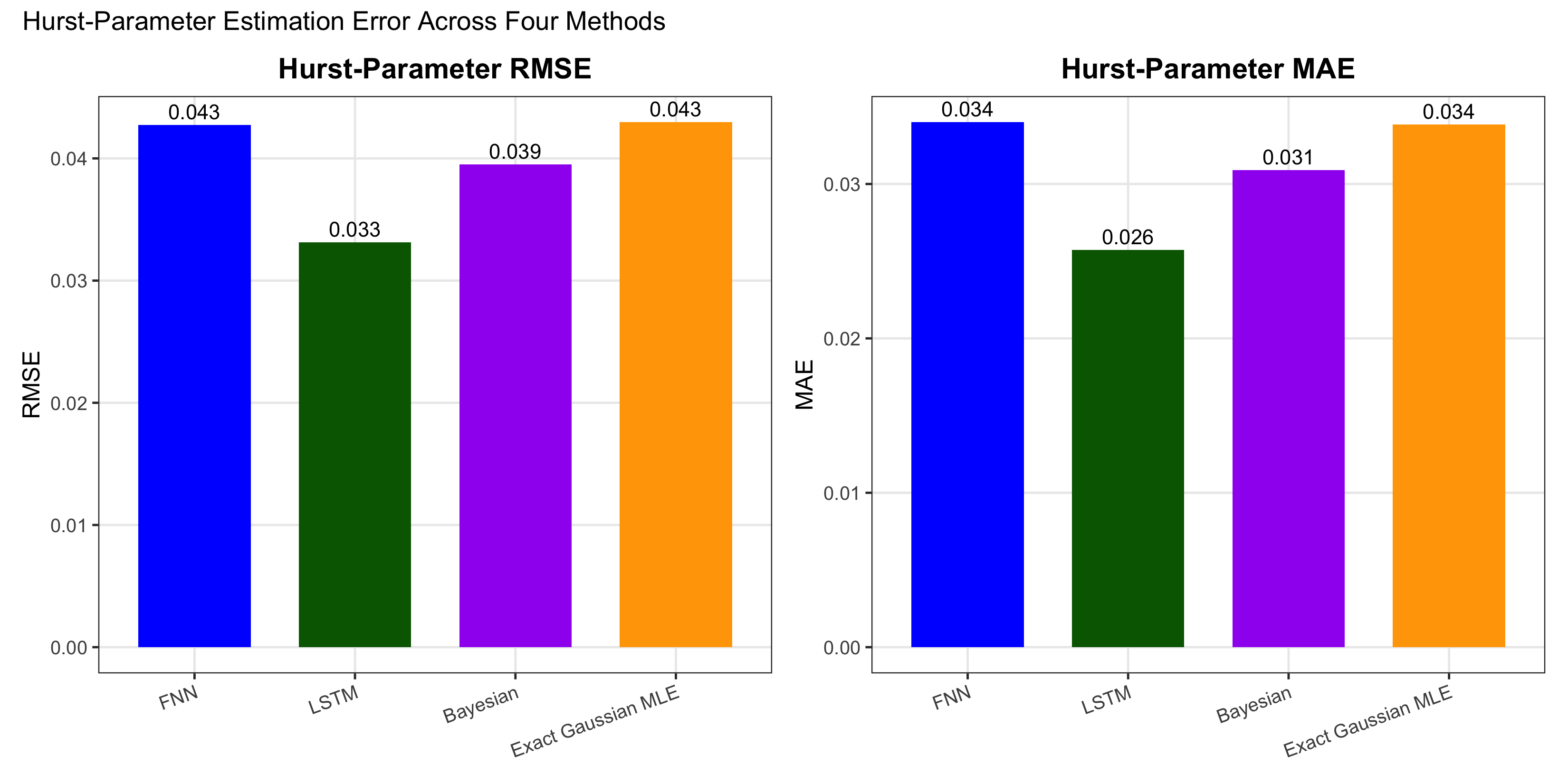}
\caption{RMSE and MAE of Hurst-parameter estimates for the FNN, LSTM,
Bayesian, and exact Gaussian MLE methods across \(1{,}000\) common
simulated test sequences.}
\label{fig:four_method_h_error}
\end{figure}

For a more direct comparison of estimation error,
Figure~\ref{fig:four_method_h_error} presents the RMSE and MAE for all
four methods.

The advantage of the LSTM is evident under both error criteria.
Relative to the FNN, the LSTM reduces Hurst-parameter RMSE from
\(0.043\) to \(0.033\), while the MAE decreases from \(0.034\) to
\(0.026\). The LSTM also outperforms both likelihood-based approaches
under these two point-estimation measures.

Point-estimation accuracy, however, provides only one aspect of the
comparison. Table~\ref{tab:four_method_h_intervals} therefore compares
the empirical coverage and average width of the \(95\%\) intervals for
the Hurst parameter.

\begin{table}[H]
\centering
\caption{Empirical coverage and average width of the \(95\%\)
Hurst-parameter intervals across the four methods.}
\label{tab:four_method_h_intervals}
\begin{tabular}{llrr}
\toprule
Method & Interval Type & Coverage (\%) & Average Width \\
\midrule
FNN
& Residual-based prediction interval
& \textbf{95.0} & 0.159 \\

LSTM
& Residual-based prediction interval
& 94.5 & \textbf{0.126} \\

Bayesian
& Posterior credible interval
& 94.5 & 0.150 \\

Exact Gaussian MLE
& Profile-likelihood confidence interval
& 94.3 & 0.148 \\
\bottomrule
\end{tabular}
\end{table}

The interval results provide additional information about the relative
performance of the four methods. The FNN achieves empirical coverage exactly equal to the nominal
\(95\%\) level, while the LSTM and Bayesian methods both achieve
\(94.5\%\) coverage and the exact Gaussian MLE achieves \(94.3\%\).
Thus, all four procedures produce Hurst-parameter coverage reasonably
close to the nominal level in this simulation experiment.

The principal difference appears in interval width. The LSTM produces
the narrowest intervals, with an average width of \(0.126\), compared
with \(0.159\) for the FNN, \(0.150\) for the Bayesian procedure, and
\(0.148\) for exact Gaussian MLE. Hence, the LSTM maintains coverage
close to \(95\%\) while producing substantially narrower intervals,
indicating greater precision in this simulation setting.

It is important to note that the intervals in
Table~\ref{tab:four_method_h_intervals} do not have identical
statistical interpretations. The FNN and LSTM intervals are empirical residual-based prediction
intervals constructed from validation-set residuals, the Bayesian
interval is a posterior credible interval, and the Exact Gaussian MLE
interval is obtained from the profile likelihood.
Accordingly, the comparison focuses on their empirical frequentist
coverage and width under the common simulation design rather than
treating the interval constructions themselves as theoretically
equivalent.

\begin{figure}[H]
\centering
\includegraphics[width=0.95\textwidth]
{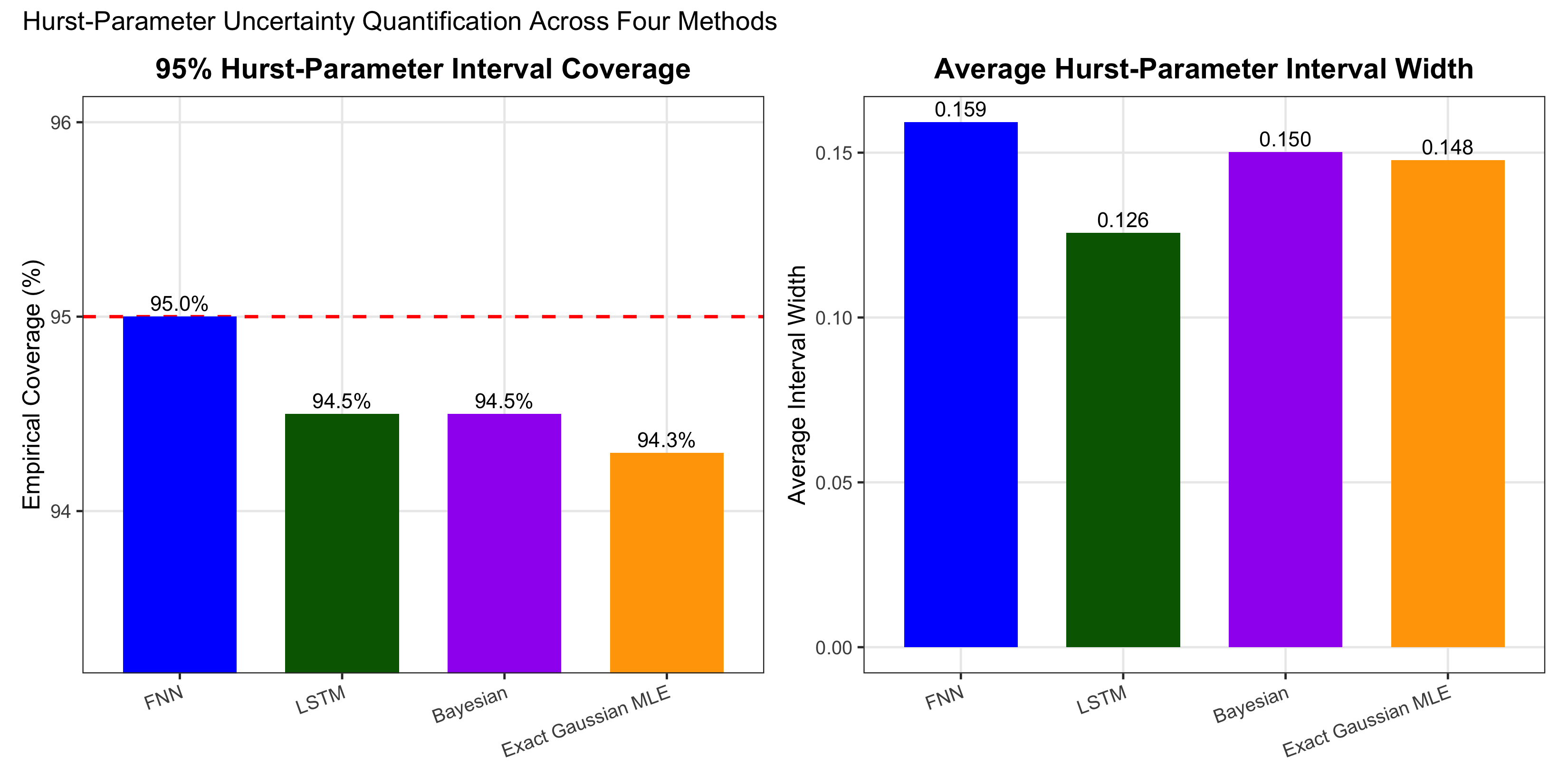}
\caption{Empirical coverage and average width of the \(95\%\)
Hurst-parameter intervals for the four methods. The dashed horizontal
line in the coverage panel indicates the nominal \(95\%\) level.}
\label{fig:four_method_h_intervals}
\end{figure}

Figure~\ref{fig:four_method_h_intervals} displays these two aspects of
interval performance graphically.

Taken together, the point- and interval-estimation results indicate that,
in the present simulation setting, the LSTM provides the strongest overall
Hurst-parameter estimation performance among the four methods considered. It achieves the lowest RMSE and MAE, the highest correlation and \(R^2\), negligible bias, and the narrowest average interval while retaining empirical coverage
close to the nominal level.

\subsubsection{Option-Price Prediction}

The comparison was next extended from estimation of the Hurst parameter
to the primary pricing objective. Table~\ref{tab:four_method_option}
reports the RMSE, MAE, bias, correlation, and \(R^2\) for call- and
put-price predictions over the same \(1{,}000\) test sequences.

\begin{table}[H]
\centering
\caption{Option-price prediction performance across the four methods
on the common set of \(1{,}000\) simulated test sequences.}
\label{tab:four_method_option}
\resizebox{\textwidth}{!}{
\begin{tabular}{llrrrrr}
\toprule
Method & Quantity & RMSE & MAE & Bias & Correlation & \(R^2\) \\
\midrule
FNN
& Call & 1.659 & 1.271 & -0.490 & 0.924 & 0.824 \\
FNN
& Put  & 1.659 & 1.271 & -0.490 & 0.924 & 0.824 \\

LSTM
& Call & \textbf{1.305} & \textbf{0.999} & \textbf{0.000}
& \textbf{0.945} & \textbf{0.891} \\
LSTM
& Put  & \textbf{1.305} & \textbf{0.999} & \textbf{0.000}
& \textbf{0.945} & \textbf{0.891} \\

Bayesian
& Call & 3.107 & 2.116 & 1.658 & 0.876 & 0.383 \\
Bayesian
& Put  & 3.107 & 2.116 & 1.658 & 0.876 & 0.383 \\

Exact Gaussian MLE
& Call & 2.695 & 1.835 & 0.938 & 0.856 & 0.536 \\
Exact Gaussian MLE
& Put  & 2.695 & 1.835 & 0.938 & 0.856 & 0.536 \\
\bottomrule
\end{tabular}
}
\end{table}

The LSTM again provides the strongest predictive performance. For the
call price, its RMSE is \(1.305\), compared with \(1.659\) for the
FNN, \(2.695\) for exact Gaussian MLE, and \(3.107\) for the Bayesian
method. The corresponding MAE values are \(0.999\), \(1.271\),
\(1.835\), and \(2.116\), respectively. The same error values are obtained for the put price under the pricing setup used in the simulation. This symmetry arises because the call and put prices are linked through the same underlying parameter realization and the corresponding put--call pricing relationship, so prediction errors are transmitted consistently across the two option-price outputs.

The LSTM also exhibits negligible systematic pricing bias, with a bias
that rounds to \(0.000\). In comparison, the FNN exhibits a negative
bias of \(-0.490\), whereas the exact Gaussian MLE and Bayesian methods
exhibit positive biases of \(0.938\) and \(1.658\), respectively.
Thus, the lower RMSE achieved by the LSTM is accompanied by predictions
that are closely centered around the true simulated option prices.

The correlation and \(R^2\) measures provide a similar conclusion.
For both call and put prices, the LSTM achieves the highest correlation,
\(0.945\), and the highest \(R^2\), \(0.891\), followed by the FNN
with values of \(0.924\) and \(0.824\), respectively. Although the
likelihood-based methods retain substantial correlations with the true
option prices, their larger RMSE and MAE values indicate lower
point-prediction accuracy than the neural-network methods under the
present simulation design.

\begin{figure}[H]
\centering
\includegraphics[width=0.95\textwidth]
{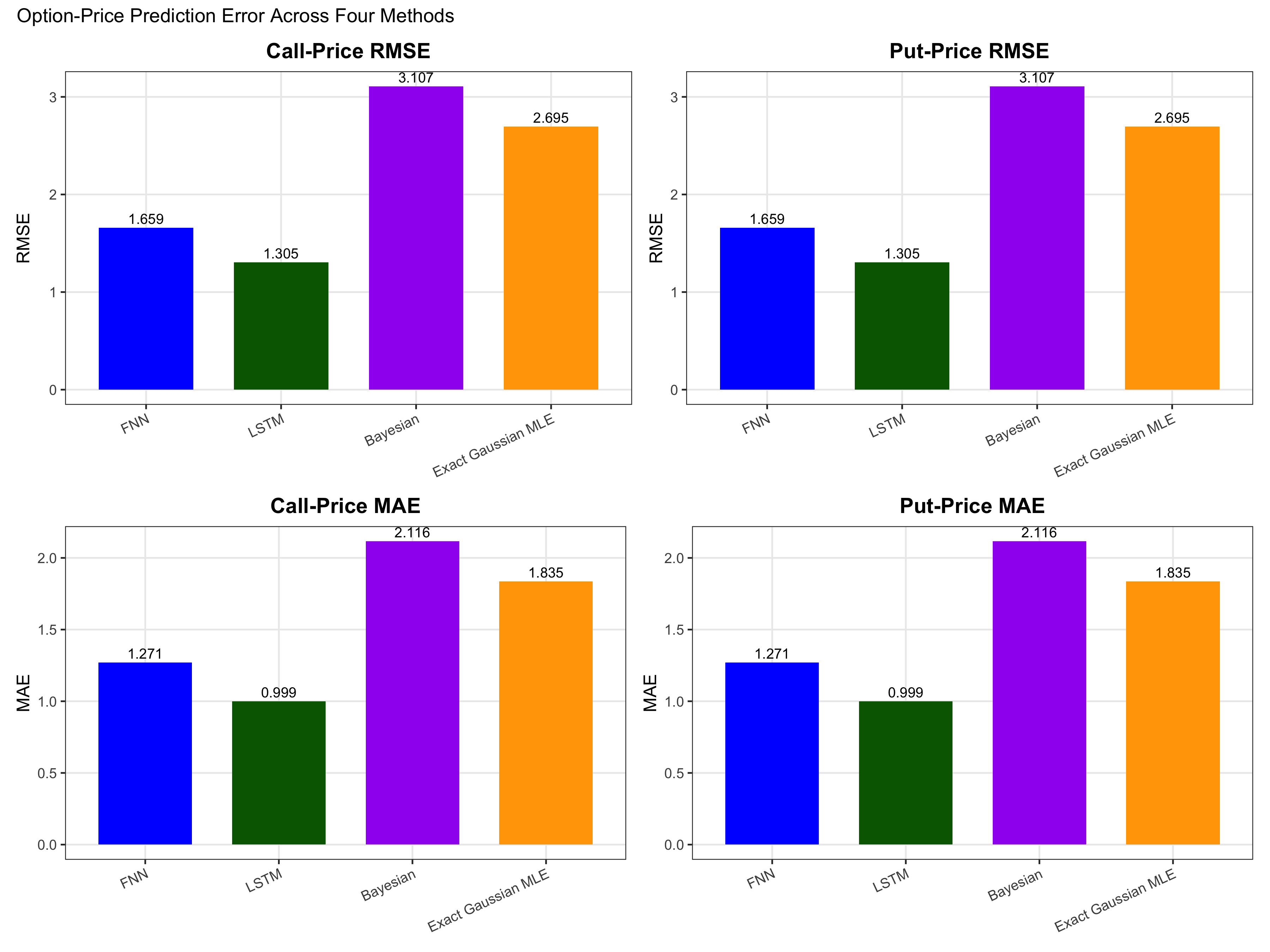}
\caption{RMSE and MAE for call- and put-price predictions from the FNN,
LSTM, Bayesian, and exact Gaussian MLE methods across \(1{,}000\)
common simulated test sequences.}
\label{fig:four_method_option_error}
\end{figure}

Figure~\ref{fig:four_method_option_error} summarizes the RMSE and MAE
comparisons graphically.

The figure shows a consistent ordering under both error measures. The
LSTM produces the smallest pricing errors, followed by the FNN, exact
Gaussian MLE, and Bayesian procedure. These results indicate that the
sequence-based LSTM is able to extract information from the simulated
return history that is particularly effective for the pricing task.

\paragraph{Option-Price Uncertainty Quantification.}

In addition to point-prediction accuracy, the four methods were compared
in terms of empirical interval coverage and average interval width on
the same \(1{,}000\) test sequences. Because the methods quantify
uncertainty differently, the resulting intervals have different
interpretations. The FNN and LSTM use empirical residual-based prediction intervals,
the Bayesian method uses posterior credible intervals, and the exact
Gaussian MLE uses \(95\%\) parametric-bootstrap confidence intervals
based on \(B=200\) bootstrap replications for each test sequence.

\begin{table}[H]
\centering
\caption{Option-price interval performance across the four methods on
the common \(1{,}000\)-sequence test set.}
\label{tab:four_method_option_intervals}
\begin{tabular}{llcc}
\toprule
Method & Quantity & Coverage (\%) & Average Width \\
\midrule

FNN
& Call Price & 94.100 & 5.891 \\
& Put Price  & 94.100 & 5.808 \\
\midrule

LSTM
& Call Price & 93.400 & \textbf{5.177} \\
& Put Price  & 93.400 & \textbf{5.021} \\
\midrule

Bayesian
& Call Price & 89.400 & 9.939 \\
& Put Price  & 89.400 & 9.939 \\
\midrule

Exact Gaussian MLE
& Call Price & \textbf{95.300} & 7.730 \\
& Put Price  & \textbf{95.300} & 7.730 \\
\bottomrule
\end{tabular}
\end{table}

The exact Gaussian MLE provides empirical coverage closest to the
nominal \(95\%\) level, attaining \(95.300\%\) coverage for both call
and put prices, with an average interval width of \(7.730\). The FNN
also provides coverage relatively close to the nominal level, with
\(94.100\%\) coverage and average widths of \(5.891\) and \(5.808\)
for the call and put prices, respectively.

The LSTM produces the narrowest intervals among the four methods, with
average widths of \(5.177\) for the call price and \(5.021\) for the
put price. Its empirical coverage is \(93.400\%\), indicating that the
gain in interval precision is accompanied by modest undercoverage
relative to the nominal \(95\%\) level. In contrast, the Bayesian
credible intervals are substantially wider, with an average width of
\(9.939\), while their empirical coverage is \(89.400\%\). Thus, the
greater interval width of the Bayesian procedure does not translate
into improved frequentist coverage under the present simulation
design.

Taken together with the point-prediction results, these findings
illustrate a trade-off between predictive accuracy, interval
sharpness, and empirical coverage. The LSTM provides the strongest
point-prediction accuracy and the narrowest option-price intervals,
whereas the exact Gaussian MLE achieves coverage closest to the
nominal \(95\%\) level. The FNN occupies an intermediate position,
providing coverage relatively close to \(95\%\) with narrower
intervals than the exact Gaussian MLE. These results therefore
highlight that no single performance criterion alone fully
characterizes the predictive and inferential behavior of the four
methods.

\subsubsection{Computational Performance}

Statistical accuracy must also be considered together with computational
cost. Table~\ref{tab:four_method_computation} reports the observed
computational performance of the four procedures. Because the neural
networks are trained once and subsequently applied to new sequences,
their reported evaluation times correspond to prediction using the
already-trained models and do not include the one-time training cost.
In contrast, the Bayesian and exact Gaussian MLE procedures require
sequence-specific estimation.

\begin{table}[H]
\centering
\caption{Per-sequence computational performance across the four methods
on the common set of \(1{,}000\) simulated test sequences.}
\label{tab:four_method_computation}
\begin{tabular}{lr}
\toprule
Method & Median Time per Sequence (s) \\
\midrule
FNN                & 0.000063 \\
LSTM               & 0.000477 \\
Bayesian           & 80.318 \\
Exact Gaussian MLE & 0.370 \\
\bottomrule
\end{tabular}
\end{table}

The computational differences are substantial. Once trained, the FNN
and LSTM require approximately \(0.000063\) and \(0.000477\) seconds
per sequence, respectively. Exact Gaussian MLE requires approximately
\(0.370\) seconds per sequence, whereas the Bayesian procedure requires
approximately \(80.318\) seconds per sequence. Thus, although the LSTM
is computationally more demanding than the FNN at prediction time, both
neural-network approaches provide extremely rapid inference relative to
the sequence-specific likelihood-based procedures.

\begin{figure}[H]
\centering
\includegraphics[width=0.90\textwidth]
{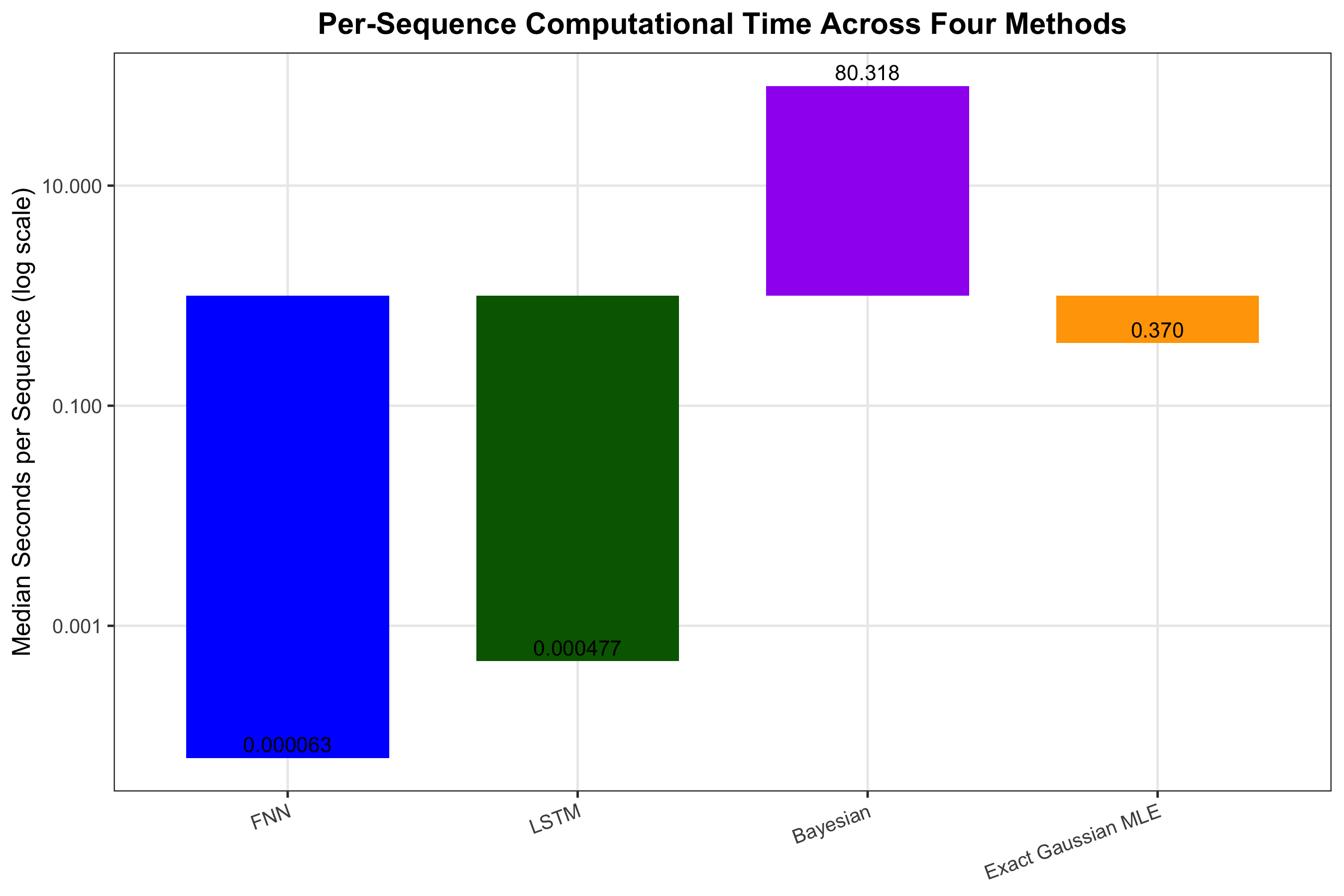}
\caption{Median computational time per sequence for the FNN, LSTM,
Bayesian, and exact Gaussian MLE methods. The vertical axis is shown
on a logarithmic scale because the computational times span several
orders of magnitude.}
\label{fig:four_method_computation}
\end{figure}

Figure~\ref{fig:four_method_computation} displays the per-sequence
computational times on a logarithmic scale, which is necessary because
the observed times span several orders of magnitude.

The computational comparison highlights an important distinction
between model development and model deployment. Training the FNN and
especially the LSTM requires a substantial initial computational
investment, as documented in the preceding subsections. Once training
is complete, however, that cost is amortized across subsequent
predictions. A new return sequence can be processed without repeating
the optimization or posterior-sampling procedure. Exact Gaussian MLE,
by contrast, must solve a new likelihood-optimization problem for each
sequence, while the Bayesian method requires posterior computation for
each new sequence.

This computational pattern is consistent with the surrogate-model
literature, in which an initial training cost is incurred to enable
rapid subsequent evaluations
\cite{liu2019pricing,chen2026deep}.

Consequently, within the present simulation design, the LSTM exhibits
a favorable balance between point-prediction accuracy, empirical
interval width, and repeated-evaluation efficiency. It provides the
lowest point-prediction errors and the narrowest reported intervals
among the methods considered, although the narrower option-price
intervals are accompanied by modest undercoverage relative to the
nominal \(95\%\) level. The FNN is computationally faster than
the LSTM but produces larger estimation and pricing errors. The Bayesian
approach provides a full posterior characterization of parameter and
pricing uncertainty, representing a distinct inferential advantage, but
at substantially greater computational cost. Exact Gaussian MLE is
considerably faster than the Bayesian procedure but, in this experiment,
does not match the point-prediction accuracy of the LSTM.

\section{Empirical Applications}
\label{sec:empirical_applications}

This section evaluates the proposed neural-network approaches
using real financial market data to examine their performance
beyond the controlled simulation setting. The empirical
applications investigate the ability of the feedforward neural
network (FNN) and long short-term memory (LSTM) models to
predict European call prices, put prices, and the Hurst
parameter from observed financial return sequences.
The results are used to assess the practical applicability
of the proposed approaches and to compare their predictive
performance with traditional statistical estimation methods.

\subsection{Empirical Application to Saudi Aramco}
\label{subsec:aramco_application}

To evaluate the ability of the proposed neural-network models to
generalize beyond simulated fractional Gaussian noise, an empirical
application was conducted using daily returns of Saudi Aramco
(\texttt{2222.SR}). Aramco provides an informative real-data setting
because its trading history encompasses substantially different market
conditions following its initial public offering, including the
COVID-19 period, the post-pandemic recovery, the subsequent oil-price
boom, and more recent market conditions.

Five empirical windows were considered. Each window contains exactly
\(252\) daily log returns so that the empirical inputs have the same
dimension as the sequences used to train the FNN and LSTM models.
The five regimes were labeled IPO and COVID-19, Post-COVID Recovery,
Oil Boom, High Oil Price Period, and Recent Market. The same return
vector within each regime was supplied to all four procedures: the
Bayesian model, Exact Gaussian maximum likelihood estimation (MLE),
the feed-forward neural network (FNN), and the long short-term memory
(LSTM) network. The FNN and LSTM predictions were obtained using the
final models trained on \(N=500{,}000\) simulated sequences, without
retraining on the empirical Aramco data.

\subsubsection{Hurst-Parameter Estimates}

Table~\ref{tab:aramco_H_comparison} reports the Hurst-parameter
estimates obtained from the four methods across the five market
regimes.

\begin{table}[H]
\centering
\caption{Estimated Hurst parameter for Saudi Aramco across five
empirical market regimes.}
\label{tab:aramco_H_comparison}
\begin{tabular}{lcccc}
\toprule
\textbf{Market Regime}
& \textbf{Bayesian}
& \textbf{Exact MLE}
& \textbf{FNN}
& \textbf{LSTM} \\
\midrule

IPO and COVID-19
& 0.514 & 0.500 & 0.549 & 0.537 \\

Post-COVID Recovery
& 0.518 & 0.500 & 0.565 & 0.540 \\

Oil Boom
& 0.642 & 0.624 & 0.597 & 0.657 \\

High Oil Price Period
& 0.666 & 0.644 & 0.585 & 0.657 \\

Recent Market
& 0.539 & 0.513 & 0.564 & 0.559 \\

\bottomrule
\end{tabular}
\end{table}

\begin{figure}[H]
\centering
\includegraphics[
    width=0.95\textwidth
]{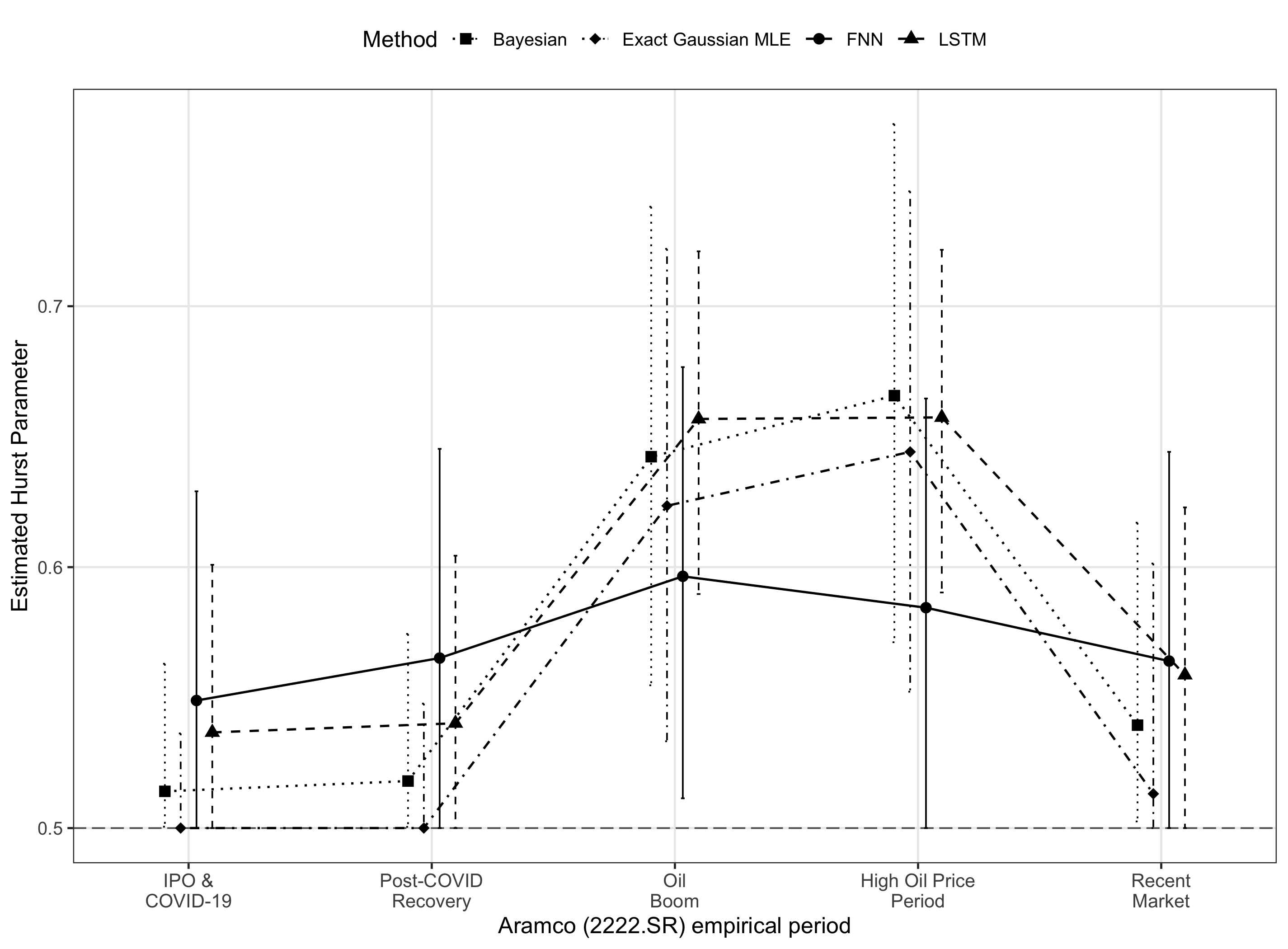}
\caption{Comparison of Hurst-parameter estimates for Saudi Aramco
across five empirical market regimes using the Bayesian, Exact Gaussian
MLE, FNN, and LSTM methods. Error bars represent the corresponding
uncertainty intervals. The horizontal dashed line at \(H=0.5\)
represents the standard Brownian-motion benchmark.}
\label{fig:aramco_H}
\end{figure}

Figure~\ref{fig:aramco_H} provides the corresponding graphical
comparison together with the uncertainty intervals produced by each
method.

Table~\ref{tab:aramco_H_comparison} and Figure~\ref{fig:aramco_H}
show substantial variation in the estimated degree of persistence
across the five market regimes. During the IPO and COVID-19 period,
the Bayesian and Exact Gaussian MLE estimates are close to the
Brownian-motion benchmark, with estimates of \(0.514\) and \(0.500\),
respectively. The FNN and LSTM produce somewhat larger estimates of
\(0.549\) and \(0.537\). A similar pattern is observed during the
Post-COVID Recovery period, where the corresponding estimates are
\(0.518\), \(0.500\), \(0.565\), and \(0.540\).

A considerably different pattern emerges during the Oil Boom and High
Oil Price regimes. During the Oil Boom, the Bayesian posterior median
increases to \(0.642\), while the Exact Gaussian MLE and LSTM estimates
increase to \(0.624\) and \(0.657\), respectively. The FNN estimate is
also above the Brownian-motion benchmark at \(0.597\). During the High
Oil Price period, the Bayesian, Exact Gaussian MLE, and LSTM estimates
remain elevated at \(0.666\), \(0.644\), and \(0.657\), respectively,
while the FNN estimate is \(0.585\). Thus, several methods identify
stronger persistence during these two periods of elevated energy-market
activity.

In the Recent Market regime, all four estimates move back toward
\(H=0.5\). The Bayesian estimate decreases to \(0.539\), while the
Exact Gaussian MLE, FNN, and LSTM estimates are \(0.513\), \(0.564\),
and \(0.559\), respectively. Overall, the empirical results indicate
that the estimated degree of persistence is not constant across market
conditions. The Bayesian, Exact Gaussian MLE, and LSTM methods show a
particularly clear increase during the Oil Boom and High Oil Price
periods followed by a decline in the Recent Market regime.

The uncertainty intervals shown in Figure~\ref{fig:aramco_H} provide
additional context for these point estimates. In periods where the
estimates are close to \(0.5\), several intervals extend toward the
Brownian-motion benchmark. The point estimates should therefore be
interpreted together with their uncertainty rather than as conclusive
evidence of long-range dependence based solely on whether
\(\widehat{H}>0.5\).

\subsubsection{Option-Price Predictions}

Table~\ref{tab:aramco_option_results} reports the call- and put-option
predictions obtained from the four methods across the five empirical
market regimes.

\begin{table}[H]
\centering
\caption{Comparison of predicted call and put option prices for
Saudi Aramco (2222.SR).}
\label{tab:aramco_option_results}

\resizebox{\textwidth}{!}{
\begin{tabular}{lcccccccc}
\toprule
& \multicolumn{4}{c}{\textbf{Call Option}}
& \multicolumn{4}{c}{\textbf{Put Option}} \\
\cmidrule(lr){2-5}
\cmidrule(lr){6-9}

\textbf{Market Regime}
& \textbf{Bayesian}
& \textbf{Exact MLE}
& \textbf{FNN}
& \textbf{LSTM}
& \textbf{Bayesian}
& \textbf{Exact MLE}
& \textbf{FNN}
& \textbf{LSTM} \\
\midrule

IPO and COVID-19
& 6.093 & 6.088 & 6.416 & 6.673
& 2.893 & 2.887 & 3.216 & 3.472 \\

Post-COVID Recovery
& 3.910 & 3.914 & 4.597 & 4.300
& 0.709 & 0.713 & 1.396 & 1.100 \\

Oil Boom
& 5.978 & 5.948 & 7.204 & 10.162
& 2.777 & 2.747 & 4.004 & 6.961 \\

High Oil Price Period
& 4.491 & 4.471 & 5.509 & 6.987
& 1.290 & 1.270 & 2.308 & 3.786 \\

Recent Market
& 4.820 & 4.837 & 5.600 & 5.778
& 1.619 & 1.636 & 2.399 & 2.577 \\

\bottomrule
\end{tabular}
}
\end{table}

\begin{figure}[H]
\centering

\begin{subfigure}[t]{0.49\textwidth}
\centering
\includegraphics[
    width=\textwidth
]{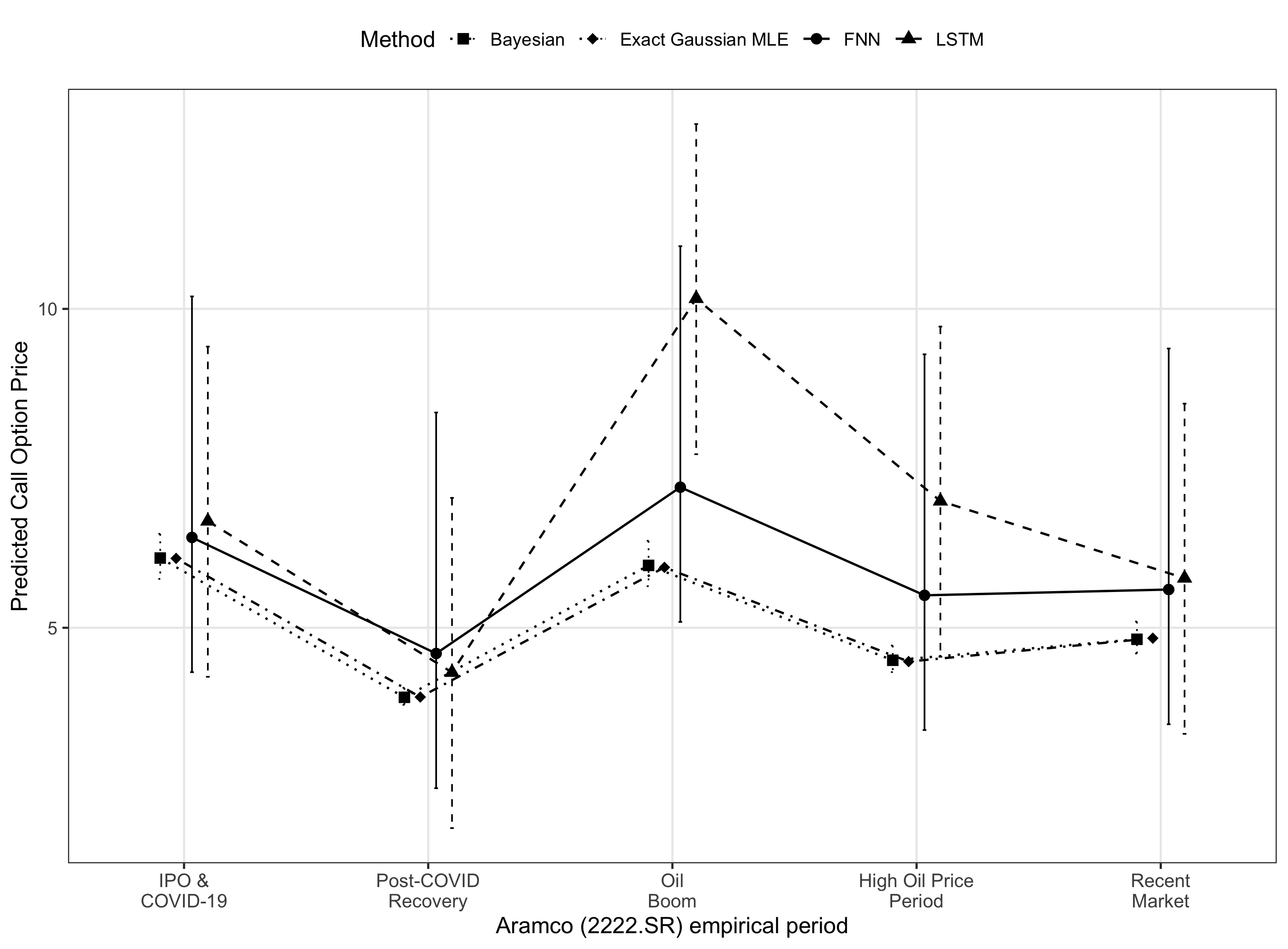}
\caption{Call-option prices.}
\label{fig:aramco_call}
\end{subfigure}
\hfill
\begin{subfigure}[t]{0.49\textwidth}
\centering
\includegraphics[
    width=\textwidth
]{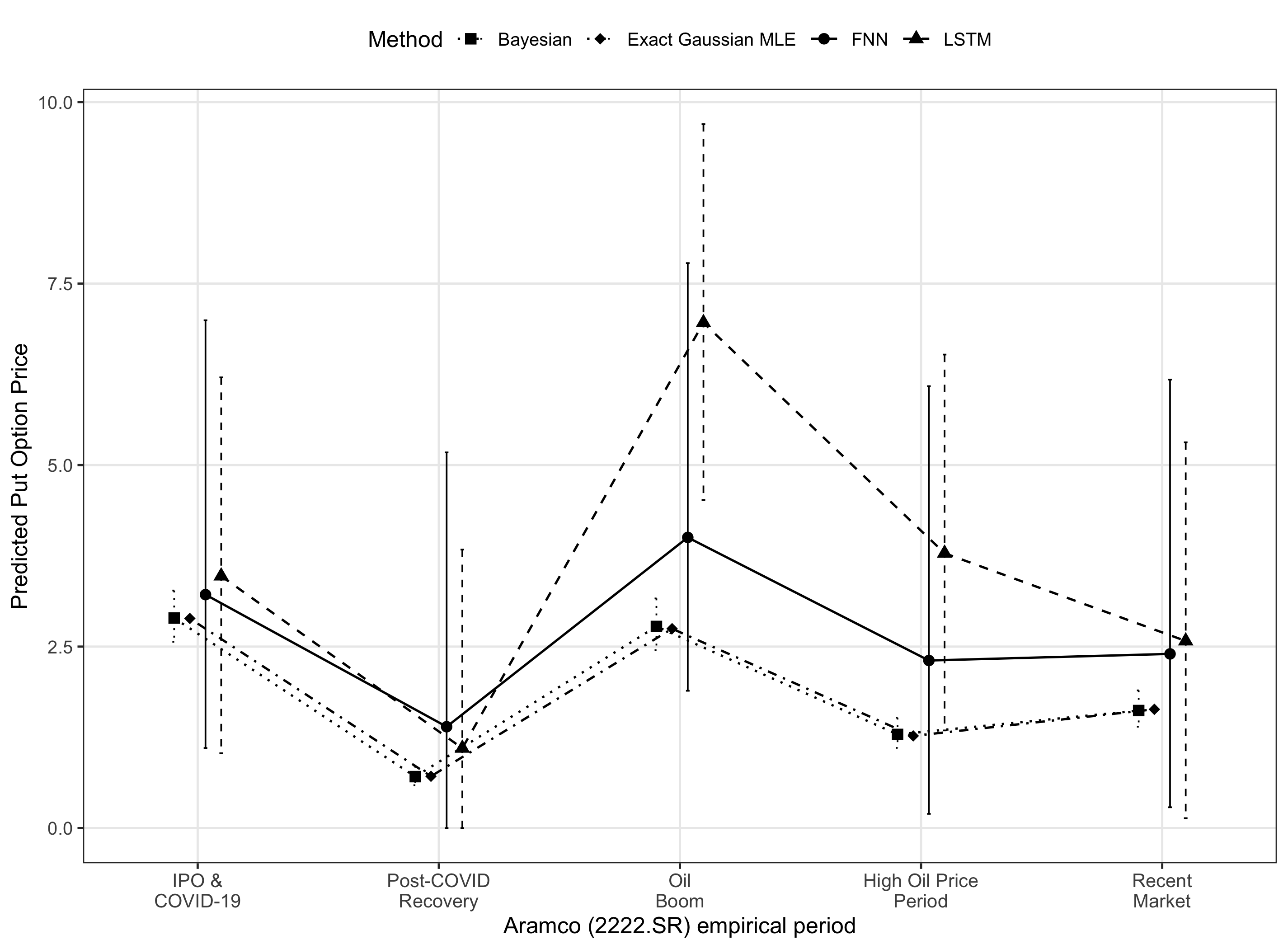}
\caption{Put-option prices.}
\label{fig:aramco_put}
\end{subfigure}

\caption{Comparison of fractional option-price predictions for Saudi
Aramco obtained from the Bayesian, Exact Gaussian MLE, FNN, and LSTM
approaches across five empirical market regimes. Error bars represent
the corresponding uncertainty intervals.}
\label{fig:aramco_option_prices}
\end{figure}

Figure~\ref{fig:aramco_option_prices} provides the corresponding
graphical comparison together with the uncertainty intervals.

Table~\ref{tab:aramco_option_results} and
Figure~\ref{fig:aramco_option_prices} show that the Bayesian and Exact
Gaussian MLE procedures produce very similar option-price predictions
throughout the five empirical regimes. During the IPO and COVID-19
period, for example, the Bayesian and Exact MLE call prices are
\(6.093\) and \(6.088\), respectively, while the corresponding put
prices are \(2.893\) and \(2.887\). During the Oil Boom period, their
call-price predictions are \(5.978\) and \(5.948\), respectively,
while during the High Oil Price period they are \(4.491\) and
\(4.471\). The corresponding put-price predictions exhibit the same
close agreement.

The neural-network predictions display greater variation across the
market regimes. During the IPO and COVID-19 period, the FNN and LSTM
call predictions are \(6.416\) and \(6.673\), respectively, compared
with \(6.093\) for the Bayesian method. The differences remain
relatively moderate in the Post-COVID Recovery period, where the FNN
and LSTM predict call prices of \(4.597\) and \(4.300\), respectively,
compared with \(3.910\) and \(3.914\) from the Bayesian and Exact MLE
procedures.

The most pronounced separation occurs during the Oil Boom period. The
Bayesian and Exact Gaussian MLE call-price predictions are \(5.978\)
and \(5.948\), whereas the FNN predicts \(7.204\) and the LSTM predicts
\(10.162\). For the corresponding put option, the four predictions are
\(2.777\), \(2.747\), \(4.004\), and \(6.961\), respectively. Thus,
the LSTM exhibits substantially stronger pricing sensitivity during
this regime.

A similar, although less pronounced, pattern occurs during the High Oil
Price period. The Bayesian and Exact MLE call prices are \(4.491\) and
\(4.471\), respectively, compared with \(5.509\) for the FNN and
\(6.987\) for the LSTM. In the Recent Market regime, the four methods
move closer together, although the neural-network predictions remain
somewhat higher than the likelihood-based prices.

The pricing differences are particularly notable because the LSTM also
produces relatively high Hurst-parameter estimates during the Oil Boom
and High Oil Price periods. More generally, the empirical results show
that agreement in the estimated direction of persistence does not
necessarily imply identical option-price predictions across the four
methods. This reflects an important methodological distinction. The
Bayesian and Exact Gaussian MLE procedures estimate the parameters of
the fractional Gaussian model and subsequently propagate those
estimates through the fractional Black--Scholes pricing equation. In
contrast, the FNN and LSTM learn a direct nonlinear mapping from the
complete return sequence to the Hurst parameter and option-price
outputs.

Consequently, the differences among the empirical option-price
predictions should not be interpreted as direct differences in
predictive accuracy. Unlike the simulation study, the true
model-implied Hurst parameter and option prices are not known for the
empirical Aramco windows. The empirical analysis therefore evaluates
the agreement, divergence, regime sensitivity, and uncertainty of the
four procedures rather than ranking them according to prediction
error.

\subsubsection{Computational Efficiency for Saudi Aramco}

An important practical distinction among the four methods is the
computational cost required to process a new empirical return window.
The Bayesian and Exact Gaussian MLE procedures require
sequence-specific statistical estimation, whereas the trained FNN and
LSTM obtain predictions through a forward pass without re-estimating
the model.

Table~\ref{tab:aramco_computation} summarizes the computational time
required by the four methods. For the five Aramco windows, the median
Bayesian computation time was approximately \(79.928\) seconds per
window. The Exact Gaussian MLE was substantially faster, requiring a
median of approximately \(0.253\) seconds per window. In contrast, the
median prediction times for the FNN and LSTM were approximately
\(0.009\) and \(0.010\) seconds per window, respectively.

\begin{table}[H]
\centering
\caption{Computational time comparison for Saudi Aramco (2222.SR).}
\label{tab:aramco_computation}
\begin{tabular}{lcc}
\toprule
\textbf{Method}
& \textbf{Median Time per Window (s)}
& \textbf{Approximate Speedup} \\
\midrule

Bayesian
& 79.928
& 1.0$\times$ \\

Exact Gaussian MLE
& 0.253
& 315.9$\times$ \\

FNN
& 0.009
& 9082.7$\times$ \\

LSTM
& 0.010
& 8155.9$\times$ \\

\bottomrule
\end{tabular}
\end{table}

The timing results demonstrate a substantial computational advantage
for the neural-network models after training. The Exact Gaussian MLE
is already considerably faster than Bayesian inference because it
avoids posterior sampling. Once trained, however, the FNN and LSTM
require only a forward pass through the fitted network and therefore
reduce the per-window evaluation time by several orders of magnitude
relative to Bayesian posterior computation.

This advantage should be interpreted specifically as an inference-time
benefit. The substantial one-time computational cost required to
generate the simulated training data and train the FNN and LSTM models
is not included in these per-window prediction times and was documented
separately in the simulation study. Consequently, the neural-network
models are particularly attractive when a trained model is used
repeatedly to evaluate many new return sequences.

\paragraph{Summary.}

The Aramco application demonstrates that the four methods respond
meaningfully to changes in empirical market conditions. The estimated
Hurst parameters are relatively close to \(0.5\) during the early
post-IPO and recovery periods, increase during the Oil Boom and High
Oil Price regimes, and subsequently decline during the Recent Market
period. The broad increase in persistence during the middle regimes is
identified by multiple estimation approaches, although the magnitude
of the estimates differs across methods.

The option-pricing analysis reveals a stronger distinction between the
likelihood-based and neural-network approaches. The Bayesian and Exact
Gaussian MLE procedures produce closely aligned fractional option
prices across all five regimes, whereas the FNN and particularly the
LSTM exhibit greater regime-dependent variation. The largest divergence
occurs during the Oil Boom period, when the LSTM produces substantially
higher call- and put-price predictions. These differences reflect the
distinct ways in which the methods extract and propagate information
from the return sequence and should not be interpreted as an empirical
ranking of predictive accuracy in the absence of known true option
prices.

The computational results provide an additional practical distinction.
Although the Exact Gaussian MLE is substantially faster than Bayesian
posterior inference, the trained FNN and LSTM reduce the per-window
evaluation time by several additional orders of magnitude. This
computational advantage becomes especially important when predictions
must be generated repeatedly for many return windows.

Combined with the simulation results, the Aramco application therefore
illustrates the complementary roles of the four approaches. The
simulation study provides a controlled environment in which estimation
accuracy, pricing accuracy, and interval performance can be evaluated
against known true values, whereas the empirical application
demonstrates how the methods respond to changing real-market
conditions. The likelihood-based approaches provide explicit
model-based statistical inference, while the trained neural networks
provide extremely rapid sequence-level predictions after the initial
training stage.

\subsection{Empirical Application to Ma'aden}
\label{subsec:maaden_application}

To further evaluate the ability of the proposed neural-network models to
generalize beyond simulated fractional Gaussian noise, an empirical
application was conducted using daily returns of Saudi Arabian Mining
Company (Ma'aden; \texttt{1211.SR}). Ma'aden provides an informative
real-data setting because of its exposure to commodity-market conditions
and the substantial changes in financial-market behavior observed over
the study period.

Five empirical windows were considered: 2014--2015, 2019--2020,
2021--2022, 2023--2024, and 2025--2026. Each window contains exactly
252 daily log returns so that the empirical inputs have the same
dimension as the sequences used to train the FNN and LSTM models.
The same return vector within each period was supplied to all four
estimation procedures: the Bayesian model, Exact Gaussian maximum
likelihood estimation (MLE), the feed-forward neural network (FNN),
and the long short-term memory (LSTM) network.

\begin{table}[htbp]
\centering
\caption{Estimated Hurst parameter for Ma'aden (1211.SR) across five
empirical periods.}
\label{tab:maaden_H_comparison}
\begin{tabular}{lcccc}
\hline
\textbf{Period}
& \textbf{Bayesian}
& \textbf{Exact MLE}
& \textbf{FNN}
& \textbf{LSTM} \\
\hline

2014--2015
& 0.5662 & 0.5465 & 0.5687 & 0.5768 \\

2019--2020
& 0.5250 & 0.5000 & 0.6190 & 0.5606 \\

2021--2022
& 0.5995 & 0.5839 & 0.6367 & 0.6247 \\

2023--2024
& 0.5541 & 0.5249 & 0.5848 & 0.5755 \\

2025--2026
& 0.5185 & 0.5000 & 0.5454 & 0.5339 \\

\hline
\end{tabular}
\end{table}

\begin{figure}[htbp]
\centering
\includegraphics[
    width=0.95\textwidth
]{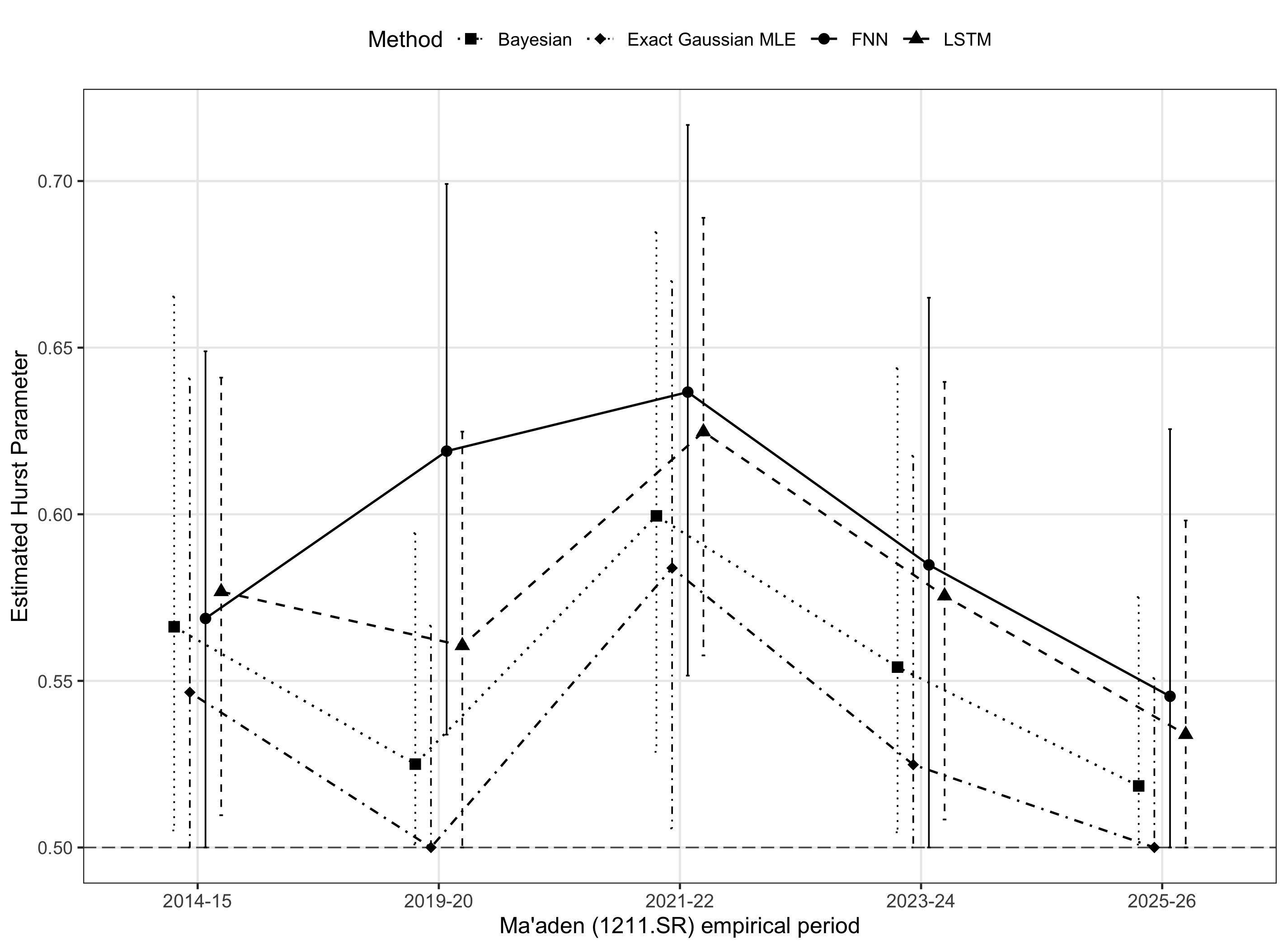}
\caption{Comparison of Hurst parameter estimates for Ma'aden
(1211.SR) across five empirical periods. The Bayesian posterior
median, Exact Gaussian MLE, FNN prediction, and LSTM prediction are
shown. The horizontal dashed line at $H=0.5$ represents the standard
Brownian-motion benchmark.}
\label{fig:maaden_H}
\end{figure}

Figure~\ref{fig:maaden_H} shows variation in the estimated degree of
persistence across the five empirical periods. During 2014--2015,
all four procedures produce estimates moderately above the
Brownian-motion benchmark. The Bayesian posterior median is
$\widehat{H}_{\mathrm{Bayes}}=0.5662$, compared with
$\widehat{H}_{\mathrm{MLE}}=0.5465$, while the FNN and LSTM predictions
are 0.5687 and 0.5768, respectively. Thus, the four methods provide
relatively similar evidence of persistence during this period.

A different pattern occurs during 2019--2020. The Bayesian estimate
declines to 0.5250 and the Exact Gaussian MLE reaches the lower
boundary of 0.5000. In contrast, the FNN produces a substantially
higher estimate of 0.6190, whereas the LSTM estimate of 0.5606 lies
between the FNN and likelihood-based estimates. Thus, the strength of
the inferred persistence during this period is more dependent on the
estimation method.

The strongest cross-method evidence of persistence occurs during
2021--2022. The Bayesian posterior median increases to 0.5995 and the
Exact Gaussian MLE to 0.5839. The neural networks indicate an even
higher degree of persistence, with estimates of 0.6367 for the FNN
and 0.6247 for the LSTM. Importantly, all four methods identify
2021--2022 as a period characterized by elevated dependence relative
to most of the remaining empirical windows.

Persistence subsequently declines. During 2023--2024, the Bayesian,
Exact MLE, FNN, and LSTM estimates are 0.5541, 0.5249, 0.5848, and
0.5755, respectively. By 2025--2026, the Bayesian estimate decreases
further to 0.5185 and the Exact MLE again reaches 0.5000, while the
FNN and LSTM predictions decline to 0.5454 and 0.5339. Overall, the
four methods reproduce a broadly similar temporal pattern, although
the neural networks generally produce somewhat higher Hurst estimates
than the likelihood-based procedures.

\begin{table}[htbp]
\centering
\caption{Comparison of predicted call and put option prices for
Ma'aden (1211.SR).}
\label{tab:maaden_option_results}

\resizebox{\textwidth}{!}{
\begin{tabular}{lcccccccc}
\hline
& \multicolumn{4}{c}{\textbf{Call Option}}
& \multicolumn{4}{c}{\textbf{Put Option}} \\
\cline{2-5}
\cline{6-9}

\textbf{Period}
& \textbf{Bayesian}
& \textbf{Exact MLE}
& \textbf{FNN}
& \textbf{LSTM}
& \textbf{Bayesian}
& \textbf{Exact MLE}
& \textbf{FNN}
& \textbf{LSTM} \\
\hline

2014--2015
& 9.8065 & 9.8048 & 11.5009 & 13.2104
& 6.6057 & 6.6040 & 8.3001 & 10.0096 \\

2019--2020
& 8.8709 & 8.9139 & 11.6559 & 11.5765
& 5.6701 & 5.7131 & 8.4550 & 8.3757 \\

2021--2022
& 7.2843 & 7.2591 & 11.1089 & 12.4577
& 4.0835 & 4.0583 & 7.9080 & 9.2569 \\

2023--2024
& 7.3818 & 7.4086 & 8.1712 & 9.3894
& 4.1810 & 4.2078 & 4.9704 & 6.1886 \\

2025--2026
& 8.0278 & 8.0447 & 9.0887 & 9.1820
& 4.8270 & 4.8439 & 5.8879 & 5.9812 \\

\hline
\end{tabular}
}
\end{table}

\begin{figure}[htbp]
\centering

\begin{subfigure}[t]{0.49\textwidth}
\centering
\includegraphics[
    width=\textwidth
]{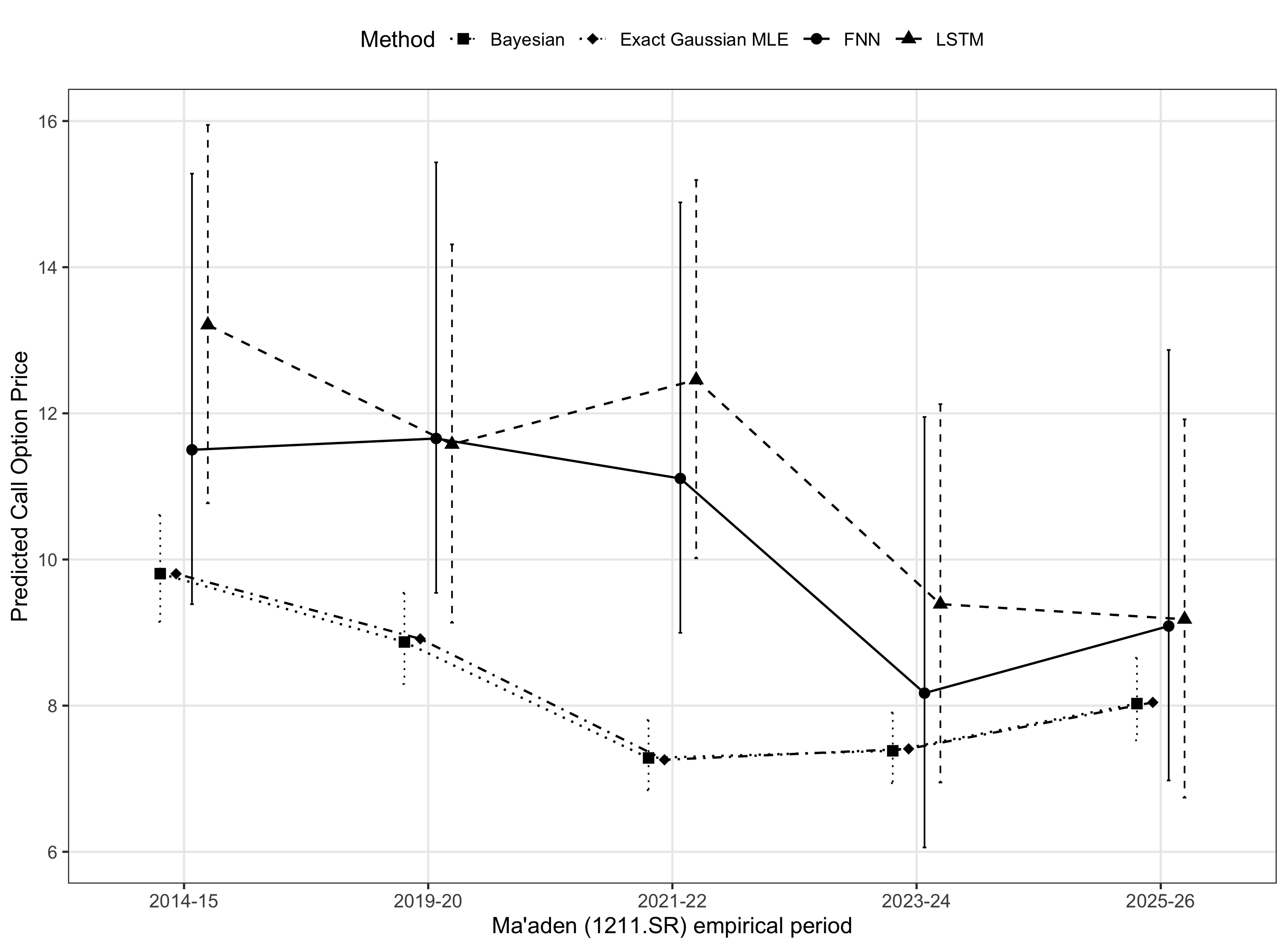}
\caption{Call option prices.}
\label{fig:maaden_call}
\end{subfigure}
\hfill
\begin{subfigure}[t]{0.49\textwidth}
\centering
\includegraphics[
    width=\textwidth
]{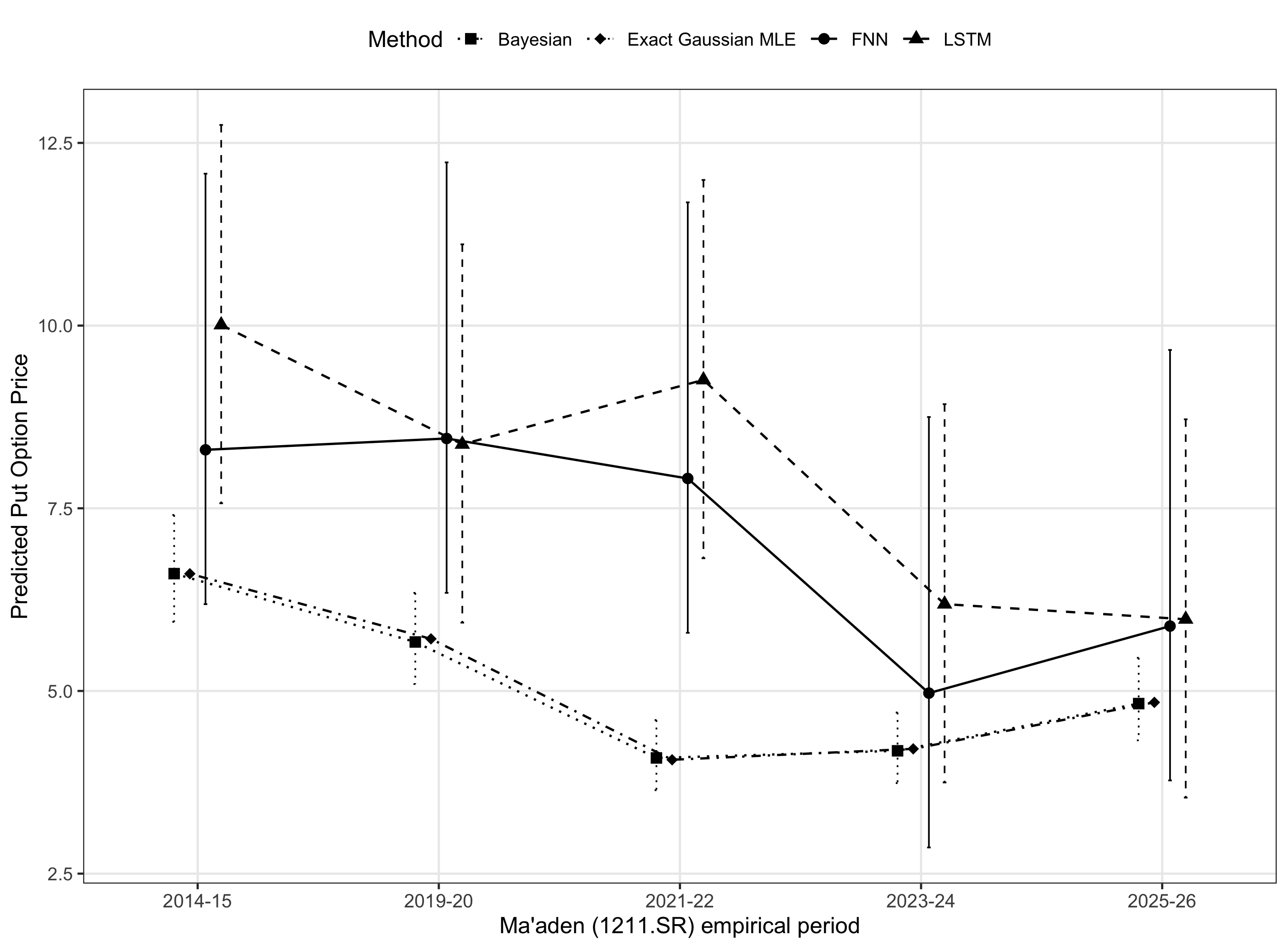}
\caption{Put option prices.}
\label{fig:maaden_put}
\end{subfigure}

\caption{Comparison of fractional option-price predictions for
Ma'aden obtained from the Bayesian, Exact Gaussian MLE, FNN, and
LSTM approaches across five empirical periods.}
\label{fig:maaden_option_prices}
\end{figure}

Figure~\ref{fig:maaden_option_prices} compares the corresponding
fractional call and put prices. The Bayesian and Exact Gaussian MLE
prices are nearly indistinguishable across all five empirical periods.
For example, during 2014--2015, the Bayesian and Exact MLE call prices
are 9.8065 and 9.8048, respectively, while the corresponding put prices
are 6.6057 and 6.6040. Similarly, during 2021--2022, the Bayesian and
Exact MLE call prices are 7.2843 and 7.2591, with corresponding put
prices of 4.0835 and 4.0583. This close agreement provides a useful
classical likelihood-based benchmark for evaluating the neural-network
predictions.

The neural-network models generally produce higher option-price
predictions than the two likelihood-based procedures. During
2014--2015, for example, the FNN predicts a call price of 11.5009,
whereas the LSTM predicts 13.2104, compared with the Bayesian value
of 9.8065. The same ordering is observed for the put price, for which
the FNN and LSTM predictions are 8.3001 and 10.0096, respectively,
compared with the Bayesian value of 6.6057.

The largest discrepancies are observed during 2021--2022. Although
all four methods identify relatively strong persistence during this
period, the Bayesian and Exact MLE call prices are approximately 7.28,
whereas the FNN and LSTM predict 11.1089 and 12.4577, respectively.
The corresponding put prices are approximately 4.08 for the
likelihood-based methods, compared with 7.9080 for the FNN and 9.2569
for the LSTM. Thus, agreement in the direction of the Hurst-parameter
movement does not necessarily translate into agreement in the
resulting option prices.

The differences become considerably smaller during the two most recent
periods. In 2023--2024, the Bayesian call price is 7.3818, compared
with 8.1712 for the FNN and 9.3894 for the LSTM. By 2025--2026, the
four call-price estimates are closer, with values of 8.0278, 8.0447,
9.0887, and 9.1820 for the Bayesian, Exact MLE, FNN, and LSTM methods,
respectively. The corresponding put-price results exhibit the same
general pattern.

These findings illustrate an important distinction between dependence
estimation and option-price prediction. Although the neural networks
capture important changes in the Hurst parameter across empirical
periods, their option-price predictions can differ substantially from
the likelihood-based benchmarks. In particular, the LSTM does not
uniformly provide the closest pricing approximation despite performing
strongly in Hurst-parameter estimation. For Ma'aden, the FNN is closer
to the Bayesian option-price benchmark in four of the five periods,
while the LSTM is slightly closer during 2019--2020. Therefore, the
relative empirical performance of the two neural-network architectures
depends on the prediction target and market period.


\subsubsection{Computational Efficiency for Ma'aden}

An important advantage of the neural-network approaches is their
computational efficiency after training. Table~\ref{tab:maaden_computation}
reports the median computational time per empirical window. Bayesian
estimation required a median of approximately \(81.6580\) seconds per
window, reflecting the computational cost of posterior simulation.
Exact Gaussian MLE was substantially faster, requiring approximately
\(0.2640\) seconds per window.

\begin{table}[htbp]
\centering
\caption{Computational time comparison for Ma'aden (1211.SR).}
\label{tab:maaden_computation}
\begin{tabular}{lcc}
\hline
\textbf{Method}
& \textbf{Median Time per Window (s)}
& \textbf{Approximate Speedup} \\
\hline

Bayesian
& 81.6580
& 1.0$\times$ \\

Exact Gaussian MLE
& 0.2640
& 309.3$\times$ \\

FNN
& 0.0088
& 9279.3$\times$ \\

LSTM
& 0.0100
& 8165.8$\times$ \\

\hline
\end{tabular}
\end{table}

Once trained, the neural-network models require only a forward pass
through the fitted architecture. The FNN required a median of
approximately \(0.0088\) seconds per window and the LSTM approximately
\(0.0100\) seconds. Relative to Bayesian inference, these timings
correspond to speedups of approximately \(9279.3\) and \(8165.8\)
times, respectively. The Exact Gaussian MLE also provides a substantial
computational improvement over Bayesian estimation, but it remains
considerably slower than either neural-network surrogate.

The timing results emphasize the different roles of the four
approaches. Bayesian inference provides full posterior uncertainty
quantification but has the greatest computational cost. Exact Gaussian
MLE offers a much faster likelihood-based alternative while retaining
explicit model-based parameter estimation. The FNN and LSTM provide
the greatest computational advantage once training has been completed,
making them particularly attractive when predictions must be generated
repeatedly across many return sequences or market periods.

\paragraph{Summary.}

The Ma'aden application provides additional evidence that the four
methods capture meaningful variation in dependence across empirical
market periods. The strongest cross-method evidence of persistence
occurs during 2021--2022, when all four Hurst estimates are clearly
above the Brownian-motion benchmark. Persistence subsequently declines,
with the Bayesian and Exact Gaussian MLE estimates approaching 0.5
during the most recent period. The FNN and LSTM reproduce the broad
regime-dependent pattern, although they generally produce somewhat
higher Hurst estimates than the likelihood-based methods.

For option pricing, the Bayesian and Exact Gaussian MLE procedures
produce remarkably similar values across all five periods, whereas the
neural-network predictions are generally higher. The magnitude of this
difference varies substantially across periods and is particularly
large during 2021--2022. The FNN is closer to the Bayesian pricing
benchmark in most of the Ma'aden periods, whereas the LSTM provides
competitive Hurst-parameter estimates. Consequently, neither neural
architecture uniformly dominates across both prediction targets.

At the same time, both neural networks provide a substantial
computational advantage. Once trained, the FNN and LSTM generate
predictions in approximately one hundredth of a second per empirical
window, compared with more than 80 seconds for Bayesian inference.
Taken together, the Ma'aden results reinforce the complementary roles
of the methods: the neural networks provide extremely fast surrogate
prediction, Exact Gaussian MLE provides a computationally efficient
likelihood-based benchmark, and Bayesian inference remains valuable
when full posterior uncertainty quantification is required.

\subsection{Empirical Application to Saudi Telecom Company (STC)}
\label{subsec:stc_application}

To further evaluate the ability of the proposed neural-network models
to generalize beyond simulated fractional Gaussian noise, an empirical
application was conducted using daily returns of Saudi Telecom Company
(STC; \texttt{7010.SR}). STC provides a useful empirical setting because
its return history spans several distinct economic and financial
conditions, including the oil-price decline, the COVID-19 shock,
the subsequent recovery and expansion, a period of elevated interest
rates, and more recent market conditions.

Five empirical windows were considered: the Oil Price Collapse,
COVID-19 Shock, Recovery and Expansion, High Interest Rate Period,
and Recent Market. Each window contains exactly 252 daily log returns
so that the empirical inputs have the same dimension as the sequences
used to train the FNN and LSTM models. The same return vector within
each period was supplied to all four estimation procedures: the
Bayesian model, Exact Gaussian maximum likelihood estimation (MLE),
the feed-forward neural network (FNN), and the long short-term memory
(LSTM) network. The FNN and LSTM predictions were obtained using the
final models trained with a total simulated sample size of
\(N=500{,}000\).

\begin{table}[htbp]
\centering
\caption{Estimated Hurst parameter for STC (7010.SR) across five
empirical market periods.}
\label{tab:stc_H_comparison}
\begin{tabular}{lcccc}
\hline
\textbf{Market Period}
& \textbf{Bayesian}
& \textbf{Exact MLE}
& \textbf{FNN}
& \textbf{LSTM} \\
\hline

Oil Price Collapse
& 0.591 & 0.572 & 0.567 & 0.610 \\

COVID-19 Shock
& 0.580 & 0.566 & 0.584 & 0.602 \\

Recovery and Expansion
& 0.524 & 0.500 & 0.564 & 0.534 \\

High Interest Rate Period
& 0.550 & 0.526 & 0.560 & 0.552 \\

Recent Market
& 0.511 & 0.500 & 0.554 & 0.528 \\
\hline
\end{tabular}
\end{table}

\begin{figure}[htbp]
\centering
\includegraphics[
    width=0.95\textwidth
]{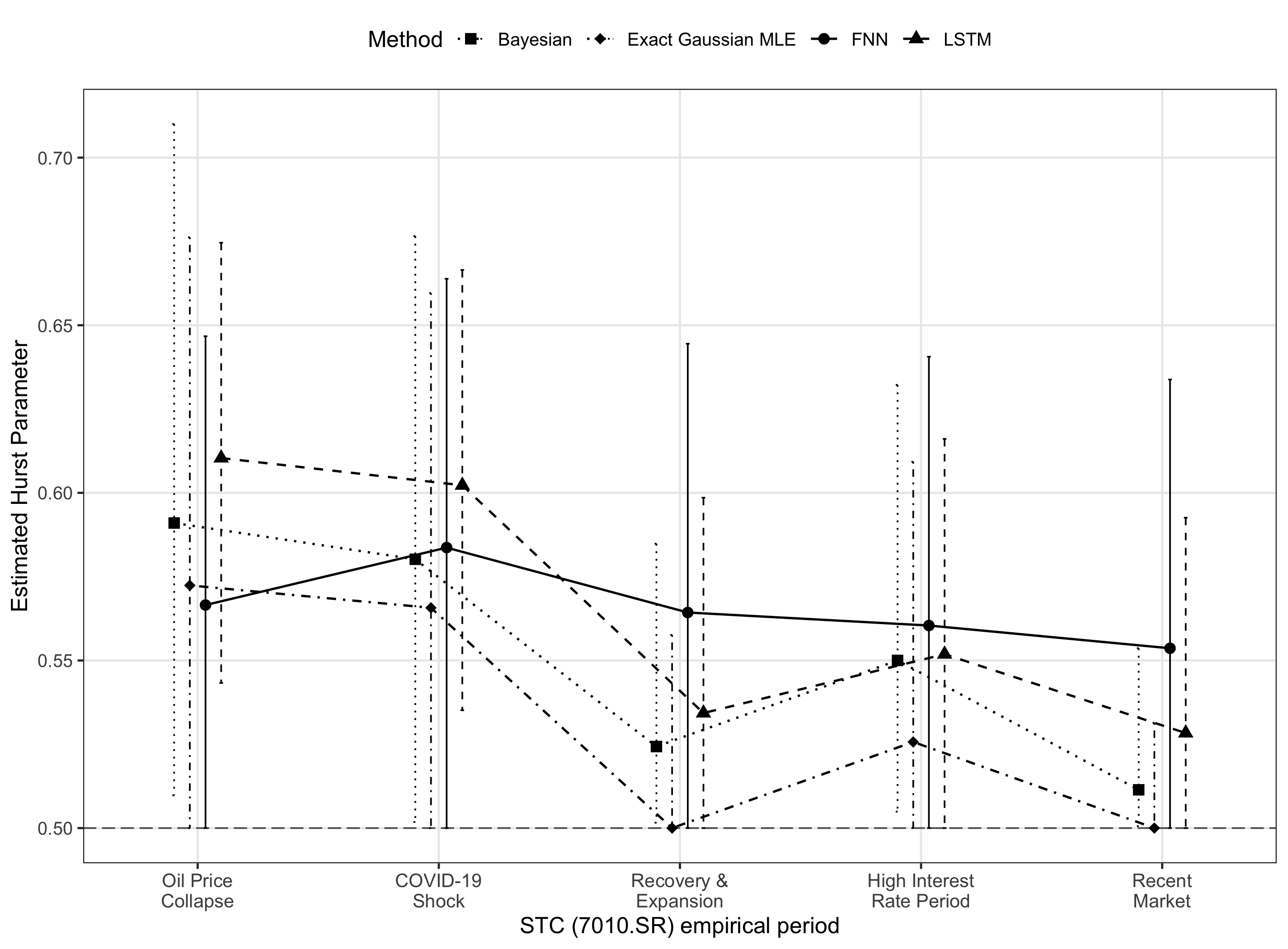}
\caption{Comparison of Hurst-parameter estimates for STC across five
empirical market periods. The Bayesian posterior median, Exact Gaussian
MLE, FNN prediction, and LSTM prediction are shown. The horizontal
dashed line at \(H=0.5\) represents the standard Brownian-motion
benchmark.}
\label{fig:stc_H}
\end{figure}

Figure~\ref{fig:stc_H} shows that the estimated degree of persistence
varies across the five STC periods. During the Oil Price Collapse,
the Bayesian posterior median is \(0.591\), while the Exact Gaussian
MLE, FNN, and LSTM estimates are \(0.572\), \(0.567\), and \(0.610\),
respectively. Thus, all four methods indicate persistence above the
standard Brownian-motion benchmark during this period.

A similar pattern is observed during the COVID-19 Shock. The Bayesian
posterior median is \(0.580\), compared with \(0.566\) for Exact
Gaussian MLE, \(0.584\) for the FNN, and \(0.602\) for the LSTM.
The four procedures therefore provide broadly similar evidence of
persistence during the first two empirical periods, although the LSTM
produces somewhat larger Hurst estimates.

The estimated persistence decreases during the Recovery and Expansion
period. The Bayesian posterior median falls to \(0.524\), while the
Exact Gaussian MLE reaches the lower boundary at approximately
\(0.500\). The FNN remains higher at \(0.564\), whereas the LSTM
estimate of \(0.534\) is closer to the Bayesian estimate. During the
High Interest Rate Period, the four estimates again become relatively
close, with values of \(0.550\), \(0.526\), \(0.560\), and \(0.552\)
for the Bayesian, Exact MLE, FNN, and LSTM methods, respectively.

The Recent Market period exhibits the weakest evidence of persistence
from the likelihood-based procedures. The Bayesian posterior median
declines to \(0.511\), while the Exact Gaussian MLE is approximately
\(0.500\). The FNN produces a somewhat larger estimate of \(0.554\),
whereas the LSTM estimate of \(0.528\) lies closer to the Bayesian
result. Overall, the results indicate a general decline in the
estimated Hurst parameter from the earlier STC periods toward the
more recent periods. The LSTM also follows the direction of the
Bayesian estimates particularly closely in the final three periods.

\begin{table}[htbp]
\centering
\caption{Comparison of predicted call and put option prices for
STC (7010.SR).}
\label{tab:stc_option_comparison}
\resizebox{\textwidth}{!}{
\begin{tabular}{lrrrrrrrr}
\hline
& \multicolumn{4}{c}{\textbf{Call Price}}
& \multicolumn{4}{c}{\textbf{Put Price}} \\
\cline{2-5}\cline{6-9}
\textbf{Market Period}
& \textbf{Bayesian}
& \textbf{Exact MLE}
& \textbf{FNN}
& \textbf{LSTM}
& \textbf{Bayesian}
& \textbf{Exact MLE}
& \textbf{FNN}
& \textbf{LSTM} \\
\hline

Oil Price Collapse
& 6.808 & 6.795 & 7.797 & 10.180
& 3.608 & 3.595 & 4.596 & 6.979 \\

COVID-19 Shock
& 6.788 & 6.780 & 8.292 & 10.469
& 3.587 & 3.579 & 5.091 & 7.268 \\

Recovery and Expansion
& 5.896 & 5.908 & 6.797 & 6.421
& 2.695 & 2.707 & 3.596 & 3.220 \\

High Interest Rate Period
& 5.615 & 5.627 & 6.451 & 6.521
& 2.415 & 2.427 & 3.250 & 3.321 \\

Recent Market
& 4.893 & 4.892 & 5.769 & 5.352
& 1.692 & 1.692 & 2.569 & 2.152 \\
\hline
\end{tabular}
}
\end{table}

\begin{figure}[htbp]
\centering

\begin{subfigure}[t]{0.49\textwidth}
\centering
\includegraphics[
    width=\textwidth
]{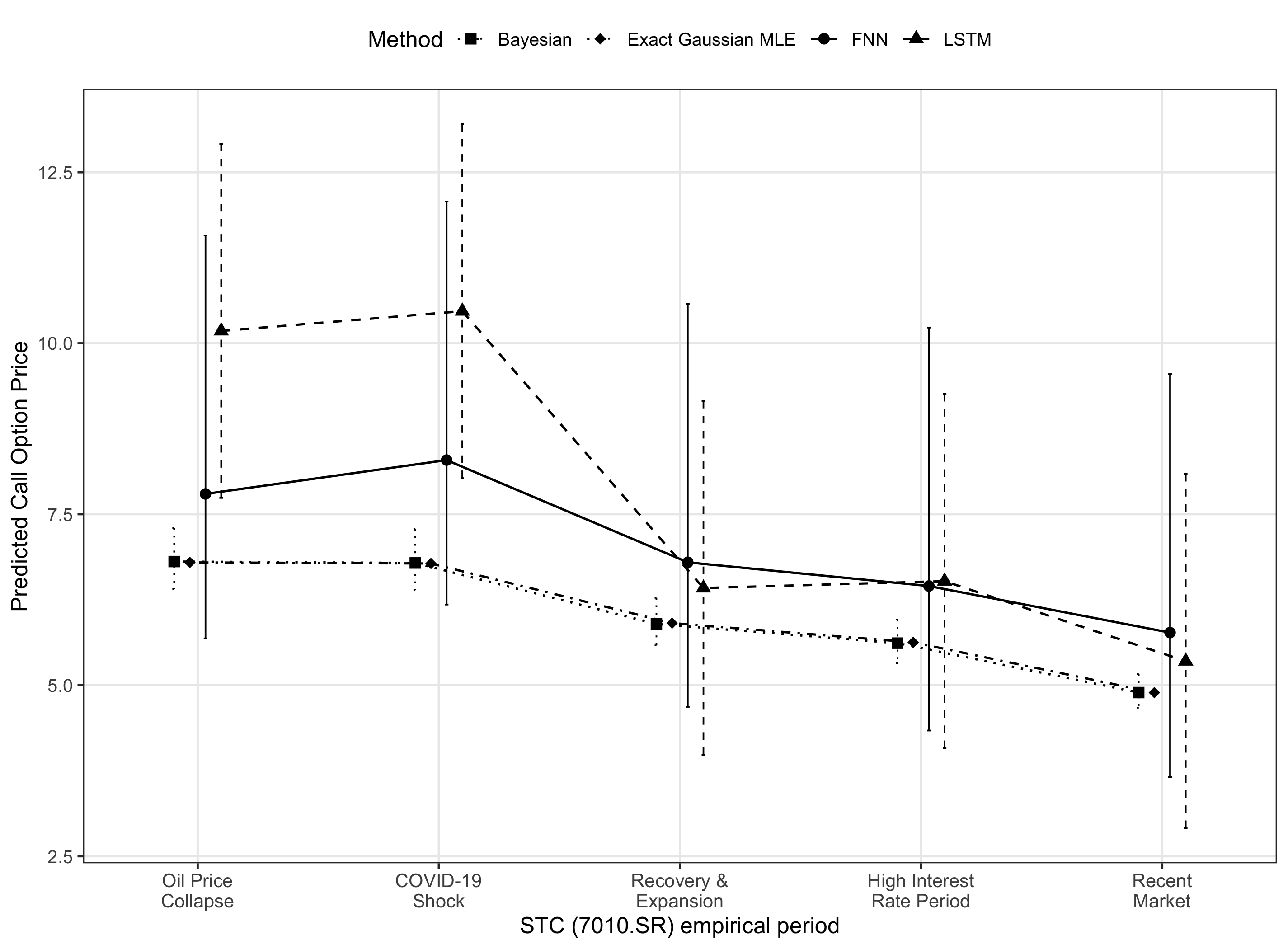}
\caption{Call option prices.}
\label{fig:stc_call}
\end{subfigure}
\hfill
\begin{subfigure}[t]{0.49\textwidth}
\centering
\includegraphics[
    width=\textwidth
]{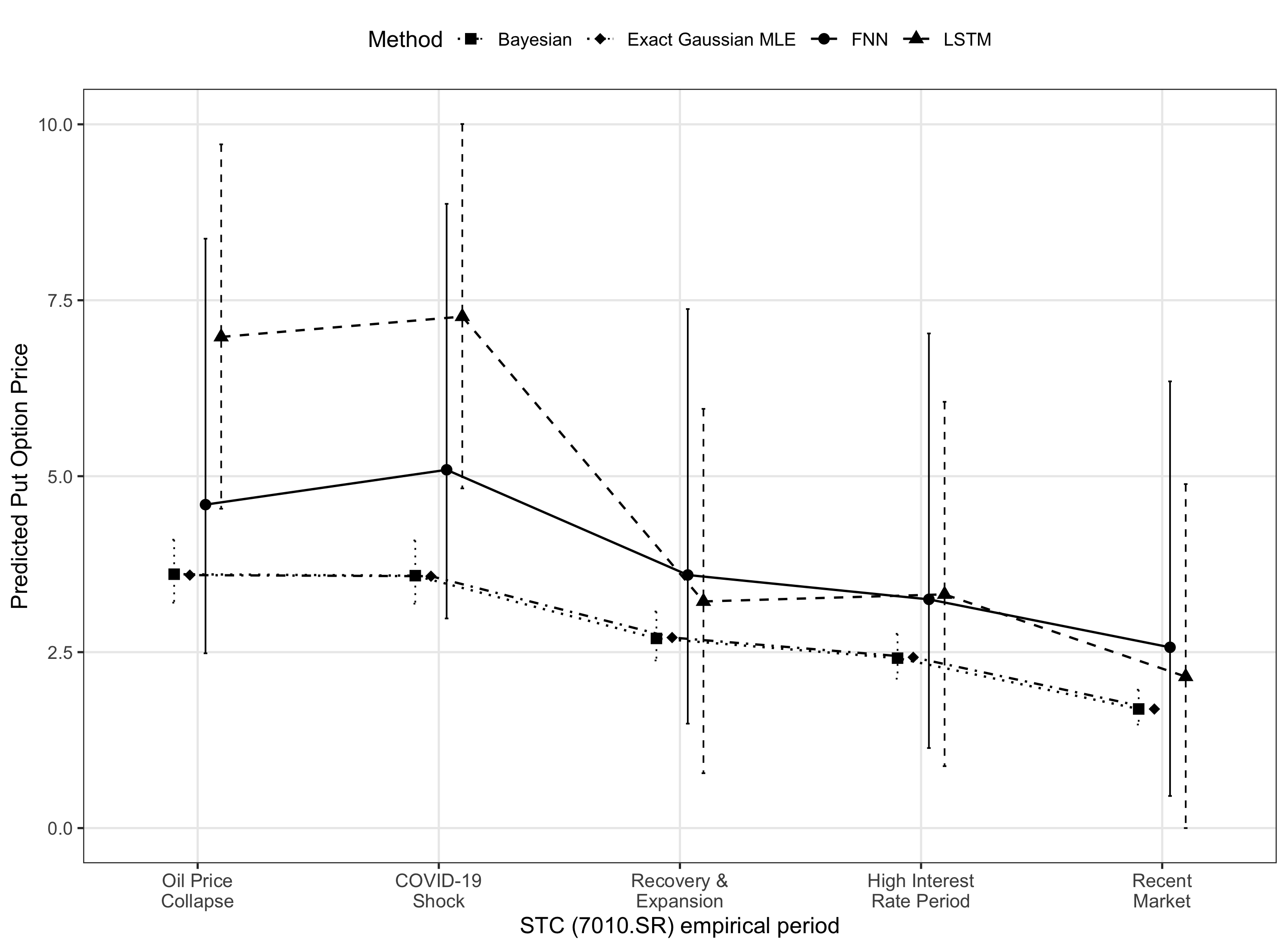}
\caption{Put option prices.}
\label{fig:stc_put}
\end{subfigure}

\caption{Comparison of fractional option-price predictions for STC
obtained from the Bayesian, Exact Gaussian MLE, FNN, and LSTM
approaches across five empirical market periods.}
\label{fig:stc_option_prices}
\end{figure}

Figure~\ref{fig:stc_option_prices} compares the corresponding
fractional call and put prices. As in the preceding empirical
applications, the Bayesian and Exact Gaussian MLE prices are remarkably
close across all five periods. During the Oil Price Collapse, for
example, the Bayesian and Exact MLE call prices are \(6.808\) and
\(6.795\), respectively, while their corresponding put prices are
\(3.608\) and \(3.595\). During the COVID-19 Shock, the corresponding
call prices are \(6.788\) and \(6.780\), and the put prices are
\(3.587\) and \(3.579\). This close agreement continues throughout
the remaining periods.

The neural-network predictions show greater variation relative to the
likelihood-based prices, particularly during the first two periods.
During the Oil Price Collapse, the FNN predicts a call price of
\(7.797\), compared with \(10.180\) for the LSTM. The corresponding
put prices are \(4.596\) and \(6.979\), respectively. Relative to the
Bayesian call price of \(6.808\), the FNN therefore provides the
closer neural-network approximation during this period.

The same pattern is particularly pronounced during the COVID-19 Shock.
The Bayesian call price is \(6.788\) and the Exact MLE price is
\(6.780\), compared with \(8.292\) for the FNN and \(10.469\) for
the LSTM. For the put option, the corresponding values are \(3.587\),
\(3.579\), \(5.091\), and \(7.268\). Thus, although the LSTM produces
a Hurst estimate reasonably close to the likelihood-based estimates
during this period, its option-price prediction is substantially
higher.

The differences become considerably smaller during the Recovery and
Expansion period. The Bayesian call price is \(5.896\), compared with
\(5.908\) for Exact MLE, \(6.797\) for the FNN, and \(6.421\) for the
LSTM. In this period, the LSTM provides the closer neural-network
approximation to the likelihood-based prices. A similar result occurs
in the Recent Market period, where the Bayesian call price is \(4.893\)
and the LSTM prediction is \(5.352\), compared with \(5.769\) for the
FNN.

During the High Interest Rate Period, the two neural-network predictions
are similar: \(6.451\) for the FNN and \(6.521\) for the LSTM, compared
with \(5.615\) and \(5.627\) from the Bayesian and Exact MLE procedures.
Overall, neither neural-network architecture uniformly provides the
closest option-price approximation across every period. The FNN is
closer during the first two periods and the High Interest Rate Period,
whereas the LSTM is closer during the Recovery and Expansion and Recent
Market periods. This reinforces the distinction between learning the
dependence parameter and learning the complete nonlinear mapping from a
return sequence to an option price.

\subsubsection{Computational Efficiency for STC}

An important advantage of the neural-network approaches is their
computational efficiency after training. Table~\ref{tab:stc_computation}
reports the median computational time per empirical window. Median
rather than mean time is reported to reduce the influence of occasional
longer runs, particularly for Bayesian posterior simulation.

\begin{table}[htbp]
\centering
\caption{Computational time comparison for STC (7010.SR).}
\label{tab:stc_computation}
\begin{tabular}{lcc}
\hline
\textbf{Method}
& \textbf{Median Time per Window (s)}
& \textbf{Approximate Speedup} \\
\hline

Bayesian
& 80.992
& 1.0$\times$ \\

Exact Gaussian MLE
& 0.244
& 331.9$\times$ \\

FNN
& 0.009
& 8803.5$\times$ \\

LSTM
& 0.011
& 7362.9$\times$ \\
\hline
\end{tabular}
\end{table}

The Bayesian procedure has the largest computational requirement, with
a median elapsed time of \(80.992\) seconds per empirical window.
Exact Gaussian MLE is substantially faster, with a median time of
approximately \(0.244\) seconds. Once trained, the neural-network
models require only a forward pass through the fitted architecture.
The median prediction times are approximately \(0.009\) seconds for
the FNN and \(0.011\) seconds for the LSTM.

These results demonstrate the substantial computational advantage of
the neural-network surrogates for repeated prediction. Bayesian
inference remains valuable because it provides full posterior
uncertainty quantification, while Exact Gaussian MLE provides a much
faster model-based alternative. The FNN and LSTM, however, reduce the
per-window prediction time by several orders of magnitude relative to
Bayesian posterior simulation, making them particularly suitable when
predictions must be generated repeatedly or rapidly.

\paragraph{Summary.}
The STC application provides further evidence that the four approaches
capture meaningful variation in dependence across empirical financial
periods. The likelihood-based estimates indicate stronger persistence
during the Oil Price Collapse and COVID-19 Shock and progressively
weaker persistence in the later periods, with the Bayesian posterior
median declining from \(0.591\) to \(0.511\). The neural-network
estimates reproduce the broad pattern, although the FNN tends to remain
somewhat higher in the later periods and the LSTM more closely follows
the Bayesian estimates during the Recovery and Expansion, High Interest
Rate, and Recent Market periods.

For option pricing, the Bayesian and Exact Gaussian MLE procedures
produce nearly identical point estimates across all five periods.
The neural-network predictions exhibit larger departures during the
earlier periods, particularly for the LSTM during the Oil Price
Collapse and COVID-19 Shock, while agreement improves in several of
the later periods. Neither neural-network architecture uniformly
dominates the other across all empirical windows. At the same time,
both provide predictions in only a small fraction of a second once
training is complete. Taken together with the Aramco and Ma'aden
applications, the STC results further support the use of neural
networks as computationally efficient surrogate models, while Bayesian
inference remains preferable when full posterior uncertainty
quantification is required.

\subsection{Empirical Application to Silver}
\label{subsec:silver_application}

To further evaluate the ability of the four methods to operate under
different empirical market conditions, an additional application was
conducted using daily returns of silver futures (\texttt{SI=F}).
Silver provides a useful complement to the equity applications because
its price dynamics are influenced by both investment demand and
commodity-market conditions and include several episodes of pronounced
financial and macroeconomic stress.

Five empirical windows were considered, each containing exactly 252
daily log returns so that the empirical inputs have the same dimension
as the sequences used to train the FNN and LSTM models. The five
periods correspond to the 2008--09 Global Financial Crisis, the 2011
Silver Peak, the 2013 Precious-Metal Crash, the COVID-19 Shock, and the
2022 Inflation Shock. The same return sequence within each period was
supplied to the Bayesian model, Exact Gaussian maximum likelihood
estimation (MLE), feed-forward neural network (FNN), and long short-term
memory (LSTM) network.

\begin{table}[htbp]
\centering
\caption{Estimated Hurst parameter for silver across five empirical
market regimes.}
\label{tab:silver_H_comparison}
\begin{tabular}{lcccc}
\hline
\textbf{Market regime}
& \textbf{Bayesian}
& \textbf{Exact MLE}
& \textbf{FNN}
& \textbf{LSTM} \\
\hline

2008--09 Crisis
& 0.522 & 0.500 & 0.561 & 0.526 \\

2011 Silver Peak
& 0.589 & 0.574 & 0.617 & 0.599 \\

2013 Metal Crash
& 0.526 & 0.500 & 0.557 & 0.539 \\

\textbf{COVID-19 Shock}
& \textbf{0.732} & \textbf{0.709}
& \textbf{0.699} & \textbf{0.657} \\

2022 Inflation Shock
& 0.523 & 0.500 & 0.566 & 0.532 \\

\hline
\end{tabular}
\end{table}

\begin{figure}[htbp]
\centering
\includegraphics[
    width=0.95\textwidth
]{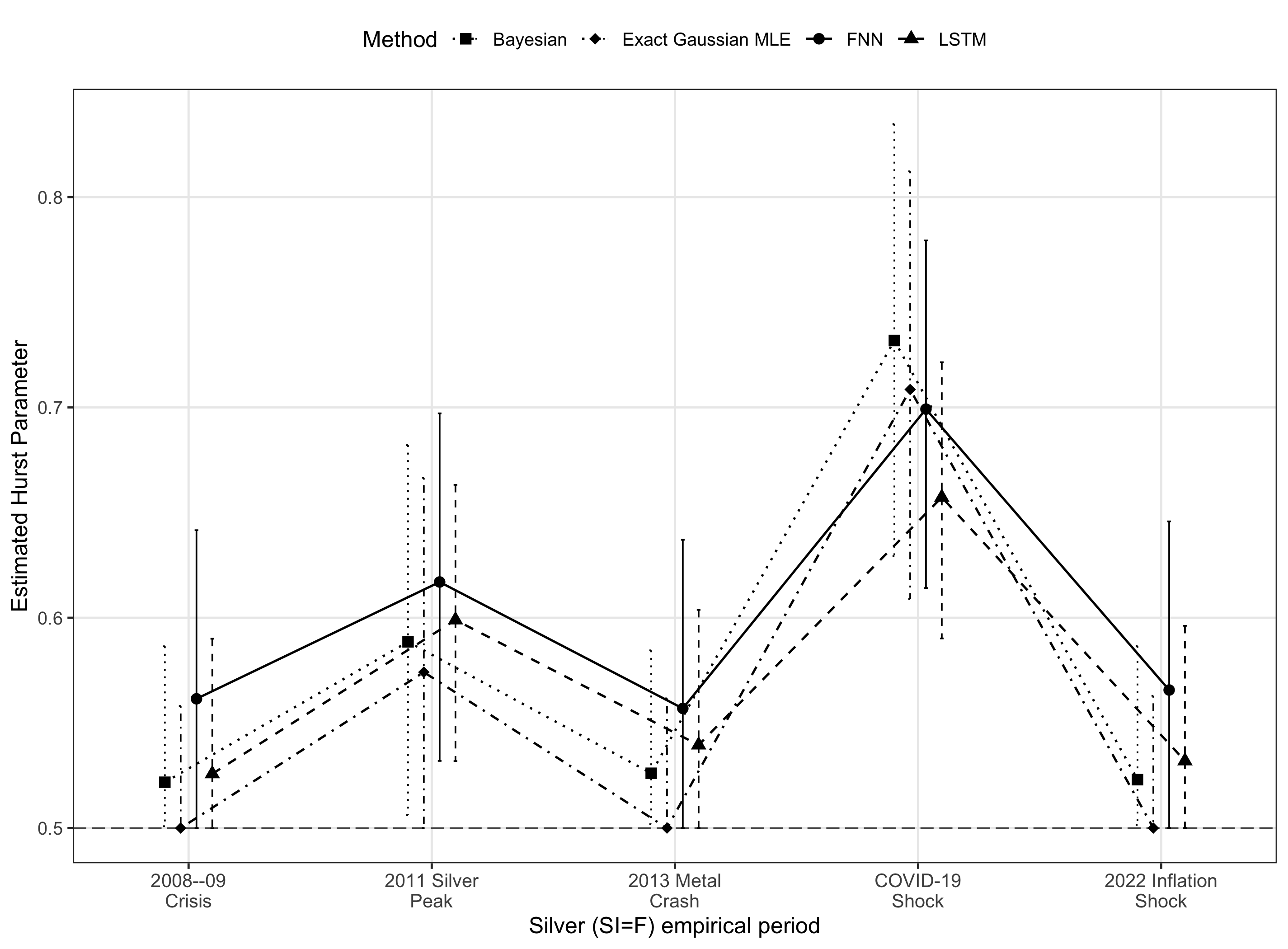}
\caption{Comparison of Hurst parameter estimates for silver across five
empirical market regimes. The Bayesian posterior median, Exact Gaussian
MLE, FNN prediction, and LSTM prediction are shown. The horizontal
dashed line at $H=0.5$ represents the standard Brownian-motion
benchmark.}
\label{fig:silver_H}
\end{figure}

Table~\ref{tab:silver_H_comparison} and Figure~\ref{fig:silver_H}
illustrate how the estimated Hurst parameter varies across the five
silver-market regimes, revealing differences in the degree of
persistence over time. During the 2008--09 Global Financial Crisis,
the Bayesian posterior median is
$\widehat{H}_{\mathrm{Bayes}}=0.522$, while the Exact Gaussian MLE
reaches the lower boundary of approximately 0.500. The FNN and
LSTM estimates are 0.561 and 0.526, respectively. These values
are relatively close to the benchmark \(H=0.5\), suggesting
limited evidence of pronounced persistence during this period.
The LSTM estimate is particularly close to the Bayesian
posterior median.

During the 2011 Silver Peak, the estimated Hurst parameter
increases across all four methods. The Bayesian posterior
median rises to 0.589, while the Exact Gaussian MLE reaches
0.574. The FNN and LSTM produce estimates of 0.617 and
0.599, respectively. These results suggest stronger persistence
than during the 2008--09 crisis, although the magnitude of
the increase differs across methods. The LSTM remains close
to the Bayesian estimate, while the FNN produces the
highest estimate for this regime.

The 2013 Precious-Metal Crash is characterized by a decline
in the estimated Hurst parameter. The Bayesian posterior
median decreases to 0.526, and the Exact Gaussian MLE
again reaches approximately 0.500. The corresponding FNN
and LSTM estimates are 0.557 and 0.539. Although the
neural-network estimates remain slightly higher than those
obtained from the statistical methods, all four approaches
indicate weaker persistence than during the 2011 Silver Peak,
with estimates again approaching the benchmark \(H=0.5\).

The most notable finding occurs during the COVID-19 Shock,
when all four methods produce Hurst parameter estimates
substantially above \(H=0.5\). The Bayesian posterior median
reaches 0.732, while the Exact Gaussian MLE increases to
0.709. Similarly, the FNN and LSTM produce estimates of
0.699 and 0.657, respectively. These are the highest
estimates obtained by each method across the five market
regimes, indicating that the COVID-19 period exhibits the
strongest estimated persistence in silver returns. The
consistent elevation of \(H\) across all four methods
suggests a pronounced change in the dependence structure
during this period, with the results providing evidence
consistent with stronger long-range dependence. Notably,
the FNN estimate is closer to the Bayesian posterior median
than the LSTM estimate in this regime.

During the 2022 Inflation Shock, the estimated Hurst
parameter declines toward \(H=0.5\). The Bayesian posterior
median decreases to 0.523, while the Exact Gaussian MLE
returns to approximately 0.500. The FNN and LSTM estimates
are 0.566 and 0.532, respectively, with the LSTM again
closely matching the Bayesian estimate. This decline
suggests that the elevated persistence observed during
the COVID-19 Shock does not extend uniformly across
subsequent market conditions.

Overall, the findings indicate that the estimated dependence
structure of silver returns varies across market regimes.
While most periods exhibit Hurst estimates relatively close
to \(H=0.5\), the COVID-19 Shock stands out as a period
of substantially elevated persistence across all four
methods. The agreement in the direction of these estimates
supports the interpretation that the strength of fractional
dependence may vary considerably under different market
conditions, although statistical significance would require
additional uncertainty assessment.

\begin{table}[htbp]
\centering
\caption{Comparison of predicted call and put option prices for silver
(\texttt{SI=F}).}
\label{tab:silver_option_comparison}
\begin{tabular}{lcccccccc}
\hline
& \multicolumn{4}{c}{\textbf{Call Price}}
& \multicolumn{4}{c}{\textbf{Put Price}} \\
\cline{2-5} \cline{6-9}
\textbf{Market regime}
& \textbf{Bayes}
& \textbf{MLE}
& \textbf{FNN}
& \textbf{LSTM}
& \textbf{Bayes}
& \textbf{MLE}
& \textbf{FNN}
& \textbf{LSTM} \\
\hline

2008--09 Crisis
& 12.185 & 12.216 & 12.914 & 14.246
& 8.985 & 9.015 & 9.713 & 11.045 \\

2011 Silver Peak
& 8.577 & 8.565 & 11.641 & 12.266
& 5.376 & 5.364 & 8.440 & 9.065 \\

2013 Metal Crash
& 7.542 & 7.563 & 8.311 & 9.140
& 4.341 & 4.363 & 5.110 & 5.939 \\

COVID-19 Shock
& 6.797 & 6.710 & 12.023 & 10.734
& 3.597 & 3.509 & 8.822 & 7.534 \\

2022 Inflation Shock
& 6.821 & 6.845 & 7.833 & 7.351
& 3.620 & 3.645 & 4.632 & 4.151 \\

\hline
\end{tabular}
\end{table}

\begin{figure}[htbp]
\centering

\begin{subfigure}[t]{0.49\textwidth}
\centering
\includegraphics[
    width=\textwidth
]{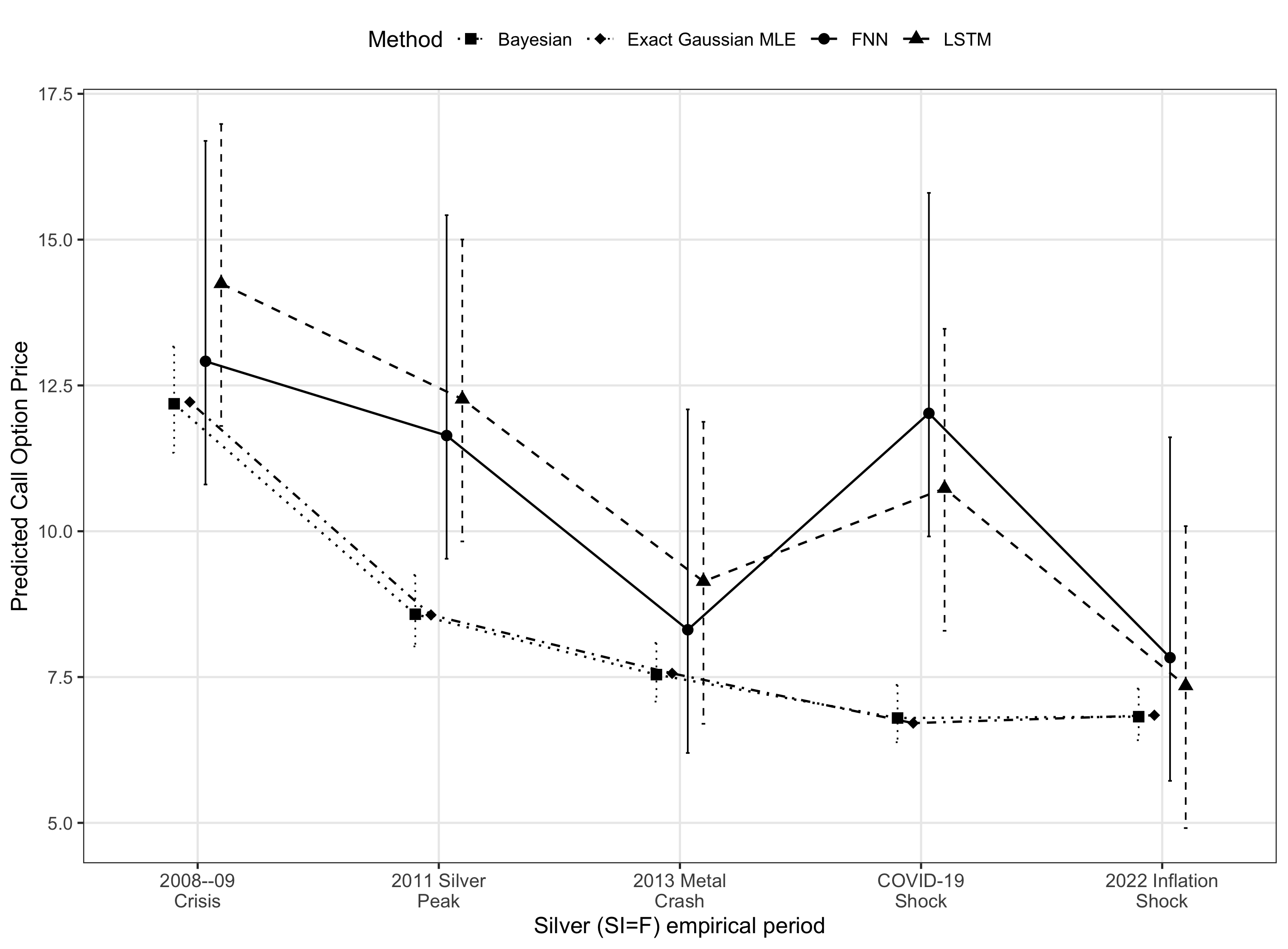}
\caption{Call option prices.}
\label{fig:silver_call}
\end{subfigure}
\hfill
\begin{subfigure}[t]{0.49\textwidth}
\centering
\includegraphics[
    width=\textwidth
]{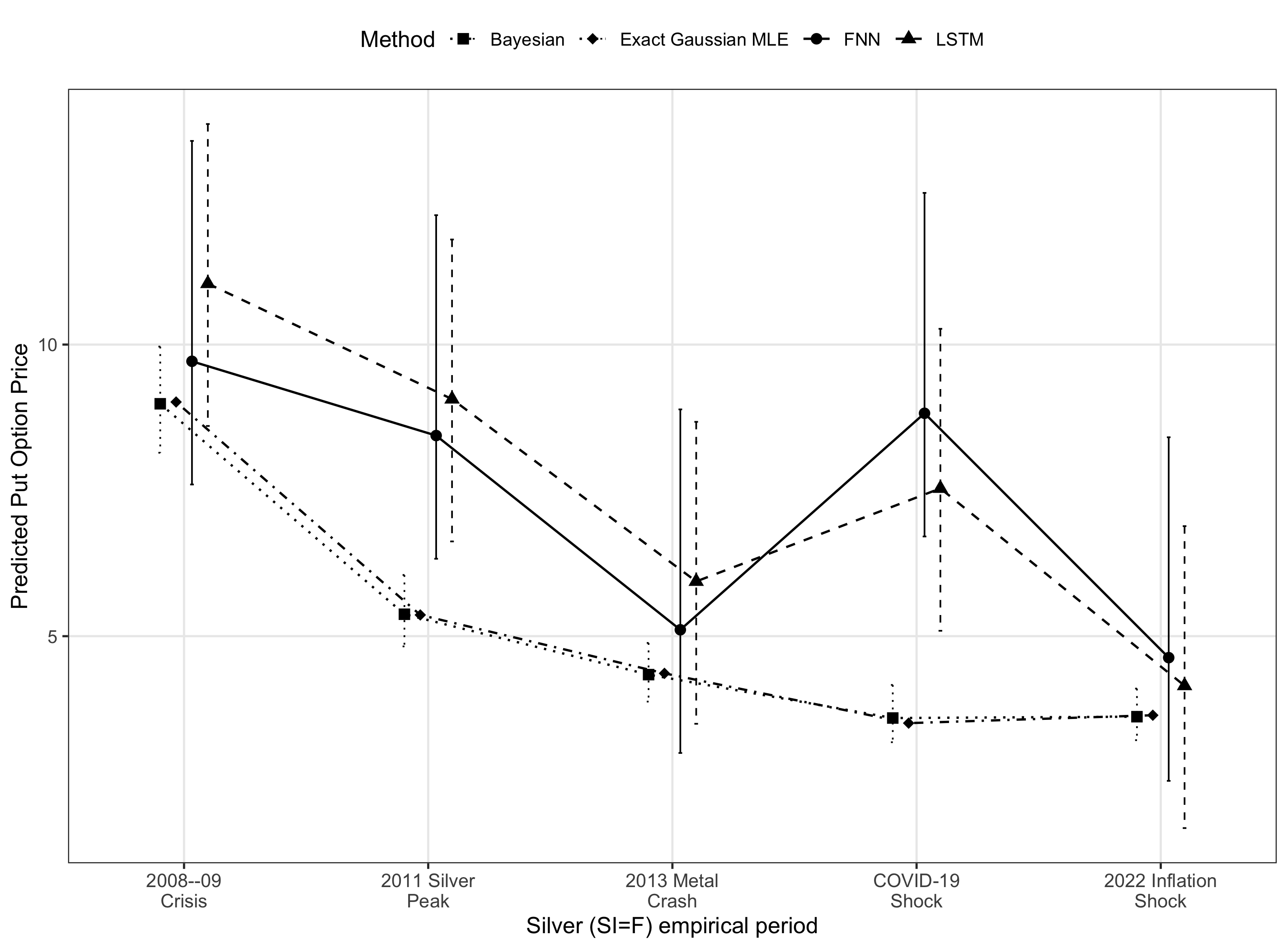}
\caption{Put option prices.}
\label{fig:silver_put}
\end{subfigure}

\caption{Comparison of fractional option-price predictions for silver
obtained from the Bayesian, Exact Gaussian MLE, FNN, and LSTM approaches
across five empirical market regimes.}
\label{fig:silver_option_prices}
\end{figure}

Table~\ref{tab:silver_option_comparison} and
Figure~\ref{fig:silver_option_prices} compare the corresponding
fractional call and put prices. As in the preceding empirical
applications, the Bayesian and Exact Gaussian MLE prices are very close
across all five periods. For example, during the 2008--09 Crisis, the
Bayesian and Exact MLE call prices are 12.185 and 12.216, respectively,
while the corresponding put prices are 8.985 and 9.015. Similarly,
during the 2022 Inflation Shock, the Bayesian and Exact MLE call prices
are 6.821 and 6.845, with corresponding put prices of 3.620 and 3.645.
This close agreement provides a consistent likelihood-based benchmark
for evaluating the neural-network predictions.

The neural-network pricing results show greater variation across
regimes. During the 2008--09 Crisis, the FNN call prediction is 12.914,
which is relatively close to the Bayesian value of 12.185, whereas the
LSTM predicts 14.246. The same pattern occurs for the corresponding put
prices, where the FNN prediction of 9.713 is closer to the Bayesian
value of 8.985 than the LSTM prediction of 11.045.

During the 2011 Silver Peak and 2013 Precious-Metal Crash, both neural
networks produce option prices above the Bayesian and Exact MLE
benchmarks. The FNN nevertheless remains closer to the likelihood-based
prices than the LSTM in both periods. For example, during the 2011
Silver Peak, the Bayesian call price is 8.577 and the Exact MLE price
is 8.565, compared with 11.641 for the FNN and 12.266 for the LSTM.

The largest differences arise during the COVID-19 Shock. The Bayesian
and Exact MLE call prices are 6.797 and 6.710, respectively, whereas
the FNN and LSTM predict 12.023 and 10.734. For the put option, the
corresponding values are 3.597, 3.509, 8.822, and 7.534. Although the
FNN reproduces the Bayesian Hurst estimate relatively closely in this
period, this agreement does not translate directly into equally close
option-price predictions. This distinction indicates that similarity
in the estimated dependence parameter alone is not sufficient to ensure
similar pricing predictions from the neural-network surrogate.

The discrepancy becomes considerably smaller during the 2022 Inflation
Shock. The Bayesian call price is 6.821 and the Exact MLE price is
6.845, compared with 7.833 for the FNN and 7.351 for the LSTM. The
corresponding put prices are 3.620, 3.645, 4.632, and 4.151. In this
period, the LSTM provides the neural-network prediction closest to the
likelihood-based benchmarks for both call and put prices.

Taken together, the silver results reinforce the distinction between
Hurst-parameter estimation and direct option-price prediction. The
neural network that most closely reproduces the Bayesian estimate of
$H$ in a particular period is not necessarily the network that produces
the closest option-price prediction. This is especially apparent during
the COVID-19 Shock, where the FNN is closer to the Bayesian Hurst
estimate but the LSTM is closer to the Bayesian option prices.

\subsubsection{Computational Efficiency for Silver}
An important advantage of the neural-network approaches is their
computational efficiency after training. Table~\ref{tab:silver_computation}
reports the median computational time per empirical window. Bayesian
estimation required approximately 82.930 seconds per window. Exact
Gaussian MLE was substantially faster, with a median computational time
of approximately 0.335 seconds per window.

\begin{table}[htbp]
\centering
\caption{Computational time comparison for silver (\texttt{SI=F}).}
\label{tab:silver_computation}
\begin{tabular}{lcc}
\hline
\textbf{Method}
& \textbf{Median Time per Window (s)}
& \textbf{Approximate Speedup} \\
\hline

Bayesian
& 82.930
& 1.0$\times$ \\

Exact Gaussian MLE
& 0.335
& 247.6$\times$ \\

FNN
& 0.0104
& 7974.0$\times$ \\

LSTM
& 0.0110
& 7539.1$\times$ \\

\hline
\end{tabular}
\end{table}

Once trained, the neural-network models require only a forward pass
through the fitted architecture. The median prediction time was
approximately 0.0104 seconds per window for the FNN and 0.0110 seconds
for the LSTM. Relative to Bayesian inference, these timings correspond
to speedups of approximately 7,974 and 7,539 times, respectively. The
Exact Gaussian MLE also provides a substantial computational improvement
over Bayesian estimation, but remains considerably slower than either
neural-network surrogate.

The timing results therefore emphasize the different roles of the four
approaches. Bayesian inference provides full posterior uncertainty
quantification but carries the greatest computational cost. Exact
Gaussian MLE provides a substantially faster model-based alternative.
The FNN and LSTM achieve the greatest computational advantage once
training has been completed, making them attractive for applications
requiring repeated predictions across many return sequences or market
periods.

\paragraph{Summary.}

The silver application provides additional evidence that the four
methods capture meaningful changes in dependence across empirical market
conditions. Persistence is relatively weak during the 2008--09 Crisis,
the 2013 Precious-Metal Crash, and the 2022 Inflation Shock, increases
during the 2011 Silver Peak, and becomes substantially stronger during
the COVID-19 Shock. All four methods identify this COVID-19 period as
the regime with the strongest persistence.

The neural-network results also demonstrate that agreement in Hurst
parameter estimation does not necessarily imply equally close
option-price predictions. The LSTM closely tracks the Bayesian Hurst
estimate in several of the lower-persistence regimes, while the FNN is
closer during the COVID-19 Shock. For option pricing, the relative
performance varies across periods, with particularly large deviations
from the likelihood-based benchmarks during the COVID-19 regime.
At the same time, both neural-network approaches reduce prediction time
by several orders of magnitude relative to Bayesian inference. These
results further support the use of the FNN and LSTM as computationally
efficient surrogate models, while Bayesian inference remains valuable
when full posterior uncertainty quantification is required.

\subsection{Al Rajhi Bank}

To further evaluate the empirical behavior of the proposed framework
within the Saudi equity market, Al Rajhi Bank (1120.SR) was analyzed
over five distinct market regimes: the 2014--2015 oil price decline,
the 2019--2020 COVID-19 shock, the 2021--2022 recovery and rate
transition, the 2023--2024 high interest rate period, and the
2025--2026 recent market period. For each regime, 252 daily log
returns were used as input to the Bayesian, Exact Gaussian MLE, FNN,
and LSTM approaches. The same option-pricing parameters used
throughout the empirical analysis were retained.

\subsubsection{Hurst Parameter Estimation}

\begin{figure}[htbp]
\centering
\includegraphics[width=0.92\textwidth]
{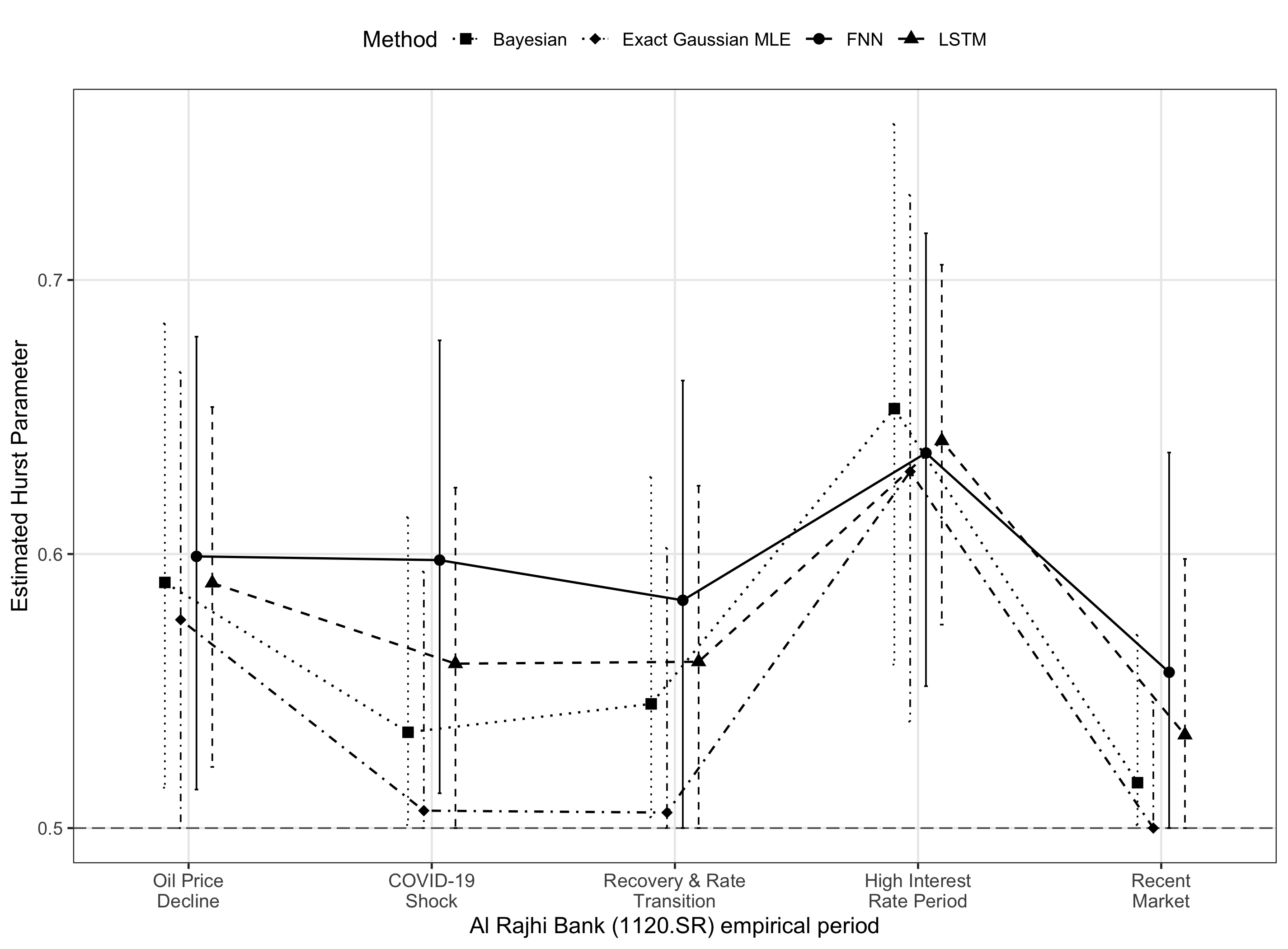}
\caption{Comparison of Hurst-parameter estimates for Al Rajhi Bank
(1120.SR) across five empirical market periods. Points represent the
Bayesian posterior median, Exact Gaussian MLE estimate, FNN prediction,
and LSTM prediction. Error bars represent Bayesian 95\% credible
intervals, Exact Gaussian MLE 95\% confidence intervals, and empirical
95\% prediction intervals for the neural-network models. The horizontal
dashed line denotes the short-memory boundary \(H=0.5\).}
\label{fig:alrajhi_h}
\end{figure}

Figure~\ref{fig:alrajhi_h} and
Table~\ref{tab:alrajhi_hurst} compare the Hurst-parameter estimates
across the four methods and five empirical market periods. The table
summarizes the point estimates, while the figure additionally displays
the corresponding uncertainty intervals. The results indicate
meaningful variation in the estimated degree of persistence across
market regimes.

\begin{table}[htbp]
\centering
\caption{Comparison of Hurst-parameter estimates for Al Rajhi Bank
(1120.SR) across five empirical market periods.}
\label{tab:alrajhi_hurst}
\begin{tabular}{lcccc}
\toprule
\textbf{Market Period}
& \textbf{Bayesian}
& \textbf{Exact Gaussian MLE}
& \textbf{FNN}
& \textbf{LSTM} \\
\midrule

\textbf{Oil Price Decline}
& \textbf{0.590} & \textbf{0.576}
& \textbf{0.599} & \textbf{0.589} \\

COVID-19 Shock
& 0.535 & 0.506 & 0.598 & 0.560 \\

Recovery \& Rate Transition
& 0.545 & 0.506 & 0.583 & 0.561 \\

\textbf{High Interest Rate Period}
& \textbf{0.653} & \textbf{0.630}
& \textbf{0.637} & \textbf{0.641} \\

Recent Market
& 0.517 & 0.500 & 0.557 & 0.534 \\

\bottomrule
\end{tabular}
\end{table}

The Bayesian posterior median of \(H\) is 0.590 during the Oil Price
Decline, suggesting moderately elevated persistence relative to the
benchmark \(H=0.5\). The estimate decreases to 0.535 during the
COVID-19 Shock and remains relatively close to 0.5 during the
Recovery and Rate Transition period, with a posterior median of
0.545. The most pronounced persistence is observed during the
High Interest Rate Period, when the Bayesian posterior median
increases to 0.653. Notably, the corresponding 95\% credible
interval, \([0.560,0.757]\), lies entirely above \(H=0.5\),
providing stronger evidence of persistent dependence under the
assumed fractional Gaussian noise model. In contrast, the Recent
Market period produces a posterior median of 0.517 with a 95\%
credible interval of \([0.501,0.571]\), indicating substantially
weaker estimated persistence.

The Exact Gaussian MLE estimates are 0.576, 0.506, 0.506, 0.630,
and 0.500 across the five periods, respectively. Although the MLE
estimates are generally lower than the corresponding Bayesian
posterior medians, both likelihood-based methods exhibit a
similar pattern across market regimes. In particular, the Oil
Price Decline shows moderately elevated Hurst estimates,
while the High Interest Rate Period produces the highest
estimates under both methods. The Exact Gaussian MLE of
0.630 during this period is substantially above \(H=0.5\),
supporting the indication of stronger persistence. During
the Recent Market period, the MLE reaches the lower
estimation boundary near \(H=0.5\), whereas the Bayesian
posterior median remains slightly above this benchmark.

The neural-network estimates generally reflect the same
regime-dependent behavior. The FNN estimates of \(H\) are
0.599, 0.598, 0.583, 0.637, and 0.557, while the
corresponding LSTM estimates are 0.589, 0.560, 0.561,
0.641, and 0.534. Both architectures produce elevated
estimates during the Oil Price Decline and their highest
estimates during the High Interest Rate Period. In the
latter regime, the FNN and LSTM estimates of 0.637 and
0.641 are substantially above \(H=0.5\) and closely
agree with the Bayesian posterior median of 0.653.
Across the five periods, the mean absolute difference
between the neural-network predictions and the Bayesian
posterior medians is approximately 0.033 for the FNN
and 0.014 for the LSTM. Thus, the LSTM demonstrates
closer overall agreement with the Bayesian Hurst estimates
across the Al Rajhi empirical windows.

The empirical residual-based prediction intervals from both
neural-network models overlap the Bayesian 95\% credible
intervals for \(H\) in all five periods. This overlap
provides a descriptive indication of agreement among
the methods but should not be interpreted as a formal
test of statistical equivalence.

Overall, the results suggest that the estimated persistence
in Al Rajhi Bank returns varies across market regimes
rather than remaining constant over time. The Oil Price
Decline exhibits moderately elevated Hurst estimates,
while the High Interest Rate Period stands out as the
regime with the strongest estimated persistence, with
all four methods producing values between 0.630 and
0.653. These findings are consistent with stronger
long-range dependence during this period under the
assumed fractional Gaussian noise model. In contrast,
the Recent Market period produces estimates much closer
to \(H=0.5\), suggesting weaker persistence. Among the
two neural-network architectures, the LSTM provides
the closest overall agreement with the Bayesian estimates.

\subsubsection{Option Price Predictions}

\begin{figure}[htbp]
\centering

\begin{subfigure}[t]{0.49\textwidth}
\centering
\includegraphics[width=\textwidth]
{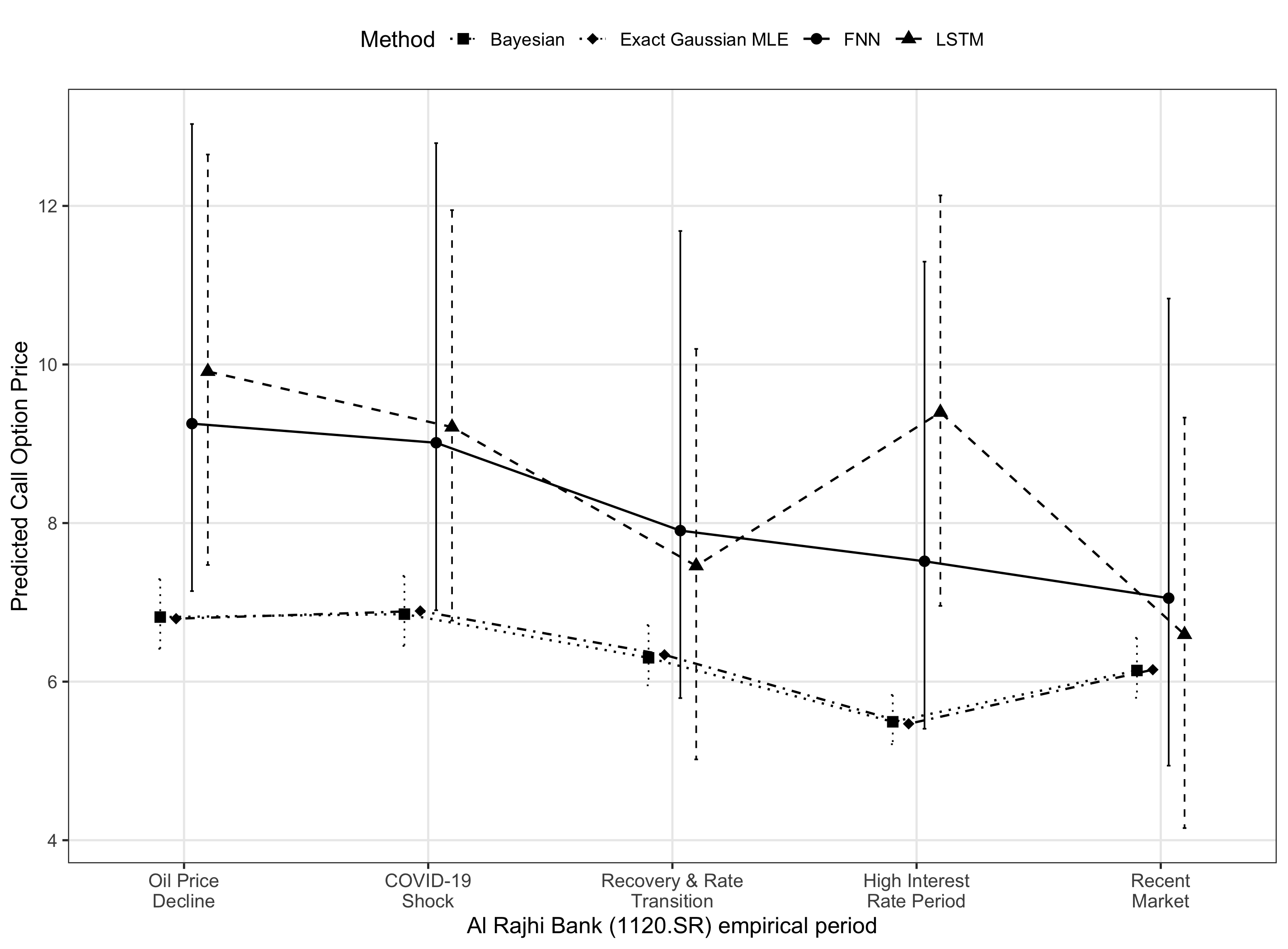}
\caption{Call option prices.}
\label{fig:alrajhi_call}
\end{subfigure}
\hfill
\begin{subfigure}[t]{0.49\textwidth}
\centering
\includegraphics[width=\textwidth]
{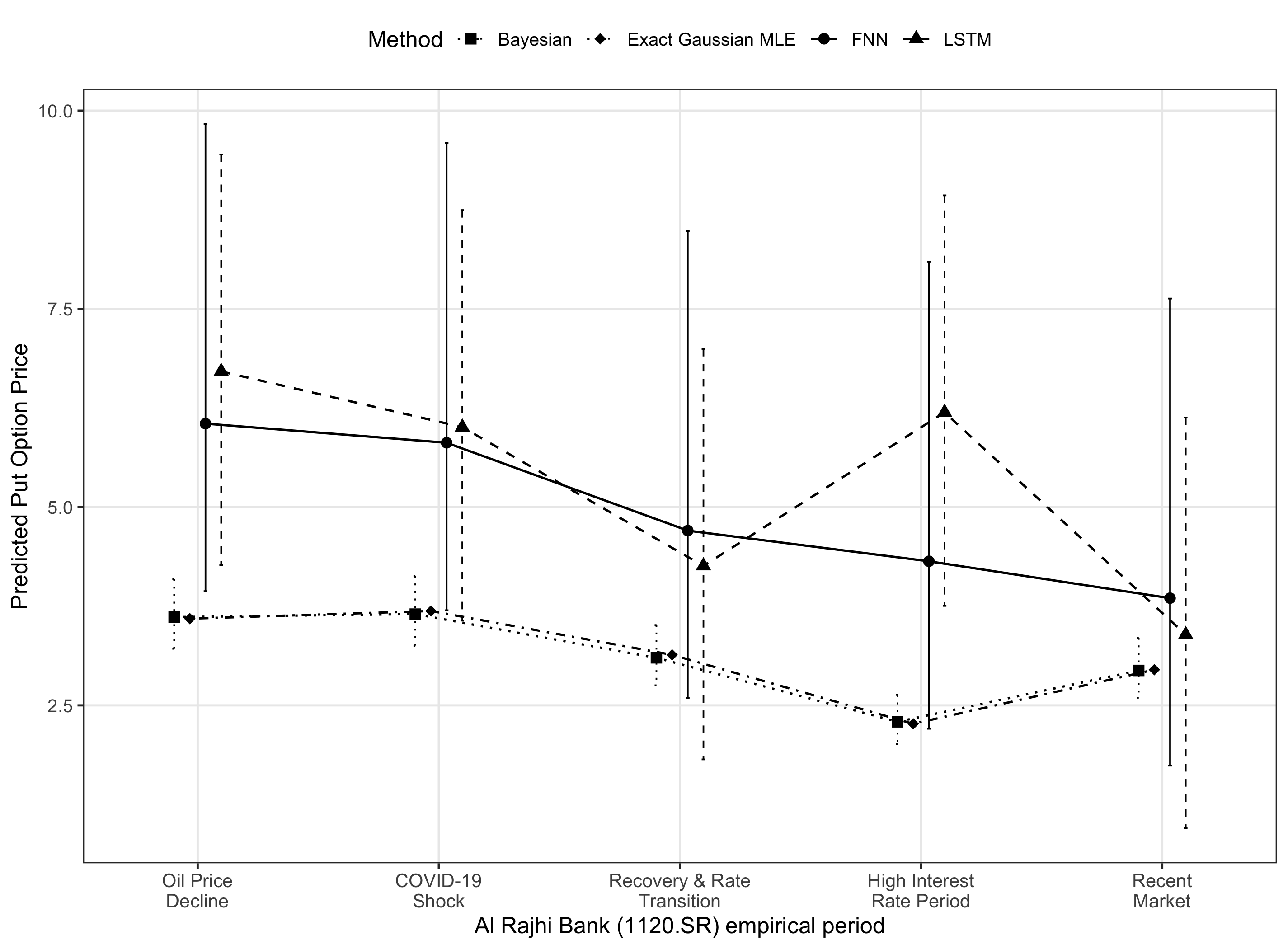}
\caption{Put option prices.}
\label{fig:alrajhi_put}
\end{subfigure}

\caption{Comparison of fractional option-price predictions for Al Rajhi
Bank (1120.SR) obtained from the Bayesian, Exact Gaussian MLE, FNN, and
LSTM approaches across five empirical market periods. Points represent
the corresponding point estimates or predictions, and error bars
represent the uncertainty intervals associated with each method.}
\label{fig:alrajhi_option_prices}
\end{figure}

Figures~\ref{fig:alrajhi_call} and~\ref{fig:alrajhi_put} compare the
call and put option prices obtained from the four approaches across the
five empirical market periods. The corresponding point predictions are
summarized in Table~\ref{tab:alrajhi_option_prices}, while the figures
additionally display the associated uncertainty intervals.

\begin{table}[htbp]
\centering
\caption{Comparison of predicted call and put option prices for
Al Rajhi Bank (1120.SR) across five empirical market periods.}
\label{tab:alrajhi_option_prices}
\resizebox{\textwidth}{!}{
\begin{tabular}{lrrrrrrrr}
\toprule
& \multicolumn{4}{c}{\textbf{Call Option Price}}
& \multicolumn{4}{c}{\textbf{Put Option Price}} \\
\cmidrule(lr){2-5}\cmidrule(lr){6-9}
\textbf{Period}
& \textbf{Bayesian}
& \textbf{MLE}
& \textbf{FNN}
& \textbf{LSTM}
& \textbf{Bayesian}
& \textbf{MLE}
& \textbf{FNN}
& \textbf{LSTM} \\
\midrule

Oil Price Decline
& 6.814 & 6.794 & 9.254 & 9.912
& 3.614 & 3.593 & 6.053 & 6.711 \\

COVID-19 Shock
& 6.851 & 6.891 & 9.013 & 9.210
& 3.651 & 3.690 & 5.812 & 6.010 \\

Recovery \& Rate Transition
& 6.301 & 6.338 & 7.905 & 7.460
& 3.100 & 3.137 & 4.704 & 4.259 \\

High Interest Rate Period
& 5.494 & 5.468 & 7.518 & 9.396
& 2.293 & 2.267 & 4.318 & 6.195 \\

Recent Market
& 6.141 & 6.151 & 7.053 & 6.594
& 2.940 & 2.950 & 3.852 & 3.393 \\

\bottomrule
\end{tabular}
}
\end{table}

The Bayesian and Exact Gaussian MLE approaches produce very similar
option-price estimates across all five periods. During the Oil Price
Decline, for example, the Bayesian median call price is 6.814 compared
with 6.794 under Exact Gaussian MLE. During the High Interest Rate
Period, the corresponding estimates are 5.494 and 5.468. Similar
agreement is observed for the put prices. Thus, despite their different
inferential frameworks, the Bayesian and Exact Gaussian MLE approaches
produce closely aligned point pricing implications in this application.

The neural-network models generally produce higher option-price
predictions than the Bayesian and Exact Gaussian MLE approaches.
During the Oil Price Decline, for example, the FNN and LSTM call
predictions are 9.254 and 9.912, respectively, compared with the
Bayesian median of 6.814. A similar pattern is observed during the
COVID-19 Shock, where the FNN and LSTM call predictions are 9.013 and
9.210, compared with the Bayesian value of 6.851.

The largest discrepancy is observed for the LSTM during the High
Interest Rate Period. The LSTM predicts a call price of 9.396 and a
put price of 6.195, compared with Bayesian medians of 5.494 and 2.293,
respectively. In contrast, the differences are considerably smaller
during the Recent Market period, where the LSTM call prediction of
6.594 is closer to the Bayesian value of 6.141.

Across the five periods, the mean absolute difference from the Bayesian
option-price median is approximately 1.829 for the FNN and 2.194 for
the LSTM for both call and put prices. Thus, although the LSTM provides
closer agreement with the Bayesian Hurst-parameter estimates, the FNN
provides closer agreement with the Bayesian option-price estimates on
average. This distinction indicates that close agreement in the
estimated dependence parameter does not necessarily imply equally close
agreement in option prices, because the pricing output also depends on
the volatility component and its interaction with the Hurst parameter.

The empirical residual-based prediction intervals provide additional information
beyond the point predictions. The FNN prediction intervals overlap the
Bayesian credible intervals for both call and put prices in all five
periods, whereas the LSTM prediction intervals overlap the corresponding
Bayesian intervals in three of the five periods. These interval
comparisons are interpreted descriptively as measures of agreement
among the methods rather than as formal tests of statistical
equivalence.

\subsubsection{Computational Efficiency for Al Rajhi Bank}

An important advantage of the neural-network approaches is their
computational efficiency after training. Table~\ref{tab:alrajhi_computation}
reports the median computational time per empirical window. Bayesian
estimation required approximately 80.355 seconds per window, reflecting
the computational cost of posterior simulation.

\begin{table}[htbp]
\centering
\caption{Computational time comparison for Al Rajhi Bank (1120.SR).}
\label{tab:alrajhi_computation}
\begin{tabular}{lcc}
\toprule
\textbf{Method}
& \textbf{Median Time per Window (s)}
& \textbf{Approximate Speedup} \\
\midrule

Bayesian
& 80.355
& 1.0$\times$ \\

Exact Gaussian MLE
& 0.231
& 347.9$\times$ \\

FNN
& 0.009
& 8928.3$\times$ \\

LSTM
& 0.010
& 7956.0$\times$ \\

\bottomrule
\end{tabular}
\end{table}

The Exact Gaussian MLE was substantially faster than Bayesian
inference, requiring a median of approximately 0.231 seconds per
window and corresponding to a speedup of approximately
\(347.9\times\). Once trained, the neural-network models require only
a forward pass through the fitted architecture. The median prediction
times were approximately 0.009 seconds per window for the FNN and
0.010 seconds for the LSTM, corresponding to speedups of approximately
\(8928.3\times\) and \(7956.0\times\), respectively, relative to
Bayesian inference.

These neural-network timings represent inference after model training
and therefore exclude the one-time computational cost required to train
the FNN and LSTM models. The timing comparison consequently reflects
the computational cost of generating predictions from already trained
networks, which is the relevant setting when the fitted models are
applied repeatedly to new return sequences.

The computational results emphasize the complementary roles of the
four approaches. Bayesian inference provides posterior distributions
and direct uncertainty quantification but incurs the greatest
computational cost. Exact Gaussian MLE provides a substantially faster
model-based alternative while retaining explicit parameter estimation.
The neural-network surrogates provide the lowest prediction times after
training, making them particularly attractive when option prices and
Hurst estimates must be generated repeatedly over many return sequences
or market periods.

\subsubsection{Overall Comparison}

The Al Rajhi Bank application highlights several important features of
the four-method comparison. First, the Bayesian and Exact Gaussian MLE
approaches produce closely aligned option-price estimates across the
five empirical periods, although their Hurst estimates differ more
noticeably in periods characterized by weaker persistence. Second, the
LSTM follows the Bayesian Hurst estimates more closely than the FNN,
with mean absolute differences of approximately 0.014 and 0.033,
respectively. Third, the FNN provides closer agreement with the Bayesian
option-price predictions on average, with a mean absolute difference of
approximately 1.829 compared with 2.194 for the LSTM.

These findings illustrate an important distinction between dependence
estimation and option-price prediction. Closer agreement in the Hurst
parameter does not necessarily imply closer agreement in option prices,
because the pricing output also depends on the volatility information
contained in the return sequence and its interaction with the Hurst
parameter. This distinction is particularly apparent during the High
Interest Rate Period, when the LSTM Hurst estimate is close to the
Bayesian posterior median but its option-price predictions differ more
substantially from the Bayesian values.

From a computational perspective, the neural-network models provide
near-instantaneous predictions once training has been completed.
However, the differences observed in several empirical periods indicate
that the FNN and LSTM are more appropriately interpreted as fast
predictive surrogates rather than exact replacements for
likelihood- or posterior-based inference. Bayesian inference retains
the additional advantage of providing a full posterior characterization
of parameter and pricing uncertainty, while the neural-network
prediction intervals quantify predictive uncertainty using the
empirical residual distributions obtained from validation data.

\subsection{Cross-Asset Synthesis}
\label{subsec:cross_asset_synthesis}

Taken together, the empirical applications reveal several recurring
patterns across the Saudi equity and commodity-market settings. First,
the estimated Hurst parameter varies across both assets and empirical
windows. Some periods produce larger estimates of \(H\), whereas
others produce estimates close to the short-memory boundary
\(H=0.5\). Under the assumed fractional Gaussian noise model, these
results indicate that the estimated degree of persistence is not
constant across the return windows considered in the empirical
analysis. However, variation in the estimated Hurst parameter does
not by itself establish the presence of genuine long-range dependence
in the underlying financial return processes, nor does it identify
the economic causes of the observed variation. The empirical results
should therefore be interpreted as evidence of variation in estimated
persistence under the assumed model rather than as evidence that
specific market events or economic conditions caused changes in
long-range dependence.

Second, the Bayesian and Exact Gaussian MLE approaches generally
produce similar qualitative patterns in the estimated Hurst parameter
and particularly close fractional option-price estimates. Across the
empirical applications, their call- and put-price predictions are
frequently nearly indistinguishable despite the different inferential
frameworks. This agreement provides a useful likelihood-based
comparison for the Bayesian point estimates. Differences between the
two approaches are more visible for the Hurst parameter in
lower-persistence periods, where the Exact Gaussian MLE frequently
approaches the lower estimation boundary near \(H=0.5\).

The neural-network results show that the relative performance of the
FNN and LSTM depends on both the asset and the prediction target.
The LSTM frequently follows variation in the Bayesian Hurst estimates
closely, particularly in several of the lower- and moderate-persistence
empirical windows. However, this agreement is not uniform across all
assets or periods. Similarly, neither neural-network
architecture consistently provides the closest option-price prediction.
The FNN is closer to the Bayesian pricing benchmark in several
applications and market regimes, whereas the LSTM is closer in others.

This distinction is important because close agreement in the Hurst
parameter does not necessarily imply close agreement in option prices.
The empirical applications contain several periods in which a neural
network closely reproduces the Bayesian estimate of \(H\) while
producing a noticeably different call or put price. Fractional option
prices depend jointly on the persistence and volatility information
contained in the return sequence. The Bayesian and Exact Gaussian MLE
approaches estimate these components explicitly before applying the
fractional Black--Scholes pricing equation, whereas the neural networks
learn the complete sequence-to-price mapping directly. Consequently,
agreement in one predicted quantity does not require equivalent
agreement in the others.

The empirical uncertainty intervals provide an additional perspective
on these point-prediction differences. In a number of empirical
windows, neural-network prediction intervals overlap the corresponding
Bayesian credible intervals even when the point predictions differ
noticeably. These overlaps indicate that some apparent differences
between the methods at the point-estimate level are smaller when
predictive and posterior uncertainty are considered. The interval
comparisons are descriptive, however, and should not be interpreted
as formal tests of statistical equivalence among the methods.

Finally, the most consistent practical advantage of the neural-network
approaches across all empirical applications is computational
efficiency. Bayesian inference requires posterior simulation for every
new return window and therefore has the largest per-window
computational cost. Exact Gaussian MLE is substantially faster while
retaining explicit model-based parameter estimation. Once training has
been completed, however, both the FNN and LSTM generate predictions
through a single forward pass and require only a small fraction of a
second per empirical window. Across the applications, this corresponds
to improvements of several orders of magnitude relative to Bayesian
inference.

Overall, the cross-asset evidence does not identify a single method
that dominates every dimension of the problem. Bayesian inference
provides full posterior uncertainty quantification, Exact Gaussian MLE
provides a substantially faster likelihood-based benchmark, and the
FNN and LSTM provide extremely rapid surrogate predictions after
training. Among the neural-network approaches, the LSTM often provides
close agreement with the model-based Hurst estimates, whereas the FNN
is competitive and in several regimes closer for option pricing.
The empirical results therefore support viewing the four approaches
as complementary: the statistical methods provide explicit
model-based inference, while the trained neural networks provide
computationally efficient approximations when repeated estimation and
pricing are required.

\section{Discussion, Conclusion, and Future Work}
\label{sec:discussion_conclusion}

This study investigated feedforward neural networks (FNN) and long
short-term memory (LSTM) networks for fractional option pricing and
compared their performance with Bayesian estimation and Exact Gaussian
maximum likelihood estimation (MLE). The simulation results showed that
increasing the training sample size generally improved the performance
of both neural-network architectures, although the gains became smaller
at the largest sample sizes. In the common four-method simulation, the
LSTM achieved the lowest Hurst-parameter RMSE of \(0.033\) and the
lowest option-price RMSE of \(1.305\). The FNN also provided competitive
predictions, while the Bayesian and Exact Gaussian MLE approaches
retained the advantage of explicit model-based parameter estimation and
uncertainty quantification.

The results also highlight a trade-off between predictive performance,
uncertainty quantification, and computational cost. Once trained, the
FNN and LSTM generated predictions through a single forward pass and
were several orders of magnitude faster than Bayesian estimation, while
Exact Gaussian MLE provided a substantially faster model-based
alternative. The uncertainty measures should nevertheless be interpreted
according to their respective frameworks. Bayesian credible intervals
represent posterior uncertainty, the MLE procedure uses likelihood-based
and bootstrap confidence intervals, and the neural-network intervals
are constructed from empirical validation residuals. Their empirical
coverage and width can therefore be compared in the controlled
simulation, but the intervals are not theoretically equivalent.

The empirical applications to Saudi Aramco, Ma'aden, STC, silver, and
Al Rajhi Bank further showed that the estimated Hurst parameter varies
across assets and market periods under the assumed fractional Gaussian
model. Bayesian estimation and Exact Gaussian MLE generally produced
closely aligned theoretical option prices, whereas the agreement of the
FNN and LSTM with the model-based approaches varied across assets and
periods. The results also demonstrate that close agreement in the Hurst
parameter does not necessarily imply close agreement in option prices,
since fractional option prices depend jointly on persistence and
volatility information. Because the empirical applications do not
provide known true Hurst parameters or observed option-price targets
corresponding to the standardized contracts considered here, these
results should be interpreted as comparisons among the methods rather
than as direct measures of empirical pricing accuracy.

Several limitations should be considered. The neural networks were
trained and tested using data generated from the same fractional model
family, with \(H\in[0.50,0.95]\), fixed sequences of \(252\) returns,
and fixed pricing parameters \(S_0=100\), \(K=98\), \(r=0.05\), and
\(T=90/365\). Performance outside these ranges or under alternative
return-generating processes is therefore not established. In addition,
the empirical analysis compares model-implied prices rather than
observed traded-option prices. The fractional Black--Scholes framework
also relies on assumptions that require careful consideration,
particularly with respect to arbitrage and market structure
\cite{Rogers1997,Cheridito2003,bjork2005note}.

Future research could extend the simulation to wider parameter ranges,
alternative sequence lengths, antipersistent processes with \(H<0.5\),
and models incorporating stochastic volatility, jumps, or time-varying
dependence. The neural networks could also include option-contract
characteristics such as moneyness, maturity, interest rates, and
dividend yields as additional inputs, allowing a more general pricing
relationship to be learned and eventually evaluated against observed
option-market data. Hybrid approaches that combine neural-network speed
with Bayesian or likelihood-based inference, together with more flexible
methods of uncertainty quantification, also provide useful directions
for future investigation.

Overall, the results demonstrate that deep-learning models can provide
computationally efficient approximations for fractional option pricing
within the framework considered in this study. The LSTM showed
particularly strong point-prediction performance in the controlled
simulation, while Bayesian and likelihood-based approaches remain
valuable when explicit parameter inference and model-based uncertainty
quantification are required. The four approaches therefore provide
complementary tools for fractional option pricing, with different
advantages depending on the objective of the analysis.

\end{document}